%% file: main.tex
\def\PaperFormat{}
\documentclass[11pt]{article}

\input{preamble}

\begin{document}

\maketitle

\begin{abstract}
\noindent
Generative AI can perform many of the tasks through which programmers build skill. When it substitutes for practice, skill development suffers; when it complements practice, learning may accelerate. Therefore, one programmer in the substitute mode and another in the complement mode can produce the same AI-assisted output yet differ in the skill they have acquired. We ask what digital practice records reveal about skill formation and whether AI-prohibited evaluations help distinguish the two modes of AI use. We follow 10,419 users of Codeforces (CF), an open, unproctored programming practice and contest platform with no enforced restriction on AI assistance, from 2018 to 2025. We link them to the rosters of the International Collegiate Programming Contest (ICPC) and the International Olympiad in Informatics (IOI), which admit entrants through qualification rounds and prohibit AI under in-person proctoring. From CF practice telemetry, we construct an AI-prompt signature (higher first-attempt acceptance rates, fewer attempts per solved problem, and fewer rapid debugging retries), consistent with AI-assisted practice. Three patterns emerge. Practice shifted toward the signature across entry cohorts spanning the releases of GitHub Copilot and ChatGPT. On CF, a stronger signature is associated with slower skill growth among users not linked to ICPC or IOI rosters, but not among roster-linked users. Among AI-era entrants who reached ICPC, larger within-user shifts toward the signature are positively associated with non-AI performance. These patterns are consistent with qualification screening on acquired skill, which admits users who complement practice with AI and excludes those who substitute AI for practice.
\end{abstract}

{\ifpaperformat\singlespacing\fi
\par}

\ifpaperformat\else\clearpage\fi

\section{Introduction}
\label{sec:intro}

Generative artificial intelligence (AI) now performs many of the tasks through which people build skill. In software development, professional education, and technical training, learners use AI in the very activities that once served as practice, and the records that organizations later evaluate, such as practice histories, portfolios, and assessment results, are produced with AI tools. Educators state the resulting problem plainly: a 2026 MIT committee on AI in teaching asks, ``When AI makes it possible to offload the cognitive work of learning, how can we assess what students actually know and understand?'' \citep{MITAICommittee2026}. Two streams of information systems (IS) research speak to this problem: how technology mediates learning and what happens when people delegate work to intelligent systems. This paper joins the two streams to address this question. We ask what digital practice records reveal about skill built with AI and whether AI-prohibited evaluations can measure what those records hide.

Those records can hide skill because the same AI tool can be used in two modes during learning. In the first mode, the learner delegates the task to the AI: the tool produces the code, the proof, or the analysis, and the learner's own practice effort is replaced. In the second, the learner practices with the AI: the tool explains, hints, and checks, and the effort is sustained. We call them the \emph{substitute} and \emph{complement} modes. The distinction draws on IS research on delegation to agentic artifacts \citep{BairdMaruping2021,FugenerEtAl2021Delegation} and on automation and augmentation as forms of task allocation \citep{FugenerWalznerGupta2025}, but it is a distinction about learning rather than task performance: delegating a task can still achieve learning through review and critique, and what matters here is whether the practice effort survives. Both modes produce polished output while the tool is present. That is the scaffold of our title: it holds up the output either way, and only when it comes off, at the moments when a person must perform without AI, does the difference in accumulated skill appear.

The mechanism has a precedent in a technology that was not generative. When robotic surgery arrived, procedures came under the control of the expert surgeon at the console, trainees' participation became less necessary, the approved ways of learning surgery stopped working, and some trainees developed alternative learning practices of their own \citep{Beane2018}. Generative AI moves the same displacement inside a single learner's practice session: the tool can do the task, and whether the learner still practices is a choice the output does not record.

We therefore ask two questions. First, in an open practice pool where both modes mix, how is a measure of AI-style practice associated with skill growth, and does that association differ between people who do and do not appear on the rosters of AI-prohibited evaluations? Second, among people on those rosters, how does a shift toward AI-style practice beforehand relate to their performance in the AI-prohibited evaluation?

These questions build on and contribute to the two IS research streams. The first is technology-mediated learning, the study of how information technology changes learning processes and outcomes \citep{AlaviLeidner2001}. It has shown that unguided access to a technology can lower learning outcomes \citep{BeloFerreiraTelang2013} while teacher-guided use raises them \citep{DengEtAl2025}, a finding our two modes build on. Like earlier studies in this stream \citep{SinghTanYoun2010,ZhangSinghGhose2019}, we work from platform traces: multi-year practice histories with AI access, which we relate to non-AI-aided performance. The second stream studies delegation to AI and its consequences for the human, such as how well people delegate and how AI advice changes their judgments \citep{FugenerEtAl2021Borgs,FugenerEtAl2021Delegation,JussupowEtAl2021}. These studies measure AI-assisted performance and the impact of AI on human knowledge while the tool is present, in short experimental sessions. In contrast, we measure performance where the tool is available and where it is prohibited, after years of practice.

Performance without the tool matters most at the skill frontier, the level of skill that only the top performers in a field reach. If AI is always available and reliable, routine skill loss looks benign; the argument breaks where the tool is weakest and where it is absent. On tasks beyond its capabilities, AI makes errors \citep{DellAcquaEtAl2026}, and catching the errors requires human expertise. Future training data also depend on whether we keep pushing the knowledge frontier \citep{AcemogluKongOzdaglar2026}. If substitute-mode use displaces the practice that builds this capability, the cost is invisible in AI-aided output but arrives years later. Such a cost has been measured in medicine: after AI-assisted colonoscopy was introduced, endoscopists' adenoma-detection rate in procedures without AI fell from 28.4\% to 22.4\% \citep{BudzynEtAl2025}.

At the skill frontier, elite competitive programming contests let us answer our two questions with data on the same people. Codeforces (CF) is an online programming practice and contest platform open to all. It has no qualification screen and no proctoring, and AI use is unrestricted in practice, so learners in both modes share the pool. The International Collegiate Programming Contest (ICPC) and the International Olympiad in Informatics (IOI) prohibit AI at in-person, proctored finals and admit entrants through qualification rounds that also prohibit AI. Our panel follows 10{,}419 CF users from 2018 through 2025, across the GitHub Copilot launch in 2021 and the ChatGPT release in 2022, and links a subset of them to ICPC and IOI rosters and scores. While we do not observe AI use directly, from practice submissions we construct an \emph{AI-prompt signature}: more problems solved on the first attempt, fewer attempts per solved problem, and fewer rapid debugging retries, the pattern a prompt-and-fix workflow with an AI assistant would leave \citep[see][]{ShihabEtAl2025}. We relate the signature to two outcomes that measure skill. On CF, results in rated contests set each user's rating, which measures skill relative to other users. Rating gains therefore track skill growth. At ICPC, the outcome is a team's score at the AI-prohibited finals.

Three patterns emerge. First, although not cleanly timed to the release of either GitHub Copilot or ChatGPT, a shift in practice toward the AI-prompt signature occurs across the entry cohorts spanning both AI tool releases. Second, among users in the open CF pool who are not linked to ICPC or IOI rosters, a one-standard-deviation stronger signature level after ChatGPT's release is associated with smaller rating gains of about eighteen points (roughly losing 6\% of their average gain during that period). Among roster-linked users, the association is smaller and statistically insignificant. This difference between the two groups grows with each later entry cohort. It is near zero among pre-AI entrants and largest among AI-era entrants. Third, within AI-prohibited ICPC, a larger shift toward the signature before the contest is associated with higher team scores among AI-era entrants. This association is positive on its own ($p = 0.055$) and larger than among pre-AI entrants ($p = 0.028$ for the difference), and the difference stays positive across robustness checks. One reading reconciles both results. Where the two modes mix, a stronger signature tracks the substitute mode. Among users who passed AI-prohibited qualification rounds, by contrast, a strengthening signature tracks the complement mode.

The results carry an implication for evaluation design. Where AI-assisted output is less diagnostic of skill \citep{CowgillHernandezLagosWright2026}, an AI-prohibited evaluation shows what practice records cannot: performance without the tool. The results also bear on two decisions that training providers and employers face. The first is how to integrate AI into learners' practice. IS evidence favors designs that induce the learner's own effort \citep{SanthanamSasidharanWebster2008} or give feedback rather than answers \citep{WambsganssEtAl2024}. The second is where to place AI-prohibited evaluations in training and hiring. For training, the MIT committee suggests unrestricted AI use on assignments in courses whose primary assessment prohibits AI \citep{MITAICommittee2026}. Our results are consistent with the same principle: where AI is allowed during practice, an AI-prohibited evaluation distinguishes the two modes of AI use.

Section~\ref{sec:lit} reviews the two IS research streams. Section~\ref{sec:setting} describes the setting and develops a framework with three predictions. Section~\ref{sec:data} describes the data, Section~\ref{sec:results} tests the predictions, and Section~\ref{sec:discussion} discusses the contributions, implications, and limitations.

\section{Theoretical background and related literature}
\label{sec:lit}

This section reviews the two IS research streams that we build on and contribute to, followed by two related literatures. The first stream, technology-mediated learning (Section~\ref{sec:lit:rcts}), supplies our outcome: the skill that forms through practice. The second, delegation to AI (Section~\ref{sec:lit:orgwork}), supplies the decision the learner faces: whether to delegate the task to AI or to practice with it. We add the setting in which the two modes can be compared: an AI-prohibited evaluation. Section~\ref{sec:lit:workplace} reviews the evidence on generative AI and productivity, which forms the backdrop to both streams. Section~\ref{sec:lit:selection} reviews research on evaluation and screening under AI, which informs the implications of our research.

\subsection{Technology-mediated learning and skill formation}
\label{sec:lit:rcts}
\label{sec:lit:obsolescence}
\label{sec:lit:chess}

Research on technology-mediated learning asks how technology shapes what people learn. We ask the same question about generative AI. \citet{AlaviLeidner2001} defined technology-mediated learning as an IS research area and called for deeper study of this question. In an early study of basic IT skills training, students in a web-based learning environment performed about as well as classroom students \citep{PiccoliAhmadIves2001}. A later study found that instruction encouraging self-regulated learning improved outcomes in e-learning-based IT training \citep{SanthanamSasidharanWebster2008}. Research on the IT workforce shows why IT skill matters: it is an investment \citep{BapnaEtAl2012,TambeYeCappelli2020} that loses value as technology changes \citep{BapnaEtAl2012,HortonTambe2025,DemingNoray2020}. These studies take for granted that training and on-the-job practice build the skill, and ask what the skill is worth. We ask a more basic question: whether the skill forms at all when AI can do the practice for the learner.

Like this paper, many studies in this stream measure learning from the digital traces that work leaves. Archival records of software work show learning curves. Specialized experience in a system raises individual productivity the most \citep{BohSlaughterEspinosa2007}, and task variety helps only up to a point \citep{NarayananBalasubramanianSwaminathan2009}. Six years of open-source project data reveal that developers learn most from peers \citep{SinghTanYoun2010}, and a 50-month panel of software-design contests shows that participants improve more when they compete against a superstar \citep{ZhangSinghGhose2019}. Coding contests and their traces are thus a familiar setting in IS research, and our signature comes from the same kind of trace: CF practice submissions. We extend this work by using such traces to ask whether the way people practice when AI is available predicts their non-AI-aided performance.

Studies in this stream show that a technology can help or hurt learning, depending on how it is used. Our two modes build on this finding. Unstructured access hurts. Across Portuguese middle schools, high broadband use lowered national exam grades \citep{BeloFerreiraTelang2013}, and in two large randomized trials in Chinese classrooms, unstructured smartphone use reduced performance \citep{DengEtAl2025}. Technology that supports the learner's own work helps. In the same trials, teacher-guided smartphone use raised performance \citep{DengEtAl2025}. A system that gives dynamic feedback rather than answers improved students' skills more than fixed scripted support or adaptive support did \citep{WambsganssEtAl2024}, and AI trainers for a strategic board game improved learning outcomes and narrowed a gender gap \citep{BaoHuangLin2024}. Unlike generative AI, none of these technologies can do the practice for the learner. In our terms, AI in the substitute mode replaces the learner's practice effort, and AI in the complement mode sustains it.

Randomized experiments provide the causal evidence for the two modes: AI access and the design of the assistance change what learners can later do without AI. When the tool does the cognitive work, learning is shallower. Students who used ChatGPT for a research task reported about 35\% less of the mental effort that builds understanding than students who used web search, and they produced worse work \citep{StadlerBannertSailer2024}. People who learned from large language model (LLM) summaries gained shallower knowledge than those who used web search, because the model did the synthesis for them \citep{MelumadYun2025}. How much is lost depends on the design of the assistance and on how the learner engages with it. In high school mathematics, access to a standard ChatGPT interface raised practice grades by 48\% but lowered grades on AI-free exams by 17\% \citep{BastaniEtAl2025}. In the same study, an AI tutor that gave hints raised practice grades without that loss in exam grades. Among developers learning a new Python library, AI lowered quiz scores on average \citep{ShenTamkin2026}. However, within the AI group, those who asked AI for explanations still learned the library, while those who delegated code generation or debugging scored far lower. Participants chose these patterns themselves, so this split is descriptive. In a three-month experiment with practicing patent lawyers, AI raised the quality of assisted drafting, but in a later task without AI only senior lawyers outperformed the control group \citep{AutorEtAl2026}. Junior lawyers showed no average gain, but compared with the control group their scores were more often either poor or good, which the authors read as AI accelerating skill acquisition among some junior lawyers and substituting for it among others. These experiments randomize AI access over sessions to months. We instead follow an elite population with career stakes through years of naturally occurring practice and link that practice to AI-prohibited evaluations.

The closest precedent to ours is chess. Training with chess computers raised players' Elo ratings, a measure of skill relative to other players, by roughly 11 points a year. The rating gains were mostly among lower-rated players, while the world-champion-level players' ratings stayed flat \citep{Gaessler2023}. Chess computers work like our complement mode: the computer replaces the opponent rather than the player's own training, and the ban on computers in contests keeps a hard boundary between training and evaluation. A coding assistant, by contrast, can deliver working code without the user grasping the algorithm. In our setting, the AI-prohibited contests restore the boundary between training and evaluation.

What we add to this stream is field evidence from practice records at platform scale. In the open CF pool, we ask whether a user's AI-prompt signature is associated with rating gains after ChatGPT's release. At ICPC, we ask whether a user's shift toward the signature is associated with the score in the AI-prohibited contest.

\subsection{Delegation to AI and its consequences for human capability}
\label{sec:lit:orgwork}

In this stream's terms, a learner who lets AI solve a practice problem delegates the task to AI. \citet{BairdMaruping2021} note that IS research has treated technology as a passive tool that people use. They argue that newer systems can take on tasks themselves, so people delegate tasks to them rather than simply using them. People often delegate poorly. AI judges what it knows and delegates to humans efficiently, whereas humans misjudge what they know and delegate knowledge work to AI poorly \citep{FugenerEtAl2021Delegation}. \citet{FugenerWalznerGupta2025} model two roles AI can take in a task: automation, in which AI replaces the human, and augmentation, in which AI works with the human. They show when each role pays off. This work asks how to divide tasks between people and AI to perform well now. We ask what delegating the practice task does to the skill a learner builds.

The studies closest to our question ask what delegation does to the person who delegates. People who defer to AI scrutinize less and bring less of their own knowledge to the task. Advice from AI makes people's answers converge \citep{FugenerEtAl2021Borgs}. Each person becomes more accurate, but people lose some of what they know that the AI does not, so groups that work with AI are far less effective than groups without it. Incorrect AI advice led to more wrong diagnoses than no advice at all, among novice physicians and experienced radiologists alike, often because they monitored their own reasoning or the system poorly \citep{JussupowEtAl2021}. Physicians reflected more critically when the system's output conflicted with their own assessment, and more shallowly when it agreed \citep{AbdelKarimEtAl2023}. Recruiters given a high-quality AI tool disengaged, deferred to the model, and stopped improving \citep{DellAcqua2022}. In preregistered experiments, people adopted AI suggestions with little scrutiny and reported higher confidence even after the AI erred, a pattern \citet{ShawNave2026} call cognitive surrender. How people see AI also shapes their effort. In a survey of 1{,}943 U.S. workers, those who reported that AI complements their own effort worked harder after its introduction, whereas those who reported that AI substitutes for their effort did not \citep{KimMesterSun2026}.

Other studies find that the effect of AI depends on how it is used. Our two modes draw the same distinction. In ad writing, an LLM that drafted the copy lowered the quality of experts' ads, whereas an LLM that gave feedback on the writer's own draft improved nonexperts' ads \citep{ChenChan2024}. In design work, generative AI raised the creativity of all users when used for brainstorming, but when used to execute the design it helped only novices and cost experts time without adding creativity \citep{HouEtAl2025}. \citet{WangLee2026} observe the delegation choice directly in 1.8 million prompts: when users can see what their use costs, they shift from letting the model search for a solution to guiding it with detailed prompts.

We add a further outcome to this stream: the skill a person acquires. We follow learners through years of practice where they could delegate to AI and measure how they later perform in evaluations without AI access.

\subsection{Backdrop: generative AI and productivity}
\label{sec:lit:workplace}

We take one premise from the productivity evidence: with the tool present, generative AI raises output. ChatGPT raised productivity on professional writing tasks and narrowed the gap between workers \citep{NoyZhang2023}. In a call center, a generative AI assistant raised productivity most among novice and lower-skilled agents \citep{BrynjolfssonLiRaymond2025}. On a task outside the capabilities of GPT-4, however, consultants using it were 19\% less likely to produce a correct solution \citep{DellAcquaEtAl2026}. In an experiment, developers with GitHub Copilot completed a programming task 55.8\% faster than developers without it \citep{PengEtAl2023}. Using Copilot also raised participation and code contributions in open-source projects \citep{SongAgarwalWen2026}. In three field experiments at large firms, an AI coding assistant raised the number of completed tasks, most for less experienced developers \citep{CuiEtAl2026}. AI also changes how programmers work. With Copilot, undergraduate computer science students moved from reading, understanding, and implementing code to prompting, viewing the response, and implementing it \citep{ShihabEtAl2025}. Our AI-prompt signature looks for the trace of this workflow. Access to ChatGPT led Stack Overflow users to give more answers, but shorter ones \citep{ShanQiu2025}. These studies run over sessions to months and mostly measure output with the tool present. The call-center study is an exception: during outages of the AI assistant, workers who had used AI still handled chats faster than before they had it \citep{BrynjolfssonLiRaymond2025}. We turn to multi-year practice histories and to performance where AI is prohibited by rule and by proctoring.

\subsection{Evaluation, signals, and screening under AI}
\label{sec:lit:selection}

Our study's implications concern where to use AI-prohibited evaluations, and two findings from research on evaluation under AI bear on that question. The first is that AI changes what signals of skill reveal. In signaling and screening theory, credentials are costly actions or tests that separate more able candidates from less able ones \citep{Spence1973,Stiglitz1975}. Records of practice can serve as such signals: in open-source projects, rank earned through a merit-based system is associated with wages up to 18\% higher \citep{HannRobertsSlaughter2013}. Such records can lose their value even without AI, such as through institutional change \citep{Yao2026Stranded}. Generative AI may exacerbate this loss, because it makes impressive signals cheaper to produce. Using experiments on hiring and start-up investing, \citet{CowgillHernandezLagosWright2026} show that screening accuracy dropped by 4 to 9\% due to applicants' access to ChatGPT. At the extreme, cheap AI-written applications destroy the signal \citep{GaldinSilbert2025}. By contrast, in a field experiment, nongenerative writing assistance raised jobseekers' hires and wages with no sign that employers were less satisfied \citep{WilesMunyikwaHorton2025}. Evaluation design can also help: when applicants can choose between an AI and a human evaluator, their choice restores the signal, because stronger and weaker applicants choose differently \citep{JabarianReshidi2025}. In short, an AI-prohibited evaluation as a screening design choice may restore a signal's value.

The second finding is that evaluation rules change how people work. After Stack Overflow banned AI-generated content, answers became longer and signaled more expertise, but participation fell \citep{BorwankarEtAl2026}. Creators who expected their AI use to be disclosed to a lay audience withdrew from the creative process and gave the AI more control, because they feared the audience would discount their own creativity \citep{JussupowEtAl2026}. Knowledge workers who know their work leaves trace data, and care how the data represent them, change their work to shape the data \citep{AaltonenStelmaszak2023}. The AI-prohibited evaluations we study are rules of this kind. They may therefore work in two ways, by screening learners or by changing how learners practice, and Section~\ref{sec:policy_levers} discusses both.

\section{Background and empirical setting}
\label{sec:setting}

\subsection{The AI shocks}

GitHub Copilot entered technical preview on June 29, 2021 and reached general availability one year later. It targeted programming directly and was among the first AI tools widely adopted by programmers. ChatGPT followed on November 30, 2022, putting generative AI into broad public use. Subsequent releases (e.g., GPT-4, Claude 2 and 3, DeepSeek-Coder) further extended AI's coding capability but did not introduce a comparably sharp adoption shock.

The two releases play different roles in our research design. Copilot was a programmer-specific tool and arrived more than a year before ChatGPT, so users who first appeared on CF after the Copilot release had already begun training with AI tools available to them. ChatGPT was a broader population-level event with a sharper date, observable across the entire panel; Copilot's reach was confined to its early adopters. We use the Copilot date to partition users by training vintage, and the ChatGPT date to define the within-user pre/post comparison; we elaborate on the motivation for this split in Section~\ref{sec:method:cohorts}.

\subsection{Competitive programming as the empirical setting}

CF is the dominant online competitive programming platform, running contests several times a week since 2010 and issuing individual Elo-style ratings, like chess, that measure each user's \emph{relative} skill based on contest performance. Anyone can create an account and practice on the platform. Practice submissions do not enter the rating system and serve as a training sandbox. The platform has no gate, our term for an AI-prohibited evaluation that a competitor must pass to advance, such as the ICPC and IOI qualifying rounds described below. Nor is practice proctored, and nothing prevents a user from running an AI tool while solving problems, so both modes of AI use mix in the practice pool. Rated contests are unproctored, and until September 14, 2024, no rule restricted AI use during contests. On that date, Codeforces introduced a rule prohibiting AI tools during contests, but it cannot verify that contestants comply and the rule is not enforced.\footnote{``Rule Restricting the Use of AI [revision 2024-09-14],'' Codeforces blog entry 133941, \texttt{codeforces.com/blog/entry/133941}; the rule applies to registered and out-of-competition participants of rated rounds and states that AI tools may be used freely outside such contests.} AI may also be a weaker substitute during contests: under time limits and a point penalty for incorrect submissions, contestants who lean on AI without the underlying skill tend to lose rating (see Section~\ref{sec:results}).

ICPC and IOI sit on the other end of the institutional spectrum. Both run multi-stage qualification rounds before the finals, and both finals are in-person, proctored, and AI-prohibited. The published rules enforce the prohibition through a closed environment rather than an AI-specific clause: they bar personal computing devices and printed materials other than a reference document, and they cut contestants off from outside help, through communication rules or a network ban.\footnote{ICPC Europe Championship 2026 Official Rules (updated March 4, 2026), \texttt{euc.icpc.global}; IOI Competition Rules (IOI 2022), \texttt{ioi2022.id}; both accessed August 26, 2026. Neither document names AI tools; the prohibition operates through the device, materials, network, and communication rules summarized here, under in-person proctoring.} ICPC builds teams of three from a single university, with qualification rounds generally held in person under rules that restrict outside assistance; IOI selects individual contestants through national olympiads whose formats and proctoring vary, and its international finals are held in person (online only in 2020 and 2021, due to COVID-19). We do not observe proctoring intensity at the earlier stages and treat screening strength as unobserved. However, users whose CF performance leans heavily on AI are unlikely to pass these qualification rounds, so the populations that reach the ICPC and IOI finals have been screened of the most AI-dependent practitioners. Medals in these high-stakes contests are recruiting signals cited on university and job applications, so the non-AI-aided scores measure frontier skill under real stakes.

On the CF platform, one further feature matters for our empirical strategy. Each submission carries a tag indicating whether it is in practice or contest mode. The behavioral signature of AI-style practice is cleanest in practice mode, where the data are also dense. Even though we do not observe a verified measure of AI use, the signature proxies for how far a user's practice has shifted toward an AI-style prompt-and-fix workflow \citep{ShihabEtAl2025}. The same signature computed on contest-mode submissions, noisier because contest behavior also reflects contest skill and time pressure, provides a robustness check against a roster-versus-skill confound (Appendix~\ref{app:section62_selection_robustness}): the cross-subpopulation pattern it shows is consistent with the practice-mode measure.

\subsection{Conceptual framework}
\label{sec:framework}

We introduce a simple framework that defines the two modes of AI use, formalizes why an AI-prohibited gate separates them, and yields the three testable predictions in Section~\ref{sec:predictions}.

Consider a learner who builds unaided frontier skill $z$ through practice. Practice effort $e$ and AI use $a$ produce that skill through a learning technology, $L$, in the human-capital tradition of \citet{BenPorath1967},
\begin{gather}
z = L(e,a), \\[3pt]
L_e > 0, \qquad L_{ea} > 0, \qquad L(0,a) = 0.
\end{gather}
Effort builds skill ($L_e>0$), and AI, used alongside practice, strictly complements it ($L_{ea}>0$, in the sense of \citealt{Hicks1970}; the complementarity central to the technology of skill formation, \citealt{CunhaHeckmanSchennach2010}): for a learner who exerts effort, more intensive AI use raises that practice's skill yield, so practicing more with AI builds strictly more skill, not less. Skill, however, comes only from the practice ($L(0,a)=0$): AI can deliver a task's output in place of the learner's own effort, but that output is not skill.

What separates learners is whether they invest effort in practice. We call practice that pairs AI with genuine effort a \emph{complement mode} and reliance on AI without effort a \emph{substitute mode}: under the same AI access, both learners get the same help with output, but the complement-mode learner accrues $z=L(e,a)$ with $e>0$, while the substitute-mode learner delegates and ends with $z=L(0,a)=0$. Throughout, skill \emph{erosion} refers to this shortfall: unaided skill that fails to form relative to the practice path, appearing as slower skill growth, not as decay of skill already formed. This gap is what our empirical strategy exploits. More AI use raises contemporaneous output in either mode, as AI route suggestions raise low-skilled taxi drivers' productivity without raising their skill \citep{KanazawaEtAl2025}, so the AI-prompt signature (proxying the magnitude of $a$) is a footprint of AI use that can reflect either mode, not a direct measure of skill. The gate, which reveals $z$, is what distinguishes the two modes.

The practice effort $e$ is costly at level $\kappa$; in the open, AI-permitted environment, leaning on AI saves $\kappa$, which is why it is attractive. The AI-prohibited gate, the qualification rounds that select the elite pool, changes the incentive: the gate scores only on realized $z$ by blocking AI, and it awards an elite credential value $V$ to entrants whose unaided skill clears a threshold, $z\ge\bar z$. Accordingly, a learner would invest the effort when the credential payoff covers its cost,
\begin{equation}
V\cdot\Pr(z\ge\bar z)-\kappa>0.
\end{equation}
Complement-mode learners pay $\kappa$, build $z$, and clear the gate; substitute-mode learners save $\kappa$ by leaning on AI, arrive with low $z$, and cannot pass. The CF contests have no such gate and are nearly free to enter, so both modes stay on the open CF platform.

The gate sorts learners by realized skill: it admits complement-mode learners with high $z$ and turns away substitute-mode learners with low $z$. How a learner came to be one type or the other is a separate question, and two explanations leave the same cross-sectional footprint. Under \emph{intrinsic traits}, the type is fixed and the gate sorts a distribution of types it did not shape. Under \emph{anticipation}, the type forms in response to the gate: a learner who foresees an AI-prohibited evaluation invests in practice, while one who does not substitutes AI for effort. Our data are consistent with either explanation, so they identify the sorting but not how the types arise. The two explanations differ in policy implications (Section~\ref{sec:policy_levers}).

This framework maps onto the two empirical margins of the signature that our design uses: the \emph{sorting margin}, the signature's level across users, and the \emph{adaptation margin}, its within-user change. The sorting margin proxies the magnitude of $a$: it shows how far practice has shifted toward AI reliance, but not which mode lies behind it. Among users who have passed the gate, an increase in $a$ raises $z$ by raising the skill yield of practice ($L_{ea}>0$). In the mixed pool, the same increase can equally be deepening substitution, so the framework predicts the direction of the link between the adaptation margin and skill \emph{only} among the screened. This reading assumes that complement-mode learners keep practicing with effort after they qualify, which is plausible because the finals, not the qualifying rounds, are the prize.

\subsection{Testable predictions}
\label{sec:predictions}

The framework of Section~\ref{sec:framework} yields three predictions, which we test in Section~\ref{sec:results}.

\textbf{Prediction 1 (universal practice shift).} After the AI releases, CF practice behavior shifts toward an AI-prompt signature (more first-attempt acceptances, fewer attempts per solved problem, and fewer rapid debugging retries) across the platform as a whole, and the shift grows with how much of a user's training postdates AI. Because AI raises output in either mode, the shift appears for substitute- and complement-users alike and does not by itself separate the user types.

\textbf{Prediction 2 (asymmetry in the unscreened pool).} In the open CF pool, where both user types coexist, a stronger AI-prompt signature, measured as its post-ChatGPT level (the sorting margin), predicts smaller rating gains among unscreened (non-rostered) users, a group that still contains substitute-users, but not among roster-linked users, whom AI-prohibited gates have screened. The signature-to-rating slope is therefore more negative in the unscreened subpopulation than in the screened one. Because the mechanism runs through modes of AI use, the asymmetry should also hold within the AI-era entry cohort alone, and the unscreened rating-gain shortfall should deepen across entry cohorts with growing AI exposure.

\textbf{Prediction 3 (positive slope inside the screened pool).} Inside the AI-prohibited ICPC environment, among users who entered competitive programming in the AI era, a larger within-user shift toward AI-intensive practice (the adaptation margin) predicts \emph{higher} non-AI-aided contest scores. Once the gate has screened substitute-users out, the remaining AI-style practice reflects complement use, which builds frontier skill rather than eroding it.

Predictions~2 and~3 together form the sharpest test. One screening mechanism predicts the direction of all three cells: a negative level gradient among the unscreened, a null one among the screened, and a positive change slope inside the AI-prohibited contest. A uniform mechanism does not explain all three. Pure erosion would put a negative gradient in both arms of the open pool and a negative change slope inside ICPC, while pure enhancement would produce no negative gradient anywhere.

All three predictions assume, rather than measure, that the signature is a proxy for AI use in either mode (Section~\ref{sec:results:patternA} and Appendix~\ref{sec:construct-validation} bound the leading alternative, that it proxies skill). Predictions~2 and~3 make this assumption testable: if the signature were not a proxy for AI use, nothing would produce the opposite signs on the two margins, the asymmetry within the AI-era entry cohort alone, or a shortfall that deepens across entry cohorts with AI exposure.

\section{Data}
\label{sec:data}

\subsection{Data sources and panel construction}
\label{sec:data:sample}

Table~\ref{tab:data_sources} lists our three data sources.\footnote{The analysis code and cleaned panels, together with the scoring-system specifications (CF Elo, ICPC team scoring, IOI individual scoring), are available from the author on request; a de-identified replication package will accompany the published article. Appendix~\ref{app:provenance} describes data provenance and research ethics.} The finals rosters of the two AI-prohibited contests, ICPC and IOI, are public. The contestants on these rosters anchor the CF panel. We build the CF panel from the CF API. We first link CF accounts to the ICPC and IOI rosters and collect the full rating and submission histories of these rostered contestants. We then add a non-rostered expansion.

\begin{table}[h!]
\centering
\caption{Data sources and roles.}\label{tab:data_sources}
\smallskip
\small
\begin{tabular}{L{3.7cm}L{3.5cm}L{3.6cm}L{3.7cm}}
\toprule
Source & Records & Contents & Role \\
\midrule
Competitive Programming Hall of Fame (\nolinkurl{cphof.org}) & 1{,}055 team-year records; 3{,}163 contestant-year records (2018--2025) & Team-level problems solved, ranks, and medals; team rosters & Outcome in an AI-prohibited contest; rosters for linking to CF accounts \\
\addlinespace[0.6em]
IOI results pages (\nolinkurl{stats.ioinformatics.org}) & 2{,}779 contestant-year records (2018--2025) & Individual-level points, ranks, and medals; country delegation rosters & Outcome in an AI-prohibited contest; rosters for linking to CF accounts \\
\addlinespace[0.6em]
CF API & 10{,}419 users; 793{,}920 user-month records; 2018-01 to 2026-04 & Rating histories, submission verdicts and timestamps, and profiles & The CF open pool panel; accounts for the roster linkage \\
\bottomrule
\end{tabular}

\vspace{0.5em}
\footnotesize\raggedright\textit{ICPC and IOI results are collected from the yearly pages \nolinkurl{cphof.org/standings/icpc/<year>} and \nolinkurl{stats.ioinformatics.org/results/<year>}.}
\end{table}

The analysis uses two unbalanced panels: the panel from the AI-prohibited ICPC and IOI contests, and the CF open pool panel. Appendix Table~\ref{tab:summary_stats} reports the sample sizes, cohort composition, and summary statistics of CF ratings for both panels. It also reports how many CF accounts are linked to the ICPC and IOI rosters.

The AI-prohibited panel comprises the ICPC and IOI contestants whose first rated CF contest falls between 2015 and 2024. ICPC contestants are linked to CF accounts through the CF usernames published on their Competitive Programming Hall of Fame profiles; IOI contestants are linked by name-and-country matching. Appendix~\ref{app:linkage} documents the linkage construction, its confidence tiers, and the full accounting from the linked accounts to the analysis panel.

The CF open pool panel contains 10{,}419 users: 2{,}480 roster-linked users and a non-rostered expansion of 7{,}939 users scraped directly from the CF API among active accounts with a maximum rating of at least 1{,}400 (the rating floor applies only to the non-rostered expansion). Of these, 6{,}399 (61.4\%) span both the pre-Copilot window (on or before 2021-06-28) and the post-ChatGPT window (on or after 2022-11-30) with at least 30 practice submissions on each side.\footnote{\label{ft-panel}The exclusions documented in Appendix~\ref{app:linkage} are based on linkage confidence, the date of a user's first rated contest, and whether the user has any practice-mode submissions. None of them is an outcome variable. The only rating-based criterion is the 1{,}400 floor on the non-rostered expansion. Imposing the same floor on roster-linked users strengthens the asymmetry of Section~\ref{sec:results:patternC} (Appendix Table~\ref{tab:c1_rating_floor}).}

Each user is assigned to a cohort by the date of their first rated CF contest. The split captures training vintage relative to the GitHub Copilot launch on June 29, 2021: pre\_AI users (first rated 2015 through 2018) trained before AI, transition users (2019 through June 28, 2021) were partway through training when AI arrived, and AI\_era users (June 29, 2021, through 2024) trained with AI available throughout. The cohort labels mark entry vintage relative to the two AI rollouts; they are not measures of individual AI exposure or use (Section~\ref{sec:results:patternA} tests, and does not support, a causal AI reading of the differences across cohorts). Appendix Table~\ref{tab:cohort_composition} reports each cohort's size, median CF rating, and median practice-submission counts, split by the pre- and post-ChatGPT windows. We observe the panel's CF practice and contest outcomes from 2018-01-01 to 2026-04-30. All estimation windows close at the end of 2025; the 2026 months enter only CF-career-total quantities, such as practice submissions per user and rated contests per user in Panel~C of Appendix Table~\ref{tab:summary_stats}.

\subsection{Outcome measurement: scoring system properties}

The CF rating is a modified, self-calibrating Elo-rating system inspired by chess \citep{Mirzayanov2015}: after each contest a competitor's rating moves according to how they ranked, with opponents' ratings entering the calculation directly, so the rating reflects field strength without additional controls. An inflation correction procedure also keeps the scale roughly stable across years. The rating is relative, measuring a contestant's standing in the pool rather than absolute skill, so it does not move when the whole pool improves by the same amount; our open-pool analysis (Section~\ref{sec:results:patternC}) therefore concerns relative performance, whether a stronger AI-prompt signature predicts losing ground against one's peers.

In contrast, ICPC and IOI provide no self-calibrating relative rating comparable to CF's: their relative measures (rank, medal) depend on the strength of the then-current contestant pool. We therefore use the absolute scores as outcomes, problems solved at the ICPC finals and total points earned at the IOI, which reflect a competitor's own performance on a fixed problem set. We then use contest-year fixed effects to absorb score variation caused by year-to-year changes in problem-set difficulty.

\FloatBarrier
\subsection{Descriptive statistics}
\label{sec:data:descriptives}

Table~\ref{tab:descriptives} reports descriptive statistics for the variables of the three analyses of Section~\ref{sec:results}, one panel each. In Panel A, the cohort differences in pre-to-post changes preview the practice-shift analysis of Section~\ref{sec:results:patternA}, which also defines the signature. In Panel B, the signature varies widely within each subpopulation, which is the variation the rating analysis of Section~\ref{sec:results:patternC} hinges on, and the two subpopulations differ sharply in baseline rating, which is why the analyses control for it. Appendix Figure~\ref{fig:cohort_signature} complements Panel A by tracing the three behaviors by cohort over the full window; its wide confidence intervals obscure any clear change at the AI releases, so the formal analysis of the shift is left to Section~\ref{sec:results:patternA}.

\begin{table}[!htbp]
\centering
\scriptsize
\renewcommand{\arraystretch}{0.92}
\caption{Descriptive statistics for the Section~\ref{sec:results} analyses. Each panel reports the variables of one analysis, sliced along the dimension that identifies its key coefficient: Panel~A by cohort $\times$ pre/post-ChatGPT period (the difference-in-differences (DiD) of Section~\ref{sec:results:patternA}), Panel~B by subpopulation (the cross-subpopulation contrast of Section~\ref{sec:results:patternC}), and Panel~C by cohort within the ICPC sample (the cohort contrast of Section~\ref{sec:results:patternB}). Cells are mean (standard deviation) unless noted.}
\label{tab:descriptives}

\begin{tabular}{l ccc cc}
\toprule
\multicolumn{6}{l}{\textit{Panel A. CF practice proxies by cohort $\times$ period (Section~\ref{sec:results:patternA}; user-quarter panel)}} \\
\midrule
 & Single-shot & Median & Quick-retry & Subs. / & $N$ \\
Cohort, period & accept rate & attempts & fraction & quarter & user-qtrs \\
\midrule
pre\_AI, pre      & $0.429$ $(0.269)$ & $1.85$ $(1.59)$ & $0.264$ $(0.193)$ & $61.6$ $(94.2)$  & $38{,}627$ \\
pre\_AI, post     & $0.450$ $(0.325)$ & $1.95$ $(2.07)$ & $0.228$ $(0.216)$ & $33.4$ $(67.7)$  & $13{,}387$ \\
\addlinespace
transition, pre   & $0.449$ $(0.248)$ & $1.73$ $(1.34)$ & $0.259$ $(0.178)$ & $88.6$ $(129.3)$ & $43{,}360$ \\
transition, post  & $0.461$ $(0.290)$ & $1.81$ $(1.78)$ & $0.247$ $(0.203)$ & $51.9$ $(87.5)$  & $32{,}751$ \\
\addlinespace
AI\_era, pre      & $0.414$ $(0.270)$ & $1.87$ $(1.45)$ & $0.264$ $(0.191)$ & $78.6$ $(129.4)$ & $6{,}558$  \\
AI\_era, post     & $0.456$ $(0.251)$ & $1.74$ $(1.43)$ & $0.272$ $(0.188)$ & $79.4$ $(112.2)$ & $14{,}780$ \\
\bottomrule
\end{tabular}

\vspace{0.8em}

\begin{tabular}{l ccc}
\toprule
\multicolumn{4}{l}{\textit{Panel B. Cross-subpopulation rating asymmetry (Section~\ref{sec:results:patternC}; $N = 6{,}690$ users)}} \\
\midrule
 & Non-rostered & Roster-linked & All \\
 & ($n = 5{,}283$) & ($n = 1{,}407$) & ($n = 6{,}690$) \\
\midrule
Rating change $\Delta r_h$ (post $-$ pre)        & $286.9$ $(272.4)$ & $350.2$ $(288.9)$ & $300.2$ $(277.1)$ \\
AI-prompt signature (post-ChatGPT PC1)                  & $-0.003$ $(1.447)$ & $-0.270$ $(1.567)$ & $-0.059$ $(1.477)$ \\
\quad p10 / p50 / p90                            & $-1.72$ / $0.11$ / $1.67$ & $-2.03$ / $-0.17$ / $1.43$ & $-1.79$ / $0.05$ / $1.62$ \\
Baseline rating $\bar r_h^{\text{pre}}$          & $1{,}377$ $(381)$ & $1{,}766$ $(451)$ & $1{,}459$ $(427)$ \\
\bottomrule
\end{tabular}

\vspace{0.8em}

\begin{tabular}{l ccc}
\toprule
\multicolumn{4}{l}{\textit{Panel C. Within-ICPC signature shift by cohort (Section~\ref{sec:results:patternB}; 579 obs, 444 users)}} \\
\midrule
 & pre\_AI & transition & AI\_era \\
\midrule
Signature shift (first-difference PC1)              & $-0.150$ $(1.324)$ & $-0.014$ $(1.868)$ & $0.161$ $(0.825)$ \\
\quad p10 / p50 / p90                            & $-1.55$ / $-0.09$ / $1.28$ & $-1.26$ / $0.12$ / $1.41$ & $-0.91$ / $0.19$ / $1.10$ \\
Baseline rating $\bar r_h^{\text{pre}}$          & $2{,}268$ $(372)$ & $1{,}946$ $(371)$ & $1{,}338$ $(337)$ \\
ICPC score (problems solved)                     & $5.9$ $(1.9)$ & $5.1$ $(2.0)$ & $4.8$ $(1.5)$ \\
$N$ observations                                  & $190$ & $309$ & $80$ \\
$N$ users                                         & $144$ & $233$ & $67$ \\
\bottomrule
\end{tabular}

\vspace{0.5em}
\scriptsize\raggedright\textit{Notes: Cohorts are defined in Section~\ref{sec:data:sample}. Panel~A cells are unadjusted cohort $\times$ period means over user-quarters; the estimated DiD (Table~\ref{tab:patternA_results}) adds user and quarter fixed effects and the $\log(1 + n_{h,t})$ submission control. ``pre'' is 2018-Q1 through 2022-Q3 and ``post'' is 2023-Q1 through 2025-Q4 (the 2022-Q4 boundary quarter is dropped); $N$ counts user-quarters in each cell on the single-shot support, and the estimation sample sizes (marginally smaller once fixed-effect singletons are dropped) are reported in Table~\ref{tab:patternA_results}. Panel~B reports the cross-subpopulation sample of Section~\ref{sec:results:patternC}; ``AI-prompt signature'' is \texttt{ai\_pc1\_post}, the post-ChatGPT first principal component of the three sign-aligned proxies. Panel~C reports the within-ICPC sample of Section~\ref{sec:results:patternB}; ``signature shift'' is the within-user first-difference principal component \texttt{shift\_pc1}. The IOI sister sample (Section~\ref{sec:results:patternB}; 345 obs, 243 contestants) has cohort signature-shift means of $0.16$, $-0.24$, and $-0.14$ and IOI-score means of $263$, $272$, and $227$ for pre\_AI, transition, and AI\_era respectively.}
\end{table}

\section{Estimation and results}
\label{sec:results}
\label{sec:method}
\label{sec:method:cohorts}

This section tests Predictions~1--3 of Section~\ref{sec:predictions} in turn. Section~\ref{sec:results:patternA} documents a cross-cohort shift in CF practice toward the AI-prompt signature. Section~\ref{sec:results:patternC} tests whether the signature predicts weaker rating gains in the open CF pool but not among roster-linked users. Section~\ref{sec:results:patternB} tests whether, inside the AI-prohibited ICPC, a shift toward the signature predicts higher scores for AI\_era entrants.

All three analyses share the cohort structure of Section~\ref{sec:data:sample}, with the pre\_AI cohort as the reference.

The three components of the signature (Section~\ref{sec:results:patternA}) enter the later analyses through two composites, each a first principal component: \texttt{ai\_pc1\_post}, the post-ChatGPT \emph{level} of the signature, compares users where the modes mix (Section~\ref{sec:results:patternC}), and \texttt{shift\_pc1}, the pre-to-post \emph{change} in a user's signature, tracks users' own shifts among the screened (Section~\ref{sec:results:patternB}). The level and the change are the sorting and adaptation margins of the framework (Section~\ref{sec:framework}).

\subsection{The behavior shift toward an AI-prompt signature in CF practice}
\label{sec:results:patternA}
\label{sec:method:a1}

We first evaluate how CF practice behavior shifts after the AI releases, focusing on three patterns that together form an AI-prompt signature (summary statistics in Table~\ref{tab:descriptives}, Panel A): \texttt{frac\_single\_shot}, the fraction of problems solved on the first attempt; \texttt{median\_attempts}, the median number of attempts to solve a problem; and \texttt{frac\_quick\_retry}, the fraction of quick retries after a failed submission, where a quick retry is a reattempt within thirty minutes of the incorrect submission.

Solving a problem without AI is iterative: the user submits, reads the verdict, locates the bug, and resubmits, often several times before the solution is accepted as correct (``AC,'' the platform's Accepted verdict). A user who relies on AI can instead submit a generated solution that often clears the problem on the first try, consistent with the workflow shift documented by \citet{ShihabEtAl2025}. If users rely on AI during practice, all three patterns should move in the AI-prompt direction. The fall in quick retries is the least obvious of the three patterns. It has two causes: more first submissions succeed and leave nothing to retry, and when an AI-generated solution fails, the user often cannot quickly debug code they did not write.

Throughout, this signature is a behavioral proxy inferred from submission timing, consistent with AI-style practice but not a direct, verified measure of AI use. Appendix~\ref{sec:construct-validation} reports construct-validation tests that bound the leading alternative, that the signature merely proxies programmer skill. Before AI was available, higher-rated (more skilled) users were less AI-like on these behaviors, not more (the correlation between CF rating and each signature component is weak, $|r| < 0.20$, and runs opposite to the AI direction), so skill does not produce the signature. As a user's own rating rises, the signature moves an order of magnitude too little for skill gains alone to produce the cohort shift, and established users did not change their practice patterns at the ChatGPT release, so the cohort shift reflects who entered the platform rather than a platform-wide trend.

To assess shifts in the three behavioral patterns after AI releases, we consider a difference-in-differences (DiD) design on the user-quarter CF practice panel. For user $h$ in quarter $t$:
\begin{equation}
\begin{split}
y_{h,t} ={}& \alpha_h + \tau_t + \beta_1 \cdot \text{post}_t \cdot \mathbbm{1}[\text{cohort}_h = \text{transition}] \\
&+ \beta_2 \cdot \text{post}_t \cdot \mathbbm{1}[\text{cohort}_h = \text{AI\_era}] + \lambda \cdot \log(1 + n_{h,t}) + \varepsilon_{h,t},
\end{split}
\label{eq:a1_did}
\end{equation}
where $y_{h,t}$ is one of the three patterns, $\text{post}_t = \mathbbm{1}[t \geq \text{2023-Q1}]$ is the post-ChatGPT indicator (ChatGPT's release quarter, 2022-Q4, is dropped from the panel), $\alpha_h$ and $\tau_t$ are user and quarter fixed effects, respectively. $n_{h,t}$ is the user's submission count in the quarter. We control for it because practice volume differs across cohorts and over time and all three patterns move with it (Table~\ref{tab:patternA_results}). Standard errors are clustered at the user level. The coefficients of interest are $\beta_1$ on the transition cohort and $\beta_2$ on the AI\_era cohort, both relative to the pre\_AI reference cohort.

The sample is the 10{,}419-user CF panel of Section~\ref{sec:data:sample}, restricted to user-quarter observations on practice-mode submissions from 2018-Q1 through 2025-Q4. Table~\ref{tab:patternA_results} reports the observation counts by outcome.

The two AI events anchor the design in different ways: the Copilot launch (2021-06-29) sets the cohort split (see Section~\ref{sec:data:sample}), and the ChatGPT release (2022-11-30) sets the within-user pre/post split. Anchoring both at Copilot would leave AI\_era users with essentially no pre-period: they enter rated competition after Copilot by construction, conflating their novice-to-competitor transition with AI exposure. Anchoring the time series at ChatGPT instead gives AI\_era users who entered before ChatGPT a clean pre-period of up to five quarters (2021-Q3 through 2022-Q3), observed after they entered rated competition. Re-anchoring the split at the Copilot launch, which is identifiable only for the pre\_AI and transition cohorts, finds the transition cohort already shifted on all three proxies (Appendix~\ref{app:copilot_did}), so the headline does not depend on the ChatGPT anchor.

The identification of $\beta_1$ and $\beta_2$ comes from each user's own change in a practice pattern from before to after the ChatGPT release, averaged within each cohort and then compared across cohorts. User fixed effects absorb time-invariant user characteristics (e.g., skill baseline, education, country), and quarter fixed effects absorb calendar-time shocks common to all users, such as platform policy changes and seasonal effects. Because cohort assignment reflects platform-entry timing rather than observed AI use, the specification does not identify a causal effect of AI; we read the coefficients as a descriptive cohort contrast.

\paragraph{Results.} The AI\_era cohort $\times$ post interaction ($\beta_2$) is significant on all three outcomes (Table~\ref{tab:patternA_results}). Each $\beta_2$ is a DiD term: the AI\_era cohort's pre-to-post change in a specific outcome measured against the pre\_AI cohort. All three changes move toward the AI-prompt signature: single-shot acceptances rise ($\beta_2 = 0.0586$, SE $0.0063$), median attempts fall ($-0.3271$, SE $0.0374$; about one-sixth of the $1.9$-attempt baseline), and quick retries fall ($-0.0190$, SE $0.0041$), all at $p < 0.001$.

\begin{table}[h!]
\centering
\small
\caption{AI-prompt signature shifts in CF practice patterns (Equation~\ref{eq:a1_did}).*}
\label{tab:patternA_results}
\begin{tabular}{lccc}
\toprule
                            & (1)                       & (2)                       & (3)                          \\
Dependent variable          & \texttt{frac\_single\_shot} & \texttt{median\_attempts} & \texttt{frac\_quick\_retry}  \\
\midrule
post $\times$ transition ($\beta_1$) & $0.0139^{**}$   & $-0.0680^{*}$             & $0.0090^{**}$               \\
                            & $(0.0046)$                & $(0.0297)$                & $(0.0029)$                   \\
post $\times$ AI\_era ($\beta_2$)     & $\mathbf{0.0586}^{***}$ & $\mathbf{-0.3271}^{***}$  & $\mathbf{-0.0190}^{***}$     \\
                            & $(0.0063)$                & $(0.0374)$                & $(0.0041)$                   \\
$\log(1 + n_{h,t})$ ($\lambda$) & $-0.0045^{***}$        & $0.0120^{**}$            & $0.0587^{***}$              \\
                            & $(0.0007)$                & $(0.0037)$                & $(0.0005)$                   \\
\midrule
User FE                     & Yes                       & Yes                       & Yes                          \\
Quarter FE                  & Yes                       & Yes                       & Yes                          \\
Observations**                & $149{,}398$               & $142{,}774$               & $149{,}441$                  \\
Users                       & $10{,}419$                & $10{,}401$                & $10{,}419$                   \\
Quarters                    & $31$                      & $31$                      & $31$                         \\
\bottomrule
\multicolumn{4}{l}{\footnotesize $^{*}\,p<0.05$, $^{**}\,p<0.01$, $^{***}\,p<0.001$.} \\
\multicolumn{4}{p{0.95\textwidth}}{\footnotesize \textit{Notes: *Sample is the CF user-quarter panel, 2018-Q1 through 2025-Q4 excluding 2022-Q4. Reference cohort: pre\_AI; the post main effect is absorbed by quarter fixed effects. Standard errors clustered at the user level in parentheses. Without the $\log(1+n_{h,t})$ control, the AI\_era coefficient on \texttt{frac\_quick\_retry} is $+0.018$; the other two coefficients keep their signs.}} \\
\multicolumn{4}{p{0.95\textwidth}}{\footnotesize \textit{**Sample sizes and user counts differ across columns because the three outcomes have different definitional supports: \texttt{median\_attempts} is undefined for user-quarters with no solved problems (dropping $\sim$6{,}600 user-quarters and 18 users); the small gap between \texttt{frac\_single\_shot} and \texttt{frac\_quick\_retry} reflects different denominators (problems attempted versus total submissions). Each regression uses the natural sample for its outcome.}}
\end{tabular}
\end{table}

The transition cohort lies between the other two, closer to the pre\_AI cohort: on single-shot submissions and median attempts, $\beta_1$ is roughly a quarter of the corresponding AI\_era effect (Table~\ref{tab:patternA_results}), so the gradient from pre\_AI to transition to AI\_era is monotonic, consistent with the signature strengthening with how much of a user's training occurred after AI arrived. For quick retries, the transition coefficient runs the opposite way ($0.0090$, SE $0.0029$, $p = 0.002$), possibly because users who adopted AI mid-career kept debugging its output by hand. The shift appears in both subpopulations: re-estimating the \texttt{frac\_single\_shot} DiD separately, the AI\_era coefficient is $0.0356$ ($p = 0.004$) among roster-linked users and $0.0617$ ($p < 0.001$) among non-rostered users.

\paragraph{The cohort contrast is not cleanly timed to AI.} Three patterns support the descriptive rather than causal reading of the cohort gradient. First, parallel pre-trends fail: quarter-by-cohort interactions (Appendix Figure~\ref{fig:cohort_event_study}) show neither cohort moving discontinuously at the ChatGPT release, and for the transition cohort a joint test rejects parallel pre-trends on all three patterns ($\chi^2(18) = 96.7$, $76.8$, and $84.6$; all $p < 0.001$). Second, at matched career stage the transition and AI\_era cohorts are statistically indistinguishable (single-shot difference $0.0002$, $p = 0.95$), even though the transition cohort entered before either AI release. Third, the gap is largest on the problems AI handles worst, not on the easy, templated problems where AI is likeliest to supply a correct first attempt: holding career stage fixed, the AI\_era-versus-pre\_AI single-shot difference is null on the easiest problems ($+0.002$, $p = 0.63$) and peaks at $+0.039$ ($p < 0.001$) on the 2{,}000--2{,}399 difficulty band. A gradual trend in practice norms unrelated to AI, together with novice-to-competitor maturation, could fit these facts as well as an AI shock, so we carry the practice shift forward as a descriptive pattern that AI alone does not explain (full specifications in the online replication package).

The shift is descriptive, but it is the premise for what follows: the AI-prompt signature is now a common feature of practice in this population. Practice patterns alone do not show whether the shift erodes or builds skill. Sections~\ref{sec:results:patternC} and~\ref{sec:results:patternB} ask that question in the CF open pool and inside the AI-prohibited ICPC.

\FloatBarrier
\subsection{A cross-subpopulation CF rating asymmetry}
\label{sec:results:patternC}
\label{sec:method:c1}

We assess CF rating changes under the AI-prompt shift discussed above, comparing two subpopulations: roster-linked users who participate in AI-prohibited ICPC and IOI contests, and non-rostered users who do not (we use subpopulations and arms interchangeably). Prediction~2 implies the asymmetry to test: the qualifiers screen substitute-users out of the roster-linked pool, so the AI-prompt signature should predict weaker rating gains only in the non-rostered pool, where substitute-users remain.

We implement this analysis in two ways: two subpopulation regressions and one pooled regression. 
\begin{itemize}
\item The subpopulation version fits two parallel cross-sectional regressions, one per subpopulation $s \in \{\text{non-rostered}, \text{rostered}\}$:
\begin{equation}
\Delta r_h = \alpha_s + \theta_s \cdot \text{ai\_pc1\_post}_h + \gamma_s \cdot \bar r_h^{\text{pre}} + \varepsilon_h,
\label{eq:c1_subset}
\end{equation}
with robust standard errors. The outcome $\Delta r_h$ is user $h$'s mean CF rating after ChatGPT minus their mean rating before, computed from the monthly rating panel as $\bar r_h^{\text{post}} - \bar r_h^{\text{pre}}$. ChatGPT was released on 2022-11-30, so the post-ChatGPT window covers months $t \in$ [2022-12, 2025-12] and the pre window covers $t \in$ [2018-01, 2022-12). The covariate $\bar r_h^{\text{pre}}$ controls for the user's baseline skill level.
\smallskip

The coefficient $\theta_s$ of \texttt{ai\_pc1\_post} is the coefficient of interest, measuring how the signature predicts the subpopulation's rating change. The covariate \texttt{ai\_pc1\_post} is the user's post-ChatGPT AI-prompt signature score: the first principal component of the three sign-aligned proxies of Section~\ref{sec:method:a1}, oriented so that all three contribute positively to the AI-prompt direction (bootstrap-validated principal component analysis (PCA) loadings $0.614$, $0.535$, and $0.580$; the three proxies are only moderately correlated, with pairwise $r$ between $0.45$ and $0.68$).

\item The pooled fully stratified version runs both subpopulations through a single regression with an indicator $\text{NR}_h$ for non-rostered users:
\begin{equation}
\begin{split}
\Delta r_h ={}& \alpha_0 + \alpha_1 \cdot \text{NR}_h + \theta_0 \cdot \text{ai\_pc1\_post}_h + \theta_1 \cdot \text{ai\_pc1\_post}_h \cdot \text{NR}_h \\
&{}+ \gamma_0 \cdot \bar r_h^{\text{pre}} + \gamma_1 \cdot \bar r_h^{\text{pre}} \cdot \text{NR}_h + \varepsilon_h.
\end{split}
\label{eq:c1_pooled}
\end{equation}
The interaction term coefficient $\theta_1$ is the coefficient of interest, measuring the cross-subpopulation asymmetry in rating changes.
\end{itemize}

The sample consists of users who remain active in the post-ChatGPT window: a user enters if they have at least two quarters between 2023-Q1 and 2025-Q4 with at least five practice submissions in each. After dropping users with missing $\Delta r_h$ or $\bar r_h^{\text{pre}}$, the analysis sample contains 6{,}690 users (5{,}283 non-rostered and 1{,}407 roster-linked).

Within a subpopulation, $\theta_s$ (Equation~\ref{eq:c1_subset}) compares users who face the same gate (or no gate). Across subpopulations, $\theta_1$ (Equation~\ref{eq:c1_pooled}) holds the signature fixed and varies the screening environment. The asymmetry test does not require roster status to be randomly assigned: it relies on the maintained assumption that the qualifying rounds have screened substitute-users out of the roster-linked pool, which the within-ICPC test of Section~\ref{sec:results:patternB} probes. The baseline control $\bar r_h^{\text{pre}}$ blocks omitted-variable bias, because high-signature users are on average lower-skill and $\bar r_h^{\text{pre}}$ independently predicts $\Delta r_h$ (coefficients near $-0.33$ in both arms, $p < 0.0001$). The specification still does not identify the \emph{causal} effect of AI on rating change, since high-signature users may differ on unobservables.

\paragraph{Results.} The within-subpopulation slopes on \texttt{ai\_pc1\_post} are starkly asymmetric (Table~\ref{tab:patternC_results}). The non-rostered slope is $-12.24$ rating points (SE $2.45$, $p < 0.0001$). The roster-linked slope is statistically null: $-2.41$ (SE $4.33$, $p = 0.58$). The interaction coefficient $\theta_1$ of the pooled regression is $-9.82$ (SE $4.98$, $p = 0.048$): the AI-prompt signature predicts around ten points weaker rating gains per signature unit in the unscreened pool than in the screened roster-linked pool. One standard deviation of \texttt{ai\_pc1\_post} is $1.48$ on the estimation sample. Accordingly, a one-SD-stronger AI-prompt signature is associated with an $18$-point lower rating gain among non-rostered users ($-12.24\times1.48$), roughly $6\%$ of the $300$-point mean post-ChatGPT gain across all users (Panel B of Table~\ref{tab:descriptives}), and with a $14.5$-point lower rating gain in the unscreened pool relative to the screened pool ($-9.82\times1.48$).

\begin{table}[h!]
\centering
\small
\caption{Cross-subpopulation CF rating asymmetry on \texttt{ai\_pc1\_post}. Sample: $N = 6{,}690$ users active in the post-ChatGPT window (5{,}283 non-rostered + 1{,}407 roster-linked). Robust standard errors. The fully stratified pooled specification tests whether $\theta_{\text{non\_rostered}}$ and $\theta_{\text{rostered}}$ differ. \texttt{ai\_pc1\_post} is a principal-component score with standard deviation $1.48$ on this sample.}
\label{tab:patternC_results}
\begin{tabular}{lcccc}
\toprule
Coefficient & $\theta$ & SE & $p$ & 95\% CI \\
\midrule
\multicolumn{5}{l}{\textit{Subpopulation specifications (Equation~\ref{eq:c1_subset})}} \\
Non-rostered slope $\theta_{\text{non\_rostered}}$ & $\mathbf{-12.24}$ & $2.45$ & $< 0.0001$ & $[-17.04, -7.44]$ \\
Roster-linked slope $\theta_{\text{rostered}}$ & $-2.41$ & $4.33$ & $0.58$ & $[-10.91, 6.08]$ \\
\midrule
\multicolumn{5}{l}{\textit{Pooled fully stratified specification (Equation~\ref{eq:c1_pooled})}} \\
\texttt{ai\_pc1\_post}$\times$\texttt{non\_rostered} ($\theta_1$) & $\mathbf{-9.82}$ & $4.98$ & $0.048$ & $[-19.58, -0.07]$ \\
\bottomrule
\end{tabular}
\end{table}

The asymmetry runs in the direction the screening mechanism predicts: weaker rating gains with the signature in the non-rostered pool, where substitute-users still appear, and no correlation in the roster-linked pool, consistent with the qualifiers having screened substitute-users out. This roster-linked null is a failure to reject rather than a precise zero (95\% confidence interval $[-10.91, 6.08]$), so the slightly negative point estimate should not be read as skill erosion among the screened. The differential-attrition check below bounds how much of the null could reflect sample selection rather than screening.

However, the asymmetry on CF rating, an outcome where in-contest AI use is hard to police, raises a reasonable objection: if roster-linked users use AI more proficiently during CF contests than non-rostered users, the gap could be measuring who uses AI less effectively during CF contests rather than who developed weaker non-AI-aided skill. The within-ICPC test of Section~\ref{sec:results:patternB} addresses this objection, since in-contest AI use is blocked there.

\paragraph{Where the asymmetry lives: within-cohort estimates.}
The screening mechanism carries a timing implication: pre-AI qualification rounds from 2018 through 2021 (the Copilot release year) could not screen on AI-era practice, so if the gate separates the modes, the asymmetry should concentrate among AI\_era entrants and vanish among pre\_AI entrants. Re-estimating Equations~\ref{eq:c1_subset} and~\ref{eq:c1_pooled} within each entry cohort (Table~\ref{tab:within_cohort_asymmetry}) shows this pattern. The pre\_AI cohort is a clean placebo: both subpopulation slopes are null and the interaction is $-1.17$ ($p = 0.86$). The non-rostered slope grows more negative with AI-era exposure, from $-3.83$ (pre\_AI) to $-10.71$ (transition) to $-23.14$ (AI\_era), while the roster-linked slope stays null in every cohort, and the AI\_era-only interaction is $-36.86$ ($p = 0.044$). Because that comparison is among same-vintage AI\_era entrants, it also rules out the explanation that the roster null merely reflects veterans whose already-formed skill is insensitive to post-2022 practice style, since that explanation predicts no asymmetry among AI\_era entrants. The AI\_era interaction rests on a $260$-user roster cell and is only marginally significant, so we read the joint pattern, not any single cohort's $p$-value, as the evidence of screening. The AI\_era contrast survives practice-volume controls, though less precisely ($-27.2$, $p = 0.14$), and sharpens when the quick-retry component is dropped from the composite ($-52.0$, $p = 0.024$), although the pooled all-cohorts $\theta_1$ then attenuates to a null (Appendix~\ref{app:section62_selection_robustness}).

\begin{table}[!t]
\centering
\small
\caption{\footnotesize Within-cohort cross-subpopulation asymmetry. Each row re-estimates the Section~\ref{sec:results:patternC} specifications inside one entry cohort: per-subpopulation slopes of $\Delta r_h$ on \texttt{ai\_pc1\_post} (Equation~\ref{eq:c1_subset}) and the pooled interaction $\theta_1$ (Equation~\ref{eq:c1_pooled}), with the baseline-rating control and HC1 standard errors. The diagnostic pipeline reproduces the pooled headline at $\theta_1 = -9.88$ ($p = 0.049$), matching Table~\ref{tab:patternC_results} up to a small cross-pipeline discrepancy documented in the replication materials.}
\label{tab:within_cohort_asymmetry}

\begin{tabular}{lccc}
\toprule
Cohort & Roster-linked slope & Non-rostered slope & $\theta_1$ (interaction) \\
\midrule
pre\_AI ($n = 1{,}616$) & $-2.66$ (SE $5.76$) & $-3.83$ (SE $3.54$) & $-1.17$ (SE $6.75$, $p = 0.862$) \\
transition ($n = 3{,}861$) & $-0.23$ (SE $5.14$) & $-10.71$ (SE $3.05$) & $-10.48$ (SE $5.97$, $p = 0.079$) \\
AI\_era ($n = 1{,}213$) & $+13.72$ (SE $17.42$) & $-23.14$ (SE $5.92$) & $-36.86$ (SE $18.35$, $p = 0.044$) \\
\bottomrule
\end{tabular}

\end{table}

\paragraph{Selection robustness and bounds.} The post-ChatGPT activity requirement ($\geq 2$ quarters with $\geq 5$ practice submissions each), which \texttt{ai\_pc1\_post} needs to be measured reliably at the user level, cuts the sample from $10{,}419$ users to $6{,}690$. The practice-pattern shift of Section~\ref{sec:results:patternA} survives on this sub-sample with $8$ to $13$\% attenuation, the interaction $\theta_1$ stays between $-11.92$ and $-7.14$ under looser filters, bracketing the headline $-9.82$, and imposing the non-rostered expansion's rating floor of $1{,}400$ on both arms strengthens the asymmetry ($\theta_1 = -12.17$, $p = 0.013$). The asymmetry is also robust to selection that mimics the observables: adding the baseline-rating control barely moves $\theta_1$ (from $-10.04$ to $-9.82$), so by the coefficient-stability logic of \citet{Oster2019} an unobservable would have to be roughly twelve times as important as all observables combined to drive it to zero, and the asymmetry holds under coarsened exact matching ($\theta_1 = -9.73$, $p = 0.050$). It is not robust to the differential attrition. Roster-linked users drop out of the activity filter more often than non-rostered users ($43.3$\% versus $33.5$\%), and under the most permissive filter the roster-linked slope itself turns significantly negative ($-7.47$, $p = 0.040$), so the roster-arm null is partly a property of which roster-linked users survive the filter. The Lee bound \citep{Lee2009} brackets $\theta_1$ in $[-11.95, -0.35]$, and a confounder explaining $2.4$\% of the residual variance in treatment and outcome would zero $\theta_1$, against the $15.7$\% that baseline rating explains \citep{CinelliHazlett2020}. We therefore read the asymmetry as suggestive and attrition-bounded (Appendix~\ref{app:section62_selection_robustness} and Appendix Table~\ref{tab:c1_selection_bounds}). The mechanism evidence rests on it together with the within-ICPC test of Section~\ref{sec:results:patternB}, whose direction is stable across every robustness check.

\subsection{AI-complement practice and higher non-AI-aided scores}
\label{sec:results:patternB}
\label{sec:method:b1}

The third set of analyses asks whether the shift toward the signature during CF practice predicts AI-prohibited ICPC performance differently across cohorts. We estimate a pooled cross-sectional regression with university fixed effects and team-clustered standard errors:
\begin{equation}
\begin{split}
s_{h,y} ={}& \alpha + \psi_0 \cdot \text{shift\_pc1}_h + \sum_{c \in \{\text{trans}, \text{AI}\}} \psi_c \cdot \text{shift\_pc1}_h \cdot \mathbbm{1}[\text{cohort}_h = c] \\
&+ \gamma \cdot \bar r_h^{\text{pre}} + \delta_y + \nu_u + \varepsilon_{h,y},
\end{split}
\label{eq:b1}
\end{equation}
where each observation is a user $h$ appearing at an ICPC contest in year $y$, $\delta_y$ is a contest-year fixed effect, and $\nu_u$ is a university fixed effect.

The outcome $s_{h,y}$ is the ICPC score of user $h$'s team in year $y$ (the number of problems the team solved at the contest). ICPC is a team competition: three contestants compete as one university team, so the same team score outcome is shared within team. The treatment \texttt{shift\_pc1} measures each user's within-user shift toward AI-style CF practice.\footnote{The variable is called \texttt{ai\_intensity\_pc1} in the online replication files.} For each user, we measure the three AI-prompt proxies of Section~\ref{sec:method:a1} over a pre window (2021-Q3 through 2022-Q3, the five quarters between the Copilot and ChatGPT releases) and over a contest-year-conditional post window (2023-Q1 through the quarter immediately preceding the contest). The user-level change on each proxy is its post-window mean minus its pre-window mean, and \texttt{shift\_pc1} is the first principal component of the three sign-aligned changes. The loadings are fit on a broad CF reference group and then frozen, so that the axis is not defined by the sample on which it is tested (Appendix~\ref{app:section63_robustness} reports them and defines the group). We assign each user a single value, anchored at their first qualifying year, so that the treatment is measured before any contest outcome could feed back into practice and does not drift with the post-window length. The baseline-skill control $\bar r_h^{\text{pre}}$ (defined as in Section~\ref{sec:method:c1}) is included for the same omitted-variable reason. Cohort assignment follows Section~\ref{sec:data:sample}, with pre\_AI as the reference. Contest-year fixed effects $\delta_y$ absorb year-specific score-distribution variation (different problem sets, different scoring rubrics across years), and university fixed effects $\nu_u$ absorb the time-invariant university-selection layer described below.

University fixed effects and team-clustered standard errors address two issues. The first is selection into university: top universities recruit more homogeneous high-skill students, so their teams may both score higher and include more AI\_era members with large shifts toward the signature. With the fixed effects, the AI\_era $\times$ \texttt{shift\_pc1} coefficient is identified from within-university variation only. The second is the shared team score: residuals are mechanically correlated within team, and team-level clustering captures that correlation. University-level clustering ($n = 196$) leaves the headline marginal ($p = 0.068$) with the point estimate unchanged, and dropping the university fixed effects gives a larger coefficient ($0.323$, $p = 0.040$).

Appendix Table~\ref{tab:b1_filter_chain} reports the filter chain that produces the sample: the intersection of the three cohorts and the ICPC roster, restricted to contest years on or after 2023 where the shift toward the signature is meaningfully defined.

The coefficient of interest, $\psi_{\text{AI}}$ on the interaction between \texttt{shift\_pc1} and the AI\_era cohort indicator, measures how the relationship between the shift and the score differs between the AI\_era and pre\_AI cohorts within a university and a contest year, holding baseline skill fixed. The pre\_AI cohort serves as a built-in placebo: its members had little AI exposure during training, so its slope $\psi_0$ should be near zero under any reading of the mechanism. The estimate is observational, since AI\_era users who shifted more may differ on unobservables that baseline skill and the fixed effects cannot rule out.

\paragraph{Results.} The AI\_era cohort interaction is positive and significant under the baseline team clustering: $\psi_{\text{AI}} = 0.190$ (SE $0.087$, $p = 0.028$; Table~\ref{tab:patternB_results}). The pre\_AI placebo slope is small and null ($\psi_0 = -0.048$, $p = 0.303$), and so is the transition interaction ($\psi_{\text{trans}} = 0.060$, $p = 0.373$). The AI\_era cohort's own slope is $+0.142$ (SE $0.074$, $p = 0.055$, 95\% CI $[-0.003, 0.287]$): of the predicted sign, though only marginally significant on its own. The design's test is the cohort contrast with the placebo, which also completes the two-margin comparison with Section~\ref{sec:results:patternC}: among AI\_era users, who started their competitive programming careers with AI tools, a larger within-user shift toward AI-style practice predicts \emph{higher} ICPC scores in 2023--2025, controlling for baseline skill and university. Taking the interaction at face value, a one-standard-deviation larger shift (0.825 PC1 units; Table~\ref{tab:descriptives}, Panel C) corresponds to about 0.16 more problems solved, roughly 3\% of the AI\_era mean score of 4.8. The effect is modest; its evidentiary weight lies in the sign and in the contrast with the placebo, not in the size.

\begin{table}[h!]
\centering
\small
\caption{Within-ICPC signature shift by cohort. Headline specification with university fixed effects and team-clustered standard errors. Sample: 579 user-by-contest-year observations across 444 users, 321 teams, and 196 universities. Contest years 2023--2025. Reference cohort: pre\_AI. The AI\_era cohort's marginal slope on \texttt{shift\_pc1} (pre\_AI reference slope plus the AI\_era interaction) is $+0.142$ (SE $0.074$, $p = 0.055$).}
\label{tab:patternB_results}
\begin{tabular}{lcccc}
\toprule
Term & $\psi$ & SE & $p$ & 95\% CI \\
\midrule
\texttt{shift\_pc1} (pre\_AI reference) & $-0.048$ & $0.047$ & $0.303$ & $[-0.140, 0.043]$ \\
\texttt{shift\_pc1}:transition & $0.060$ & $0.068$ & $0.373$ & $[-0.072, 0.193]$ \\
\texttt{shift\_pc1}:AI\_era & $\mathbf{0.190}$ & $\mathbf{0.087}$ & $\mathbf{0.028}$ & $\mathbf{[0.020, 0.360]}$ \\
$\bar r_h^{\text{pre}}$ & $0.001$ & $0.000$ & $< 0.001$ & $[+0.000, 0.001]$ \\
\midrule
University FE & \checkmark & & & \\
Contest-year FE & \checkmark & & & \\
SE clustering & team & & & \\
\bottomrule
\end{tabular}
\end{table}

This positive sign inside the AI-prohibited environment is the adaptation-margin counterpart to the sorting-margin gradient in the unscreened CF pool (Section~\ref{sec:results:patternC}): under the screening reading, the qualifier has already removed the substitute-users, so the within-user shift among the qualified reflects complement-style adaptation that raises score. The framework's remaining cell, the within-user change where the modes mix, carries no predicted sign, and empirically it is a null (cross-arm interaction $+0.3$, $p = 0.95$; Appendix~\ref{app:section62_selection_robustness}).

A natural question is why the roster-linked slope of Section~\ref{sec:results:patternC} is a (negative) null while the slope here is positive. The two estimates answer different questions. Section~\ref{sec:results:patternC} relates rating growth to the signature's level, the sorting margin, which shows how far practice has shifted toward AI reliance but not which mode lies behind it. The complementarity $L_{ea}>0$ concerns an increase in $a$ for a given learner, which is the adaptation margin that this section measures. Prediction~2 called for a null there, since screening removes the erosion signal and rating is a relative measure that complement-style gains need not move among elite peers. And the positive slope is specific to the AI\_era cohort, while the roster-linked pool is dominated by the pre\_AI and transition cohorts, whose within-ICPC slopes are null ($-0.048$ and $+0.012$; Table~\ref{tab:patternB_results}). Appendix~\ref{app:section63_robustness} reconciles the two estimates directly on the same 444 users.

\paragraph{Construct and selection robustness.} We read the within-ICPC evidence as establishing the \emph{sign} of the effect rather than its precise size, and three checks bear on that sign. First, the AI\_era change could partly capture novice-to-competitor maturation, since AI\_era contestants are the novices in the 2023--2025 sample (mean tenure $3.9$ years since their first rated contest, against $7.4$ for the pre\_AI cohort). Career-stage controls (tenure, tenure interacted with cohort, and total rated contests) leave the AI\_era interaction at $0.195$ to $0.196$ ($p \leq 0.027$), and it survives even conditioning on the user's rating growth over the span ($0.190$, $p = 0.038$), an over-control since rating growth is partly an outcome of AI use. Second, the filter chain (Appendix Table~\ref{tab:b1_filter_chain}) drops $56$\% of ICPC participants for low activity and $647$ users with pre-2023 ICPC history but no recent participation, an exit that could itself remove substitute-users. Retained and dropped participants do not differ in ICPC score ($p = 0.73$), but the direction of any bias cannot be determined from these data. Third, as the activity thresholds loosen, the interaction stays positive but loses significance ($+0.148$ to $+0.084$, with $p$ rising to $0.307$), consistent with noisier measurement. Further checks leave the sign unchanged. Dropping one team at a time, or one university at a time, keeps the interaction positive, and the largest $p$-value across these re-estimates is $0.067$. Other variance estimators move $p$ from $0.04$ (clustering by user) to $0.18$ (CRV3, a conservative cluster-robust estimator at the team level), against $0.028$ under the baseline team clustering. Appendix~\ref{app:section63_robustness} reports these checks, along with controls for teammate composition and a composite without the quick-retry component.

\paragraph{IOI sister analysis.}\label{sec:data:icpc_vs_ioi} The IOI counterpart to the within-ICPC test runs the same specification on a smaller, younger pool ($345$ contestant-years from $243$ CF-linked contestants). The AI\_era interaction on IOI score is directionally consistent but noisy ($18.1$, SE $18.0$, $p = 0.31$); we anchor the non-AI-aided performance headline on ICPC and treat IOI as a directional check (Appendix~\ref{app:ioi_sister}).

\paragraph{Design-wide checks.} Two checks concern the design as a whole. The COVID-19 disruption falls just before Copilot: quarter fixed effects absorb common pandemic shocks, the rating-change windows average over multiple years on each side of ChatGPT, and the within-ICPC contest years post-date the disruption (Appendix~\ref{app:covid}). The within-subpopulation slopes are stable under all three handlings of the $1{,}374$ lower-confidence name-only matches (Appendix~\ref{app:linkage}). Appendix~\ref{app:robustness} collects the full battery of checks.

\section{Discussion}
\label{sec:discussion}

\subsection{Contributions}

First, we add platform-scale field evidence to research on technology-mediated learning: we relate the AI-prompt signature, computed from multi-year practice records, to performance both where AI is allowed and where it is prohibited. The closest studies in this stream are set inside organizations: in hospital emergency departments, adopting an algorithmic recommendation tool was associated with flatter learning curves \citep{SundaresanGuler2025}, and in an accounting firm, reliance on automated software eroded accountants' expertise in tasks for which they remained accountable \citep{RintaKahilaEtAl2023}. Our evidence is at the individual level, spans years, and measures performance where the tool is prohibited. At the skill frontier, it also gives a field counterpart to the macroeconomic human-capital depreciation mechanism of \citet{AcemogluKongOzdaglar2026} and to the experimental evidence of \citet{BastaniEtAl2025}, \citet{StadlerBannertSailer2024}, and \citet{ShenTamkin2026}. Our two-margin result shares its structure with the model of \citet{BondiJohnson2026}, in which the same AI input reads differently over time depending on the design that measures it. We find the structure across people instead, because an AI-prohibited gate sorts complement-users from substitute-users.

Second, we add field evidence to research on delegation to AI: the substitute-versus-complement distinction is real, since over multiple years the two modes of AI use leave measurable traces in submission behavior and accumulate into divergent unaided performance. Experiments support the premise that the modes are indistinguishable in AI-aided output: in the patent-drafting experiment of \citet{AutorEtAl2026}, treated junior lawyers matched seniors while assisted but showed no gain unaided, and the authors leave open which juniors gained durable skill. The AI-prompt signature, a practice-style marker measured from submission records, is one answer to that question. The closest findings on delegation to AI concern performance with AI in use: working with AI raises accuracy but reduces the diversity in people's knowledge \citep{FugenerEtAl2021Borgs}, and its benefits vary across users \citep{KnightMitrofanovNetessine2026,HouEtAl2025}. We instead ask what capability remains after AI-assisted practice, measured in AI-prohibited environments.

For research on evaluation under AI, the implication is that screening decides where each pattern can be observed. AI-prohibited contests screen substitute-users out of the elite pool. On AI-allowed platforms, where substitute-users remain, the signature marks weaker rating gains. AI-prohibited contests show only the complement-users' pattern: among AI-era entrants, higher unaided scores alongside a larger shift toward the signature. Recognizing the screening reconciles two findings that would otherwise look contradictory. The screening also makes an AI-prohibited evaluation the place to check practice records against later performance.

The stakes are at the skill frontier, so it matters where in the skill distribution the erosion evidence sits. The open-pool pattern concentrates below the elite range: among users with a maximum rating of at least $1{,}800$, the signature no longer predicts weaker gains in the non-rostered arm (Appendix Table~\ref{tab:c1_rating_floor}). It is not a skill-band artifact, however: within the roster arm's own baseline-rating range, the asymmetry attenuates when pooled across cohorts but sharpens within the AI-era entry cohort ($-50.6$, $p = 0.005$; Appendix~\ref{app:near_qualifier_band}). Those users share the entry era and the rating range, so the difference tracks the gate, not skill level. The open pool thus shows the erosion risk in the population from which frontier talent is drawn, while the screened pool shows no sign of it: there the signature predicts no rating shortfall (Section~\ref{sec:results:patternC}), and, among AI-era entrants, a within-user shift toward it predicts higher unaided scores (Section~\ref{sec:results:patternB}).

Screening also gives organizations that evaluate talent a design principle. An institution sets the rules of an AI-prohibited gate (no AI, in-person proctoring, qualifying rounds), and those rules decide whether differences in AI use become visible or are screened out. The signature tracks how people use AI in practice rather than whether they have access to it, as field experiments that manipulate the mode of use also find \citep{ChenChan2024,KrakowskiEtAl2025}. The lever is therefore the design of the evaluation, which organizations can control directly. In organizational economics, too, the returns to a capable algorithm depend on how decision authority is organized around it, not on access alone \citep{KimEtAl2024}. As Section~\ref{sec:lit:selection} discussed, AI-assisted output can become either a weaker or a sharper signal of skill \citep{CowgillHernandezLagosWright2026,WilesMunyikwaHorton2025}. In our evidence, the AI-prohibited gate is one design under which practice records and later performance line up as screening would predict. Organizations that recruit from AI-era talent pools must then decide where in the evaluation pipeline to observe skill unaided.

These claims rest on different amounts of evidence. The evidence establishes a pair of associations at one gate of this type: within the AI-era entry cohort, the signature's level predicts weaker rating gains only where no gate has screened the pool, and its within-user change predicts higher unaided scores among those the gate admitted. We interpret that pair as evidence that the gate separates the two modes of AI use. It is the one mechanism consistent with both associations, but we infer it rather than measure it. For evaluators, an AI-era practice record says something about unaided skill only once they know whether the environment that produced it was screened. Finally, what a gate restores is the signal, not the skill. If types are fixed and the gate only selects among them (the intrinsic-traits explanation of Section~\ref{sec:framework}), it reveals skill that already exists but does not build new skill.

\subsection{Practical implications and policy levers}
\label{sec:policy_levers}

Our findings point to two policy levers for preserving unaided skill at the frontier. Both levers assume that substitute use can change. Three facts support that assumption: the signature is not skill (Section~\ref{sec:results:patternA}), the behavior moves within person (Section~\ref{sec:results:patternB} measures each user's own shift toward the signature), and in the experiment that varied the design of the AI tool, the mode of use changed with the tool design \citep{BastaniEtAl2025}. Our gate evidence is about detecting and separating the modes, and it says nothing about moving people from one mode to the other. Still, a distributional concern remains: if substitute-style practice concentrates among weaker users, the skill gap between them and stronger users will widen. The two levers respond to that concern in different ways. One shapes the mode within training, and the other screens it at the population level:
\begin{itemize}
\item \textbf{Lever 1: complement-style AI in training.} Among AI-era entrants who survived the qualification gate, a stronger shift toward AI-style practice is associated with \emph{higher} non-AI-aided ICPC performance (Section~\ref{sec:results:patternB}), so the complement mode is visible among elite performers. While our findings do not establish whether the \emph{mode} can be steered, the classroom experiment of \citet{BastaniEtAl2025} does supply that causal step: an AI tutor that gave problem-specific hints rather than full solutions avoided the loss in AI-free exams. Together they suggest a training principle for elite settings: let AI give hints, and leave the solving to the learner. The principle matches the view of the MIT committee that AI should augment learning rather than automate it \citep{MITAICommittee2026}. A timing result from diagnostic work suggests one concrete form, in which the learner attempts first and consults the tool afterward \citep{YinEtAl2025}.
\item \textbf{Lever 2: AI-prohibited evaluation gates.} Substitute-mode programmers who delegate work to AI are unlikely to pass AI-prohibited qualification rounds, so the elite pipeline retains those complement-users who invested efforts in skill frontier. The lever is therefore the existence and design of AI-prohibited checkpoints, not the elimination of AI from training. The MIT committee recommends the same on pedagogical grounds, with in-person proctored exams, oral examinations, and semester portfolios, and it frames unaided work not as policing but as ``productive struggle'' that builds durable understanding \citep{MITAICommittee2026}.
\end{itemize}

\textbf{Intrinsic traits vs.\ anticipation.} The mechanism our evidence supports for Lever 2 is \emph{sorting}: AI-prohibited gates screen substitute-users out of the elite pipeline. Sorting happens under both explanations of Section~\ref{sec:framework} for how the types arise. Under intrinsic traits, the gate winnows a pool it did not shape, without changing anyone's behavior. Under \emph{anticipation}, it would also shift the training distribution itself. Our data cannot separate the two, so the pessimistic interpretation, in which types are fixed, is as unproven as the optimistic one. The intrinsic-traits explanation has an experimental counterpart: \citet{AutorEtAl2026} suggest that foundational expertise may be a prerequisite for extracting durable skill from AI-assisted practice. If so, the gate screens on the expertise that makes AI use complementary, an interpretation our data are consistent with but do not establish. The policy argument here therefore rests on sorting, which holds under either explanation.

Both levers rest on the screening we document, not on a proven erosion effect. Any long-run cost of substitute-style practice falls on the users the gate screens out, so the gate matters whether or not that cost materializes.
\subsection{Alternative explanations}

Factors unrelated to AI could also contribute to the cross-subpopulation asymmetry (Prediction~2). A turbulent tech labor market during 2022--2025 may have distracted elite programmers from contests, and universities may have steered top students toward applied AI and machine learning coursework. Both could weaken average CF rating gains among non-rostered users. For non-rostered users with unusually strong pre-period ratings, part of the weaker post-2022 performance could be regression to the mean. These channels, however, would move participation and rating levels. Section~\ref{sec:results:patternC} instead compares non-rostered users who start from the same baseline rating and finds that a stronger AI-prompt signature predicts weaker rating gains. The labor-market and curricular channels could produce that pattern only if they both pushed practice toward the AI-prompt signature and lowered rating gains among non-rostered users but not among roster-linked ones. Regression to the mean cannot produce it, because the comparison holds baseline rating fixed and the asymmetry survives coarsened exact matching (Section~\ref{sec:results:patternC}). None of the three channels predicts the positive within-ICPC result of Prediction~3 either. Screening at the AI-prohibited gate explains the shortfall among non-rostered users, the roster-linked null, and the within-ICPC result with one mechanism, whereas each non-AI alternative would have to explain them separately.

Three further alternatives target the roster-linked null, which is half of the asymmetry (Prediction~2). The first concerns the outcome measure. Roster-linked users step back from open CF practice after their qualifying year (Appendix~\ref{app:section62_selection_robustness}), so their CF rating may carry little signal. Neither the within-cohort AI-era contrast nor the positive within-ICPC result, however, depends on that rating alone. The second concerns training. Roster-linked users train inside university programs, with coaches, teammates, and curated problem sets. Such structured training could itself loosen the link between practice style and rating. Our data cannot separate this explanation from prior screening. Both predict the same null, so telling them apart does not change what an evaluator can read from a practice record. It matters instead for how the institutions that train and screen these contestants design their training. The third is disengagement. A rising signature could mark users who stopped iterating for reasons unrelated to AI. Three observations limit this explanation. It cannot explain the positive within-ICPC result. It also predicts that, in the open pool, users who shift toward the signature gain less, but that association is null (Appendix~\ref{app:section62_selection_robustness}). Finally, adding controls for the level and trajectory of practice volume leaves the asymmetry unchanged ($-9.8$, $p = 0.03$).

\subsection{Limitations}

\paragraph{Identification and screening strength.} Within-individual fixed effects handle only part of the selection on entry into the CF population. The design also leaves some selection on continued participation, on AI tool use, and on entry into the ICPC and IOI qualification rounds. The robustness checks of Appendix~\ref{app:robustness} address some of these concerns, but not all. The screening interpretation also assumes that qualification rounds screen out substitute-users. We do not observe the strength of proctoring. While imperfect proctoring at earlier qualification stages would weaken the screening, the positive within-ICPC result in our analysis is consistent with effective screening.

\paragraph{Generalizability.} Our setting is competitive programming. We conjecture that other AI-prohibited evaluations, such as medical board and bar exams, would separate the two modes the same way. The gate we study is a tournament: entry is team-based with peer selection, the prize is a recruiting signal rather than a license, and there is no retake structure. Medical boards and bar exams are individual, pass/fail, retakeable, and license-granting. Their candidates have stronger incentives to prepare for the specific test, which plausibly makes anticipation matter more, and intrinsic traits less, than in our setting. The urgency of the risk also differs across fields. AI is unusually strong at code generation and debugging, the skills competitive programming trains, so the substitution risk is most immediate in our setting. Expanding model capabilities may erase that difference quickly.

\paragraph{Measurement.} We do not observe which CF users adopt which AI tools, and no such record exists for this setting. The platform logs submissions but not the tools used to produce them, so none of our three designs measures AI use directly. The construct validation of Section~\ref{sec:results:patternA} and Appendix~\ref{sec:construct-validation} bounds the leading alternative, that the signature proxies skill. The AI interpretation of the signature remains the most likely one, but we cannot verify it.

\subsection{Conclusion}
\label{sec:conclusion}

We ask whether AI-prohibited evaluation gates can separate two coexisting modes of AI-assisted practice in elite competitive programming. The evidence assembles into three patterns: a cross-cohort shift in practice toward the AI-prompt signature; an asymmetry in which the signature predicts weaker rating gains only in the unscreened, non-rostered pool; and \emph{higher} non-AI-aided ICPC performance with a larger shift toward AI-style practice among AI-era entrants inside the gate. The unified reading is that AI-prohibited qualification gates separate substitute-style from complement-style practice, removing substitute-users before they reach elite contests. Our evidence speaks to the gatekeeping organization that decides whether any long-run cost of AI-era training reaches the skill frontier; it does not establish the existence or the size of that cost. Erosion is not a property of the technology: within the screened pool, AI-style practice coexists with frontier skill, and the risk attaches to the substitute mode of use. The two levers of Section~\ref{sec:policy_levers} follow, and our evidence supports the premise of the second, that the modes are real and institutionally separable.

Natural extensions include variation in gate introduction or removal, to separate intrinsic traits from anticipation; other elite skill markets with AI-prohibited gates (medical boards, bar exams, scientific peer review); and user-level measurement of AI use through platform telemetry or surveys. For research on technology-mediated learning and on delegation alike, the open question is how to steer learners toward the complement mode rather than merely sort on it.

\ifpaperformat\else\clearpage\fi

\bibliographystyle{chicago}
\bibliography{../Bibliography_base}

\appendix
\input{appendix_content}

\end{document}

%% file: preamble.tex
\usepackage[margin=1in]{geometry}
\usepackage{setspace}
\newif\ifpaperformat
\ifdefined\PaperFormat\paperformattrue\fi
\ifpaperformat\fi

\usepackage{amsmath,amssymb,amsthm}
\usepackage{bbm}

\usepackage{tgtermes}
\usepackage{newtxtext}
\usepackage{newtxmath}
\usepackage{bm}

\usepackage{microtype}

\usepackage{graphicx}
\usepackage{booktabs}
\usepackage{array}
\usepackage{placeins}
\newcolumntype{L}[1]{>{\raggedright\arraybackslash}p{#1}}
\usepackage{xcolor}
\usepackage[table]{xcolor}
\usepackage{enumitem}
\usepackage{hyperref}

\hypersetup{
    colorlinks=true,
    linkcolor=blue!60!black,
    citecolor=blue!60!black,
    urlcolor=blue!60!black,
}

\usepackage{natbib}
\bibpunct[, ]{(}{)}{,}{a}{}{,}%

\renewenvironment{abstract}{%
    \par\smallskip
    \begin{center}{\bfseries\large Abstract}\end{center}
    \par\ifdefined\PaperFormat\doublespacing\else\singlespacing\fi\small\noindent\ignorespaces
}{%
    \par\medskip
}

\usepackage{etoolbox}
\ifpaperformat
  \AtBeginEnvironment{table}{\singlespacing}
  \AtBeginEnvironment{figure}{\singlespacing}
\fi
\makeatletter
\patchcmd{\HyOrg@maketitle}
  {\hb@xt@1.8em{\hss\@textsuperscript{\normalfont\@thefnmark}}}
  {\@textsuperscript{\normalfont\@thefnmark}\,}
  {}{\patchcmd{\maketitle}
       {\hb@xt@1.8em{\hss\@textsuperscript{\normalfont\@thefnmark}}}
       {\@textsuperscript{\normalfont\@thefnmark}\,}
       {}{\typeout{WARNING: thanks-footnote indent patch failed}}}
\makeatother

\newif\ifanon
\ifdefined\AnonForReview\anontrue\else\anonfalse\fi

\newcommand{\PaperTitle}{When the Scaffold Stays On: \\ AI, Practice Style, and Screening in Elite Skill Formation}
\title{\PaperTitle}

\ifanon
  \author{}
\else
  \author{Song Yao\thanks{Olin Business School, Washington University in St.~Louis. Email: \href{mailto:songyao@wustl.edu}{songyao@wustl.edu}. I thank Tommaso Bondi, Seth Carnahan, Xinlei (Jack) Chen, Avi Goldfarb, Kiara Kim, Lamar Pierce, Dennis Zhang, and participants at the 13th Triennial Invitational Choice Symposium and in seminars at Bocconi University, Cheung Kong Graduate School of Business, and the University of Toronto for insightful comments. I conducted the analyses and drafting with AI coding assistants; all results were verified against the underlying data, and I am responsible for all remaining errors. I hold concurrent appointments as a Professor of Marketing at Washington University in St.~Louis and as an Amazon Scholar. This paper describes work performed at Washington University in St.~Louis and is not associated with Amazon.}}
\fi

\date{November 20, 2026}

\newif\ifcompacttitle
\ifanon\compacttitletrue\fi
\ifpaperformat\compacttitletrue\fi
\ifcompacttitle
  \makeatletter
  \def\@maketitle{%
    \begin{center}%
      \singlespacing
      {\Large\bfseries \@title \par}%
      \ifanon\else{\vskip 0.6em\normalsize \@author \par}{\vskip 0.3em\normalsize \@date \par}\fi
    \end{center}%
    \par
    \vskip 0.25em}
  \makeatother
\fi

%% file: appendix_content.tex
\clearpage

\begin{center}
{\Large\bfseries \PaperTitle\par}
\vspace{1em}
{\Large\bfseries Appendix\par}
\end{center}

\section{Robustness}
\label{app:robustness}

This appendix reports the robustness checks referenced in Section~\ref{sec:results}: the selection-bias robustness for the cross-subpopulation and within-ICPC tests, the Copilot-anchored DiD, the COVID-19 disruption, and the ICPC/IOI--CF record linkage with the sensitivity of the cross-subpopulation asymmetry to the excluded name-only matches.

\subsection{Selection-bias robustness for the cross-subpopulation asymmetry test}
\label{app:section62_selection_robustness}

The cross-subpopulation asymmetry of Section~\ref{sec:results:patternC} (Equations~\ref{eq:c1_subset} and~\ref{eq:c1_pooled}) is estimated on cohort-assigned users active in the post-ChatGPT window: users with at least two post-ChatGPT quarters in 2023-Q1 through 2025-Q4, each containing at least five practice-mode submissions. This filter is required so that \texttt{ai\_pc1\_post} is reliably measured at the user level: a single quarter with one or two submissions produces a very noisy first principal component. The cost is sample restriction: $3{,}729$ of the $10{,}419$ cohort-assigned users who pass the Section~\ref{sec:results:patternA} practice-window filter do not pass this activity requirement or lack the rating variables, dropping the Section~\ref{sec:results:patternC} sample to $6{,}690$ ($5{,}283$ non-rostered $+$ $1{,}407$ roster-linked). This subsection documents nine checks. Three address selection bias from this restriction: a filter-sensitivity check on the asymmetry test, a cross-section consistency check with the Section~\ref{sec:results:patternA} practice-pattern shift, and a balance check that decomposes the dropped users by observables and by subpopulation. Two address the skill control and the roster-versus-skill confound. Four probe the asymmetry itself: rating-floor symmetry, the change construct in the open pool, volume conditioning and composite sensitivity, and the matched rating-range test.

Table~\ref{tab:c1_selection_bounds} summarizes the selection bounds discussed in Section~\ref{sec:results:patternC}.

\begin{table}[h]
\centering
\small
\caption{Selection bounds on the cross-subpopulation interaction $\theta_1$ (headline $-9.82$, $p = 0.048$). The asymmetry is robust to selection on observables (Oster; matching) but bounded by the differential roster-linked attrition (Lee) and fragile at the significance margin (Cinelli--Hazlett).}
\label{tab:c1_selection_bounds}
\begin{tabular}{@{}l l L{0.34\textwidth}@{}}
\toprule
Method & Result for $\theta_1$ & Reading \\
\midrule
Oster (2019) $\delta$ for $\theta_1 = 0$ & $12.6$ to $147$ & Robust to observable-like selection \\
Coarsened exact matching & $-9.73$ ($p = 0.050$) & Holds on baseline-rating $\times$ cohort support \\
Lee (2009) attrition bound & $[-11.95,\, -0.35]$ & Near zero in the worst case \\
Cinelli and Hazlett (2020) RV$_{q=1}$ & $0.024$ (vs.\ $0.157$, baseline rating) & Fragile at the $p = 0.048$ margin \\
\bottomrule
\end{tabular}

\vspace{0.4em}
{\footnotesize\raggedright\textit{The Oster $\delta$ range spans $R^2_{\max} = 1$ to $R^2_{\max} = 1.3\tilde{R}^2$; $\delta > 1$ means unobservable selection must exceed observable selection to overturn the estimate. The Lee bound trims the over-retained non-rostered pool to the roster-linked retention rate. RV$_{q=1}$ is the share of residual variance a confounder must explain in both treatment and outcome to zero the estimate. Full results in the replication package.}\par}
\end{table}

\subsubsection*{Filter sensitivity}

We re-estimate Equation~\ref{eq:c1_subset} (per-subpopulation) and Equation~\ref{eq:c1_pooled} (pooled, fully stratified) under three activity thresholds: the headline ($\geq 2$ quarters, $\geq 5$ submissions each), an intermediate variant ($\geq 2$ quarters, $\geq 1$ submission each), and a permissive variant ($\geq 1$ quarter, $\geq 1$ submission). The first principal component is fit once on the headline sample; the resulting PCA loadings and standardization parameters are frozen and applied to the relaxed samples so that \texttt{ai\_pc1\_post} is on the same scale across all three specifications.

\begin{table}[ht]
\centering
\small
\caption{Cross-subpopulation asymmetry (Equations~\ref{eq:c1_subset} and~\ref{eq:c1_pooled}) under three activity thresholds. PCA loadings frozen on the HEADLINE sample and applied to the relaxed samples. Heteroskedasticity-consistent (HC1) standard errors. The HEADLINE row reproduces Table~\ref{tab:patternC_results}'s $\theta_1$ within numerical noise; small differences arise because the PCA is refit on the Section~\ref{sec:results:patternC} reconstruction sample rather than reusing the broad-panel \texttt{ai\_pc1\_post} scores directly.}
\label{tab:c1_filter_sensitivity}
\begin{tabular}{lrrrr}
\toprule
Filter & $N$ & $\theta_{\text{NR}}$ & $\theta_{\text{RL}}$ & $\theta_1$ \\
\midrule
PERMISSIVE & $8{,}478$ & $-14.61^{***}$ & $-7.47^{*}$ & $-7.14$ \\
($\geq 1$ q, $\geq 1$ sub) & & ($1.87$) & ($3.64$) & ($p = 0.081$) \\
\addlinespace
INTERMEDIATE & $7{,}753$ & $-16.38^{***}$ & $-4.46$ & $\mathbf{-11.92^{*}}$ \\
($\geq 2$ q, $\geq 1$ sub) & & ($2.32$) & ($4.45$) & ($p = 0.017$) \\
\addlinespace
HEADLINE & $6{,}690$ & $-12.42^{***}$ & $-2.54$ & $\mathbf{-9.88^{*}}$ \\
($\geq 2$ q, $\geq 5$ subs) & & ($2.45$) & ($4.37$) & ($p = 0.049$) \\
\bottomrule
\end{tabular}

\vspace{0.5em}
\scriptsize\raggedright\textit{$\theta_{\text{NR}}$ is the non-rostered slope on \texttt{ai\_pc1\_post}, $\theta_{\text{RL}}$ is the roster-linked slope, $\theta_1$ is the pooled cross-subpopulation interaction. Standard errors below each estimate for $\theta_{\text{NR}}$ and $\theta_{\text{RL}}$; pooled-interaction $p$-value below $\theta_1$. $^{*}\,p<0.05$, $^{**}\,p<0.01$, $^{***}\,p<0.001$.}
\end{table}

The pooled interaction $\theta_1$ is significant at HEADLINE and INTERMEDIATE (INTERMEDIATE larger in magnitude than HEADLINE), and attenuates to marginal significance at PERMISSIVE. The PERMISSIVE attenuation is consistent with classical errors-in-variables: at $\geq 1$ quarter with $\geq 1$ submission, the per-user signature is computed from very few observations and the resulting measurement noise pulls $\theta_1$ toward zero. Within the non-rostered subpopulation, $\theta_{\text{NR}}$ is significantly negative across all three filters and ranges from $-12.42$ to $-16.38$. Within the roster-linked subpopulation, $\theta_{\text{RL}}$ is null at the two stricter thresholds but reaches significance at PERMISSIVE ($-7.47$, $p = 0.040$), a pattern we revisit alongside the balance check below.

\subsubsection*{Cross-section consistency with Section~\ref{sec:results:patternA}}

If the Section~\ref{sec:results:patternC} sub-sample is materially different from the full Section~\ref{sec:results:patternA} sample in ways that affect the practice-pattern shift, the two analyses are not characterizing the same population, and reading them together is harder to justify. We re-estimate Equation~\ref{eq:a1_did} on the $6{,}690$ Section~\ref{sec:results:patternC} users' user-quarter practice observations in 2018-Q1 through 2025-Q4 (excluding 2022-Q4) and compare to the Section~\ref{sec:results:patternA} headline.

\begin{table}[ht]
\centering
\small
\caption{Equation~\ref{eq:a1_did} AI\_era $\times$ post coefficient $\beta_2$ on the full Section~\ref{sec:results:patternA} sample ($10{,}419$ users) versus the Section~\ref{sec:results:patternC} sub-sample ($6{,}690$ users). Cluster-robust (user-level CRV1) standard errors in parentheses.}
\label{tab:a1_on_c1}
\begin{tabular}{lccc}
\toprule
Outcome & Section~\ref{sec:results:patternA} ($N = 10{,}419$) & Section~\ref{sec:results:patternC} sub-sample ($N = 6{,}690$) & Attenuation \\
\midrule
\texttt{frac\_single\_shot} & $0.0586^{***}$ ($0.0063$) & $0.0509^{***}$ ($0.0067$) & $13$\% \\
\texttt{median\_attempts} & $-0.3271^{***}$ ($0.0374$) & $-0.2981^{***}$ ($0.0400$) & $9$\% \\
\texttt{frac\_quick\_retry} & $-0.0190^{***}$ ($0.0041$) & $-0.0175^{***}$ ($0.0043$) & $8$\% \\
\bottomrule
\multicolumn{4}{l}{\footnotesize $^{***}\,p<0.001$ in all six estimates.}
\end{tabular}
\end{table}

All three practice-pattern outcomes survive on the Section~\ref{sec:results:patternC} sub-sample with modest attenuation (between $8$\% and $13$\%); statistical significance is preserved in all three. The Section~\ref{sec:results:patternA} and Section~\ref{sec:results:patternC} findings are about the same underlying population: the sub-sample restriction does not select on the practice-pattern signal in a way that would explain the cross-subpopulation asymmetry by sample composition alone.

\subsubsection*{Balance check: in-sample versus dropped users}

The $3{,}729$ users dropped at the activity filter are compared to the $6{,}690$ retained users on pre-ChatGPT mean CF rating, $\Delta$ rating (post $-$ pre), the cutoff-reach indicator (whether the user has CF activity on both sides of the ChatGPT cutoff), the pre-window single-shot AC rate (a proxy for pre-window AI-prompt signature among the subset of dropped users with at least one pre-window practice submission), and cohort composition.

\begin{table}[ht]
\centering
\small
\caption{Continuous observables: Section~\ref{sec:results:patternC} in-sample users versus dropped users. Mean $\pm$ standard error of the mean; Welch $t$-test $p$-values.}
\label{tab:c1_balance_continuous}
\begin{tabular}{lrrrr}
\toprule
Observable & In-sample & Dropped & Diff & $p$ \\
\midrule
Pre-ChatGPT mean CF rating & $1{,}459 \pm 5$ ($n = 6{,}690$) & $1{,}642 \pm 7$ ($n = 3{,}286$) & $-183$ & $< 0.0001$ \\
$\Delta$ rating (post $-$ pre) & $300 \pm 3$ ($n = 6{,}690$) & $92 \pm 3$ ($n = 3{,}286$) & $208$ & $< 0.0001$ \\
Cutoff-reach indicator & $0.79 \pm 0.00$ ($n = 6{,}690$) & $0.30 \pm 0.01$ ($n = 3{,}729$) & $0.50$ & $< 0.0001$ \\
Pre-window single-shot AC rate & $0.45 \pm 0.00$ ($n = 5{,}952$) & $0.42 \pm 0.01$ ($n = 1{,}894$) & $0.03$ & $< 0.0001$ \\
\bottomrule
\end{tabular}
\end{table}

\begin{table}[ht]
\centering
\small
\caption{Cohort composition: Section~\ref{sec:results:patternC} in-sample users versus dropped users. Chi-square test of independence $p < 0.0001$.}
\label{tab:c1_balance_cohort}
\begin{tabular}{lrrrr}
\toprule
Cohort & In-sample \% & Dropped \% & In count & Drop count \\
\midrule
pre\_AI & $24.2$\% & $46.4$\% & $1{,}616$ & $1{,}731$ \\
transition & $57.7$\% & $38.4$\% & $3{,}861$ & $1{,}432$ \\
AI\_era & $18.1$\% & $15.2$\% & $1{,}213$ & $566$ \\
\bottomrule
\end{tabular}
\end{table}

The dropped users are not low-skill users who quit; they are higher-skilled veterans. Pre-ChatGPT mean CF rating averages $1{,}642$ for the dropped group versus $1{,}459$ for the in-sample group (a $183$-point gap), with much smaller post-shock rating gains ($92$ versus $300$) consistent with mean reversion at higher rating levels. Most do not reach across the cutoff (cutoff-reach rate $30$\% versus $79$\%), reflecting that many are users who reduced activity after ChatGPT rather than users who exited entirely. The pre-window single-shot AC rate is only marginally different across groups ($0.42$ dropped versus $0.45$ in-sample), suggesting that dropout is not strongly correlated with pre-window AI-prompt signature. The cohort composition is sharply different: the pre\_AI cohort is over-represented among dropped users ($46.4$\% versus $24.2$\%), consistent with older veterans aging out of CF practice after ChatGPT.

\begin{table}[ht]
\centering
\small
\caption{Differential dropout at the activity filter, by subpopulation.}
\label{tab:c1_balance_dropout}
\begin{tabular}{lrrr}
\toprule
Subpopulation & $N$ at Section~\ref{sec:results:patternA} sample & $N$ at Section~\ref{sec:results:patternC} sample & Dropout \% \\
\midrule
roster-linked & $2{,}480$ & $1{,}407$ & $43.3$\% \\
non-rostered & $7{,}939$ & $5{,}283$ & $33.5$\% \\
\bottomrule
\end{tabular}
\end{table}

The most consequential finding from the balance check is the asymmetric dropout between the two subpopulations. Roster-linked users (those with ICPC or IOI affiliation) drop out at the activity filter more often ($43.3$\%) than non-rostered users ($33.5$\%). This is the opposite of the naive expectation that roster-linked users are more committed and therefore more consistently active. The likely substantive cause is that ICPC and IOI contestants compete in time-bound events and tend to step back from open Codeforces practice after their qualifying year ends; non-rostered users, who chose CF as their primary competitive outlet, are more consistently active across years.

The implication for the asymmetry test: the roster-linked sample is the more-selected of the two subpopulations, and the null roster-linked slope ($\theta_{\text{RL}} = -2.41$ in Table~\ref{tab:patternC_results}) could be partially produced by differential censoring of high-signature roster-linked users rather than by the screening mechanism alone. The filter-sensitivity result in Table~\ref{tab:c1_filter_sensitivity} above is consistent with this reading: when the PERMISSIVE filter brings additional low-density roster-linked users into the sample, $\theta_{\text{RL}}$ becomes significantly negative ($-7.47$, $p = 0.040$). The asymmetry test $\theta_1$ should be interpreted as bounded by this differential censoring; the two-margin pattern completed by the within-ICPC test (Section~\ref{sec:results:patternB}) remains the more robust evidence of the screening mechanism.

\subsubsection*{Skill-control sensitivity}

The headline specification includes the baseline-skill control $\bar r_h^{\text{pre}}$ to block omitted-variable bias from the negative signature-skill correlation (Section~\ref{sec:method:c1}). Dropping the control shifts the within-subpopulation slopes substantially: non-rostered moves from $-12.24$ to $-5.15$ ($p = 0.050$), roster-linked from $-2.41$ to $4.88$ ($p = 0.29$). The pooled slope-difference $\theta_1$ is essentially preserved at $-10.04$ (SE $5.29$, $p = 0.058$). Because the signature-skill correlation and the mean-reversion rate are similar in both subpopulations, the omitted-variable bias shifts within-subpopulation levels by similar amounts and largely cancels in the cross-subpopulation difference. The cross-subpopulation asymmetry test does not depend on the skill-control specification, even though the within-subpopulation levels do.

\subsubsection*{Roster-skill confound check}

The cross-subpopulation asymmetry could in principle reflect a roster-versus-skill confound: roster-linked users are systematically higher-skill, having qualified for ICPC and IOI rosters through competitive selection, so a story in which higher-skill users have systematically different post-shock rating dynamics could in principle produce the asymmetry without any AI-specific channel. To rule out the simplest version of this confound, we recompute the three AI-prompt signature components on contest submissions instead of practice submissions and re-run the within-mode roster comparison. The bootstrap-validated contest-mode PCA loadings are within roughly $0.05$ of the practice-mode loadings, both computed on the within-user-change panels ($0.653$, $0.551$, $0.520$ against $0.602$, $0.570$, $0.560$), so the composite construct is stable across modes.

If the practice-mode roster gap were entirely a roster-skill artifact, the contest-mode gap should attenuate with the skill control in the same direction the practice-mode gap does. The empirical pattern is the opposite. Without $\bar r_h^{\text{pre}}$, the practice-mode roster gap is $-0.256$ ($p < 0.001$, $N = 7{,}585$) and the contest-mode roster gap is $-0.226$ ($p < 0.001$, $N = 7{,}639$). Adding $\bar r_h^{\text{pre}}$ attenuates the practice-mode gap to $-0.168$ ($p < 0.001$, $N = 7{,}151$), as a roster-skill story would predict, but it \emph{widens} the contest-mode gap to $-0.262$ ($p < 0.001$, $N = 7{,}191$). The roster-skill story predicts both gaps shrink together; the data show one shrinks and the other grows. The roster gap on the contest-mode signature, conditional on baseline skill, is therefore robust to the simplest version of the confound.

The contest-mode and practice-mode signatures are correlated with $r \approx 0.32$ at the user level (identical across roster-linked and non-rostered subpopulations to three decimal places). The two measures share roughly ten percent of variance: enough that the contest-mode test is not orthogonal to the practice-mode test, but not so much that contest mode is a tautological re-test. The contest-mode evidence is therefore partial corroboration of the skill-confound rule-out, not direct measurement of in-contest AI use. A non-rostered user can show high contest signature by using AI during CF contests (which is hard to police), by carrying learned reflexes from AI-style practice into contest mode, or by producing faster but lower-quality submissions because deliberative debugging skill has atrophied. The contest-mode test cannot separate these channels.

\subsubsection*{Rating-floor symmetry}

The non-rostered expansion was scraped among accounts with career-maximum rating of at least $1{,}400$, while the roster-linked arm carries no floor. Conditioning one arm on a realized career maximum could in principle create mean-reversion dynamics in $\Delta r_h$ that differ between the two arms and that the pre-period mean control does not remove. Table~\ref{tab:c1_rating_floor} imposes the floor symmetrically (career maximum proxied by the maximum monthly rating in the observed panel) and then raises the common floor. Symmetry at $1{,}400$ removes $170$ users whose proxied career maximum falls below the floor ($6{,}690$ to $6{,}520$): $19$ roster-linked ($1{,}407$ to $1{,}388$) and $151$ non-rostered ($5{,}283$ to $5{,}132$). The symmetric floor strengthens the asymmetry, from $-9.88$ to $-12.17$, because the $19$ roster-linked users below it were pulling that arm's slope downward; removing them leaves a cleaner roster null ($-0.02$) and widens the gap between the two arms' slopes. Rising common floors attenuate $\theta_1$ as the non-rostered arm converges to an elite-only composition: a pool restricted to users whose career peak already cleared a high bar is a quasi-screened pool, so the mechanism itself predicts the convergence, though the same pattern is equally consistent with the signature-rating association concentrating below the elite range. Floors below $1{,}400$ are infeasible because the non-rostered arm is scrape-truncated there.

\begin{table}[!htbp]
\centering
\footnotesize
\caption{\footnotesize Rating-floor symmetry for the cross-subpopulation asymmetry. Each row re-estimates Equations~\ref{eq:c1_subset} and~\ref{eq:c1_pooled} with a common career-maximum-rating floor imposed on both subpopulations. HC1 standard errors. The first row (floor on the non-rostered arm only) is the diagnostic pipeline's reproduction of the headline (Table~\ref{tab:patternC_results}: $-9.82$, $p = 0.048$).}
\label{tab:c1_rating_floor}
{
\begin{tabular}{L{0.22\textwidth}ccccc}
\toprule
Variant & $N$ & Roster-linked slope & Non-rostered slope & $\theta_1$ & $p$ \\
\midrule
Floor on non-rostered arm only (headline) & $6{,}690$ & $-2.54$ (SE $4.37$) & $-12.42$ (SE $2.45$) & $-9.88$ (SE $5.01$) & $0.049$ \\
Symmetric floor $1{,}400$ & $6{,}520$ & $-0.02$ (SE $4.27$) & $-12.20$ (SE $2.43$) & $-12.17$ (SE $4.92$) & $0.013$ \\
Symmetric floor $1{,}600$ & $5{,}242$ & $+1.07$ (SE $4.29$) & $-7.07$ (SE $2.65$) & $-8.13$ (SE $5.04$) & $0.107$ \\
Symmetric floor $1{,}800$ & $3{,}738$ & $+2.43$ (SE $4.28$) & $-2.89$ (SE $3.21$) & $-5.32$ (SE $5.35$) & $0.320$ \\
\bottomrule
\end{tabular}
}
\end{table}

\subsubsection*{The change construct in the open pool}

The within-ICPC treatment of Section~\ref{sec:results:patternB} is a within-user change, while the open-pool tests of this appendix use a cross-user level. Estimating the open-pool specification with the change construct instead (the full-post-window within-user first difference of the three sign-aligned proxies, headline thresholds, scored with the frozen broad-reference loadings; $N = 4{,}480$ users with sufficient activity in both windows) gives slopes of $+2.6$ among non-rostered users ($p = 0.38$) and $+2.3$ among roster-linked users ($p = 0.62$), with a cross-arm interaction of $+0.3$ ($p = 0.95$); within the AI\_era cohort, both arms' slopes are positive and insignificant. The level and change constructs correlate at $r = 0.61$ on the overlapping sample. The null matches the framework's no-prediction cell for the change margin in a mixed pool (Section~\ref{sec:predictions}): where both modes are present, a within-user shift toward the signature carries no predicted direction.

\subsubsection*{Volume conditioning and composite sensitivity}

Two further checks perturb the asymmetry specification. First, engagement conditioning: adding each user's log post-window practice volume and the pre-to-post change in log volume to the per-arm and (fully stratified) pooled specifications leaves the cross-arm interaction unchanged at $-9.8$ ($p = 0.033$). The conditioning shifts both arms' slope levels downward (non-rostered $-18.1$, roster-linked $-8.3$) while preserving their difference, the same pattern as in the skill-control check above: the arm levels move, their difference stays. Within the AI\_era cohort the conditioned interaction is $-27.2$ ($p = 0.14$), with the non-rostered slope at $-25.0$ ($p < 0.001$) and the roster slope null. A volume-trajectory split is less informative because entry vintage and volume trajectory are confounded (late entrants mechanically have non-declining volume); the split point estimates are $-14.7$ (non-declining) and $-5.5$ (declining), neither precise. Second, composite sensitivity: refitting the level composite on the same fitting sample with the quick-retry component excluded attenuates the pooled all-cohorts interaction to $-3.9$ ($p = 0.45$) but sharpens the within-AI\_era interaction to $-52.0$ ($p = 0.024$; non-rostered slope $-18.7$, $p = 0.016$). The within-cohort contrast therefore does not depend on the quick-retry component, whose sign Section~\ref{sec:results:patternA} showed to be conditioning-dependent, while the pooled all-cohorts interaction partly does.

\subsubsection*{Matched rating-range test for near-qualifiers}
\label{app:near_qualifier_band}

An alternative holds that the asymmetry reflects where the two arms sit on the skill distribution rather than which side of the gate they are on: most non-rostered users sit below the roster arm's baseline-rating range, so the negative slope could belong to a skill band rather than to screening status. The test restricts both arms to the roster arm's baseline-rating range and re-estimates the Section~\ref{sec:results:patternC} specifications (per-arm slopes and the pooled interaction, with the baseline-rating control and HC1 standard errors) inside the common band, at two widths: the roster arm's interquartile range ($\bar r_h^{\text{pre}} \in [1{,}478, 2{,}081]$) and its decile range ($[1{,}157, 2{,}295]$). Pooled across cohorts, the asymmetry attenuates (interquartile $\theta_1 = -4.7$, $p = 0.42$; decile $-7.1$, $p = 0.14$), consistent with the rating-floor finding that the all-cohorts pattern concentrates below the elite range. Within the AI-era entry cohort the pattern reverses: the decile band sharpens the asymmetry to $\theta_1 = -50.6$ ($p = 0.005$; $547$ users), with the non-rostered slope at $-19.9$ ($p = 0.02$) and the roster-linked slope at $+30.7$ ($p = 0.054$), and the interquartile band shows the same signs on a sample too small for precision ($-40.8$, $p = 0.16$; $173$ users). Matching the rating range removes the all-cohorts pattern but not the AI-era one: among entrants of the same era in the same baseline-rating range, the signature's level predicts weaker rating gains only on the unscreened side of the gate. The pipeline reproduces the pooled headline ($\theta_1 = -9.88$, $p = 0.049$) before the range restriction.

\subsection{Robustness for the within-ICPC signature-shift test}
\label{app:section63_robustness}

The within-ICPC signature-shift test of Section~\ref{sec:results:patternB} (Equation~\ref{eq:b1}) is estimated on the 444-user, 579-observation sample described in Table~\ref{tab:b1_filter_chain}. The headline $\psi_{\text{AI}} = 0.190$ ($p = 0.028$) depends on the activity filter that defines the qualifying practice window. This subsection documents six checks: filter sensitivity; a threshold check with the treatment pipeline held fixed, with influence and clustering diagnostics; a balance check on filter dropout by cohort; a cross-check with the Section~\ref{sec:results:patternA} practice-pattern shift; teammate composition and the two-component composite; and the IOI sister analysis.

\begin{table}[h]
\centering
\caption{Sample filter chain for the within-ICPC regression (Equation~\ref{eq:b1}). The within-ICPC regression and the practice-pattern DiD (Equation~\ref{eq:a1_did}) share the same analysis sample of 10{,}419 users; the within-ICPC regression retains 444 users after the ICPC-specific filters, yielding 579 user-by-contest-year observations across 321 teams in 196 universities.}
\label{tab:b1_filter_chain}
\small
\begin{tabular}{c p{6.6cm} r r r}
\toprule
Step & Filter & Users & Drop & Retention \\
\midrule
0 & The three cohorts in the CF panel (shared with Equation~\ref{eq:a1_did}) & 10{,}419 & --- & --- \\
1 & Linked to any ICPC team roster entry & 1{,}673 & $-8{,}746$ & 16.1\% \\
2 & ICPC participation in contest year $\geq 2023$ & 1{,}026 & $-647$ & 61.3\% \\
3 & \texttt{shift\_pc1} defined (qualifying pre- and post-window practice activity) & 447 & $-579$ & 43.6\% \\
4 & $\bar r_h^{\text{pre}}$ also defined (rated CF history before ChatGPT) & 444 & $-3$ & 99.3\% \\
\bottomrule
\end{tabular}
\end{table}

\subsubsection*{Filter sensitivity}

We re-estimate Equation~\ref{eq:b1} under three activity-filter thresholds for measuring within-user changes in the three behavioral proxies. The treatment is built per (user, contest year), matching the contest-year-conditional window of the main specification, but aggregating at the user-year level rather than collapsing each user to a single value as the headline does. PCA loadings are fit on the headline-threshold sample and frozen across the INTERMEDIATE and PERMISSIVE thresholds so that the treatment is on a comparable scale across thresholds.

\begin{table}[h]
\centering
\small
\caption{Within-ICPC signature-shift test under three activity-filter thresholds, with the treatment built per (user, contest year) and aggregated as within-user changes. AI\_era $\times$ \texttt{shift\_pc1} coefficient with team-clustered standard errors. PCA loadings are fit on the headline-threshold sample and frozen.}
\label{tab:b1_filter_sensitivity}
\begin{tabular}{lrrrrr}
\toprule
Filter & $N_{\text{obs}}$ & $N_{\text{users}}$ & $\psi_{\text{AI}}$ & SE & $p$ \\
\midrule
PERMISSIVE ($\geq 1$ quarter, $\geq 1$ sub) & 541 & 465 & $\mathbf{0.434}$ & $0.257$ & $0.091$ \\
INTERMEDIATE ($\geq 2$ quarters, $\geq 1$ sub) & 405 & 355 & $0.162$ & $0.246$ & $0.512$ \\
HEADLINE ($\geq 2$ quarters, $\geq 5$ subs each) & 386 & 336 & $0.108$ & $0.213$ & $0.611$ \\
\bottomrule
\end{tabular}
\end{table}

Direction is positive across all three thresholds and grows as the filter relaxes ($0.434$ at the loosest, $0.108$ at the headline threshold), so the Step-3 activity filter that drops $56$\% of ICPC participants is not producing the positive direction. None of the relaxed cells reaches conventional significance given the smaller per-user-year samples (smallest $p = 0.091$). This table builds the treatment per user-year to vary the filter cleanly, whereas the manuscript headline collapses each user to a single per-user value. The per-user-year magnitudes here are therefore not directly comparable to the headline $0.190$.

\subsubsection*{Threshold check with the treatment pipeline held fixed; influence and clustering diagnostics}

The per-user-year table above varies the construction and the thresholds at once. Table~\ref{tab:b1_fixed_sweep} separates them: each variant re-runs the headline treatment pipeline itself, holding the ICPC participant pool, the first-qualifying-year anchor, and the frozen broad-reference loadings fixed, and changing only the three activity thresholds (minimum active pre-window quarters / post-window quarters / submissions per active quarter). The headline row is the canonical per-user treatment. The interaction stays positive at every threshold; significance is specific to the headline filter, and the smooth attenuation as low-activity users enter is the pattern classical measurement error predicts when sparser windows measure the practice shift more noisily. The AI\_era-observation column counts observations in the regression sample; the construction stage has four to six more rows per variant, removed because a few AI\_era users lack the baseline-rating control. For the headline row, that is four rows from three users, leaving the $80$ observations and $67$ AI\_era users reported in Table~\ref{tab:descriptives}, Panel~C, in the main text.

\begin{table}[!htbp]
\centering
\footnotesize
\caption{\footnotesize Fixed-construction threshold sweep for the within-ICPC test, with influence and clustering diagnostics on the headline sample. Spec: Equation~\ref{eq:b1} (team-clustered standard errors unless stated). The AI\_era $\times$ \texttt{shift\_pc1} interaction is reported.}
\label{tab:b1_fixed_sweep}
{
\begin{tabular}{lccccc}
\toprule
Thresholds (pre / post / subs) & $N$ obs & $N$ users & AI\_era obs & $\psi_{\text{AI}}$ & $p$ \\
\midrule
$3$ / $2$ / $5$ (headline) & $579$ & $444$ & $80$ & $+0.190$ (SE $0.087$) & $0.028$ \\
$3$ / $2$ / $1$ & $718$ & $554$ & $90$ & $+0.148$ (SE $0.086$) & $0.084$ \\
$3$ / $1$ / $1$ & $748$ & $582$ & $90$ & $+0.084$ (SE $0.082$) & $0.307$ \\
$2$ / $2$ / $5$ & $675$ & $517$ & $101$ & $+0.122$ (SE $0.077$) & $0.111$ \\
\bottomrule
\end{tabular}
}
\end{table}

On the headline sample, influence diagnostics show no single cluster drives the estimate: across $321$ leave-one-team-out re-estimates, one per team cluster in the regression sample, $\psi_{\text{AI}}$ ranges over $[+0.156, +0.227]$ with $p$ in $[0.011, 0.067]$ (eight of the re-estimates push $p$ above $0.05$, none above $0.10$); across $196$ leave-one-university-out re-estimates, the range is $[+0.158, +0.233]$ with $p$ in $[0.012, 0.066]$ (four push $p$ above $0.05$). Alternative clusterings on the same sample give team $p = 0.028$ (baseline), university $p = 0.068$, and user $p = 0.04$; the conservative CRV3 estimator, a jackknife-type small-sample correction at the team level, gives $p = 0.18$. Clustering simultaneously by team and by user returns a standard error smaller than either one-way estimate ($p = 0.0003$). Because $69.6$\% of the $444$ user clusters are singletons on $579$ observations, a setting in which two-way cluster-robust variance estimates are unreliable, we read that number as a small-sample artifact and do not rely on it. The CRV3 result supports the main text's reliance on the direction rather than the size of the within-ICPC estimate.

\subsubsection*{Balance check on filter dropout by cohort}

Step 3 of Table~\ref{tab:b1_filter_chain} drops $579$ of the $1{,}026$ ICPC participants at Step 2, retaining $447$. The first-difference construct is stricter than a post-window-levels filter because it requires both a pre-window (2021-Q3 to 2022-Q3) and a post-window (2023+) with the qualifying activity thresholds. Dropout rates differ substantially across cohorts:

\begin{table}[h]
\centering
\small
\caption{Within-ICPC sample dropout at Step 3 of Table~\ref{tab:b1_filter_chain} (\texttt{shift\_pc1} defined under the within-user first-difference construct), by cohort.}
\label{tab:b1_balance_cohort_dropout}
\begin{tabular}{lrrrr}
\toprule
Cohort & $N$ at Step 2 & $N$ at Step 3 & Dropped & Dropout \% \\
\midrule
pre\_AI & 431 & 144 & 287 & $66.6$\% \\
transition & 412 & 233 & 179 & $43.4$\% \\
AI\_era & 183 & 70 & 113 & $61.7$\% \\
\bottomrule
\end{tabular}
\end{table}

Transition users drop out the least ($43.4$\%), pre\_AI users the most ($66.6$\%); AI\_era ($61.7$\%) sits between. The pattern is consistent with the first-difference construct's both-windows requirement: pre\_AI users are more likely to have stopped active CF practice by 2023, AI\_era users are more likely to lack pre-2022 practice history (their rated career started post-Copilot by construction), and transition users span both periods most fully. Dropped users have higher pre-ChatGPT mean rating ($1{,}932$ versus $1{,}802$, Welch $p < 0.0001$) and lower cutoff-reach rate ($0.74$ versus $0.85$, $p < 0.0001$). ICPC score itself does not differ between retained and dropped users ($p = 0.73$), so the dropout is not directly outcome-correlated; pre-window single-shot AC rate is modestly higher among retained users ($0.46$ versus $0.42$, $p < 0.001$). The cohort interaction in the headline is estimated on a sample whose AI\_era cell is the second-most-attrited; whether this biases the AI\_era $\times$ \texttt{shift\_pc1} slope upward or downward depends on the within-cohort relationship between the within-user change in AI-prompt signature and dropout, which we do not directly observe.

\subsubsection*{Cross-check with the practice-pattern shift}

If the within-ICPC sub-sample participates in the broader practice-pattern shift documented in Section~\ref{sec:results:patternA}, the AI\_era $\times$ post DiD coefficients on the three behavioral proxies should survive on these users' user-quarter observations. We re-run Equation~\ref{eq:a1_did} on the $446$ users that the reconstruction retains at Step 4 (within $\pm 2$ of the headline $444$-user sample, an acceptable reconstruction tolerance).

\begin{table}[h]
\centering
\small
\caption{Equation~\ref{eq:a1_did} AI\_era $\times$ post coefficient $\beta_2$ on the full Section~\ref{sec:results:patternA} sample ($10{,}419$ users) versus the within-ICPC sub-sample ($446$ users). Cluster-robust (user-level) standard errors in parentheses.}
\label{tab:a1_on_b1}
\begin{tabular}{lrr}
\toprule
Outcome & Full sample ($10{,}419$ users) & Within-ICPC sub-sample ($446$ users) \\
\midrule
\texttt{frac\_single\_shot} & $0.0586$ ($0.0063$, $p < 0.001$) & $0.0736$ ($0.0195$, $p < 0.001$) \\
\texttt{median\_attempts} & $-0.3271$ ($0.0374$, $p < 0.001$) & $-0.2705$ ($0.1130$, $p = 0.017$) \\
\texttt{frac\_quick\_retry} & $-0.0190$ ($0.0041$, $p < 0.001$) & $-0.0193$ ($0.0134$, $p = 0.15$) \\
\bottomrule
\end{tabular}
\end{table}

The single-shot effect is \emph{amplified} on the within-ICPC sub-sample to $126$\% of its full-panel magnitude ($0.0736$ versus $0.0586$, $p < 0.001$). The median-attempts effect retains $83$\% of the full-panel magnitude ($-0.2705$ versus $-0.3271$, $p = 0.017$). The quick-retry effect has essentially the same magnitude as the full panel ($-0.0193$ versus $-0.0190$) but loses statistical significance on the smaller sample ($p = 0.15$). The amplified pattern is consistent with the within-ICPC sub-sample being more active by construction (the first-difference construct requires qualifying practice in both the 2021-Q3 to 2022-Q3 pre-window and the 2023+ post-window), so the single-shot shift is larger in magnitude on this sub-sample than on the full panel.

\subsubsection*{Teammate composition and the two-component composite}

ICPC score is a team outcome, so Equation~\ref{eq:b1}'s interaction could reflect teammate composition rather than the user's own practice change. Of the 579 estimation observations, 447 have at least one CF-linked teammate whose shift toward the signature is measurable (228 have two). On the covered sub-sample the interaction is $+0.137$ without and $+0.137$ with teammates' mean shift added as a control (the teammate term is $+0.02$, $p = 0.88$), so the attenuation from the headline $+0.190$ reflects the sub-sample, not the control; a full-sample variant that fills missing teammate means and adds a missing indicator gives $+0.20$ ($p = 0.016$). In the sub-sample where all three team members have a measurable shift, the teammate term is larger ($+0.67$, $p = 0.05$) and the interaction attenuates to $+0.07$ ($p = 0.06$), remaining positive; that cell is heavily selected on linkage and saturated with fixed effects, so we read it as bounding rather than headline evidence. Separately, rescoring the same builder rows with a two-component composite that excludes quick-retry (broad-reference loadings refit on the two remaining sign-aligned changes; the three-component rescoring reproduces the stored treatment at $r = 1.000$) gives $+0.198$ ($p = 0.076$): the interaction's sign and size do not depend on the quick-retry component.

\subsubsection*{Construction details for the change composite}
We fit the \texttt{shift\_pc1} loadings on a broad reference group rather than on the 444 within-ICPC users so that the axis is estimated from population-level co-variation and is not defined by the same sample on which it is then tested. The loadings on \texttt{frac\_single\_shot}, the negation of \texttt{median\_attempts}, and the negation of \texttt{frac\_quick\_retry} are $0.630$, $0.530$, and $0.568$. The group comprises the $4{,}959$ practice-panel users with $\geq 3$ active quarters in the pre-window and $\geq 2$ active quarters in the full post-window (2023-Q1 through 2025-Q4), where a quarter is active if it has $\geq 5$ practice submissions; about 91\% of these users are in the analysis panel, the remainder being out-of-window-cohort or name-only-matched users the panel excludes. Each user's single value is anchored at the user's earliest year in which both the pre- and post-windows contain sufficient practice activity; for users with multiple 2023+ ICPC contests this can differ from the year of the literal earliest contest when that earliest year lacked qualifying activity.

\subsubsection*{Reconciling the roster-linked rating null with the within-ICPC slope}
Weighting the three within-ICPC cohort slopes ($-0.048$, $+0.012$, and $+0.142$) by their user counts (144, 233, and 67; Table~\ref{tab:descriptives}, Panel~C) gives a pooled slope of about $+0.01$, indistinguishable from zero. That is what the roster-linked regression of Section~\ref{sec:results:patternC} finds in its own units: $-2.41$ with a standard error of $4.33$ ($p = 0.58$). Restricting the Section~\ref{sec:results:patternC} rating regression to exactly this 444-user ICPC sample makes the reconciliation direct rather than weighted. The open-pool slope on \texttt{ai\_pc1\_post} is null ($+4.31$, SE $5.88$, $p = 0.46$) on the very users for whom the within-ICPC AI\_era interaction is $+0.190$, while the cross-subpopulation asymmetry estimated against this screened group is $-16.55$ ($p = 0.009$), if anything sharper than the headline $-9.82$. The 444-user slope is noisy and its positive sign is not distinguishable from zero, and part of the sharper gap reflects this group being younger and more AI\_era than the full roster pool.

\subsubsection*{The IOI sister analysis}
\label{app:ioi_sister}

The IOI counterpart to the within-ICPC signature-shift test is weaker than ICPC's in two ways. First, precision: the IOI is a smaller pool, with only $345$ contestant-year observations from $243$ CF-linked contestants over 2023--2025, against $579$ from $444$ users for ICPC. Second, age and timing: IOI participants are typically high-schoolers aged 17--19, while ICPC participants are college students aged 18--23, so a higher share of an IOI contestant's skill-formation window is post-AI, leaving less pre-AI training to anchor the cohort contrast.

Nevertheless, we estimate the same shift-by-cohort specification on the IOI individual contestants linked to CF accounts, clustered at the contestant level. The AI\_era cohort interaction on IOI score is $18.1$ (SE $18.0$, $p = 0.31$): directionally consistent with the ICPC result but far noisier, with a $95\%$ confidence interval including zero. We anchor the non-AI-aided-performance headline on ICPC and treat IOI as a directional check only.

\subsection{Copilot-anchored DiD}
\label{app:copilot_did}

The Section~\ref{sec:results:patternA} headline pins $\text{post}_t$ to ChatGPT. A natural question is whether the practice-pattern shift is detectable around the first AI shock (Copilot, 2021-06-29) before ChatGPT was released. We re-run Equation~\ref{eq:a1_did} with $\text{post}_t$ pinned to the Copilot date and drop the AI\_era cohort, whose first rated contest is post-Copilot by construction, so the cohort $\times$ post-Copilot interaction is unidentified for that cohort. The Copilot-isolation specification restricts the post window to 2021-Q3 through 2022-Q3, five quarters during which Copilot is the only AI shock ($N = 76{,}035$ observations across $8{,}586$ users). The transition $\times$ post-Copilot coefficient on \texttt{frac\_single\_shot} is $0.0398$ (SE $0.0053$, $p < 0.001$), on \texttt{median\_attempts} is $-0.1768$ (SE $0.0306$, $p < 0.001$), and on \texttt{frac\_quick\_retry} is $-0.0100$ (SE $0.0033$, $p = 0.002$), all three significant in the AI-prompt direction. The transition cohort's signature shift is therefore already present around the first AI shock, before ChatGPT, so the headline DiD is not a ChatGPT-specific artifact.

The Copilot-isolation quick-retry coefficient ($-0.0100$, $p = 0.002$) is also notable on its own: it is negative here (the AI-prompt direction), in contrast to the positive coefficient ($0.0090$, $p = 0.002$), opposite to the AI-prompt direction, for the transition cohort in the ChatGPT-anchored headline. One reading is that Copilot's integration into the code editor, as autocomplete-style suggestions, reduced human retries directly, while ChatGPT's generate-and-debug workflow reintroduced debugging cycles among transition users. The shift is more concentrated around Copilot than around ChatGPT for this mid-career cohort: the ChatGPT-anchored transition coefficient on \texttt{frac\_single\_shot} ($0.0139$ in Table~\ref{tab:patternA_results}) is smaller because its pre-period baseline already absorbs the post-Copilot shift. A second specification that extends the post window through 2025-Q4 ($N = 127{,}124$) recovers similar magnitudes on single-shot and median-attempts ($0.0339$ and $-0.1569$), with the quick-retry coefficient attenuating to null ($-0.0027$, SE $0.0028$, $p = 0.333$) as the ChatGPT-era observations rejoin the post window.

\begin{figure}[htp]
\centering
\includegraphics[width=\textwidth]{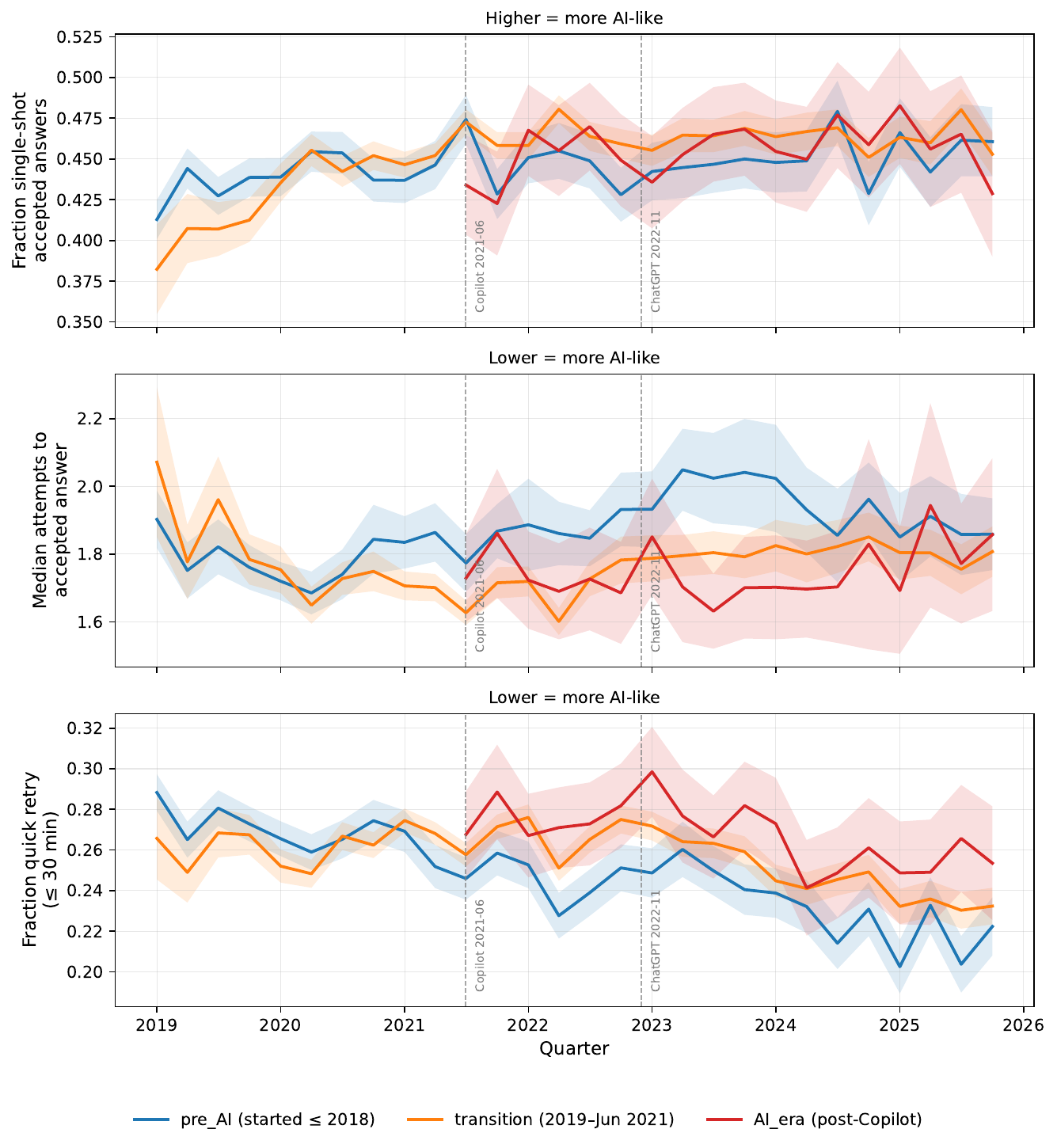}
\caption{AI-prompt signature components by cohort and quarter. Each panel plots the cohort mean of one practice proxy by quarter with 95\% confidence bands; the vertical axis names the metric and the panel title gives its AI-prompt direction. Vertical dashed lines mark the Copilot (2021-06) and ChatGPT (2022-11) releases, and the AI\_era series begins after the Copilot launch by cohort construction. To hold the sample fixed across the time series, the figure uses spanning users (those with at least 30 practice submissions in both the pre-Copilot and post-ChatGPT windows; Table~\ref{tab:summary_stats}) in the three focal cohorts.}
\label{fig:cohort_signature}
\end{figure}

Figure~\ref{fig:cohort_signature} shows the raw signature-component levels by cohort and quarter across both release dates, the descriptive backdrop for the two-shock timing discussed above.

\begin{figure}[!t]
\centering
\includegraphics[width=\textwidth]{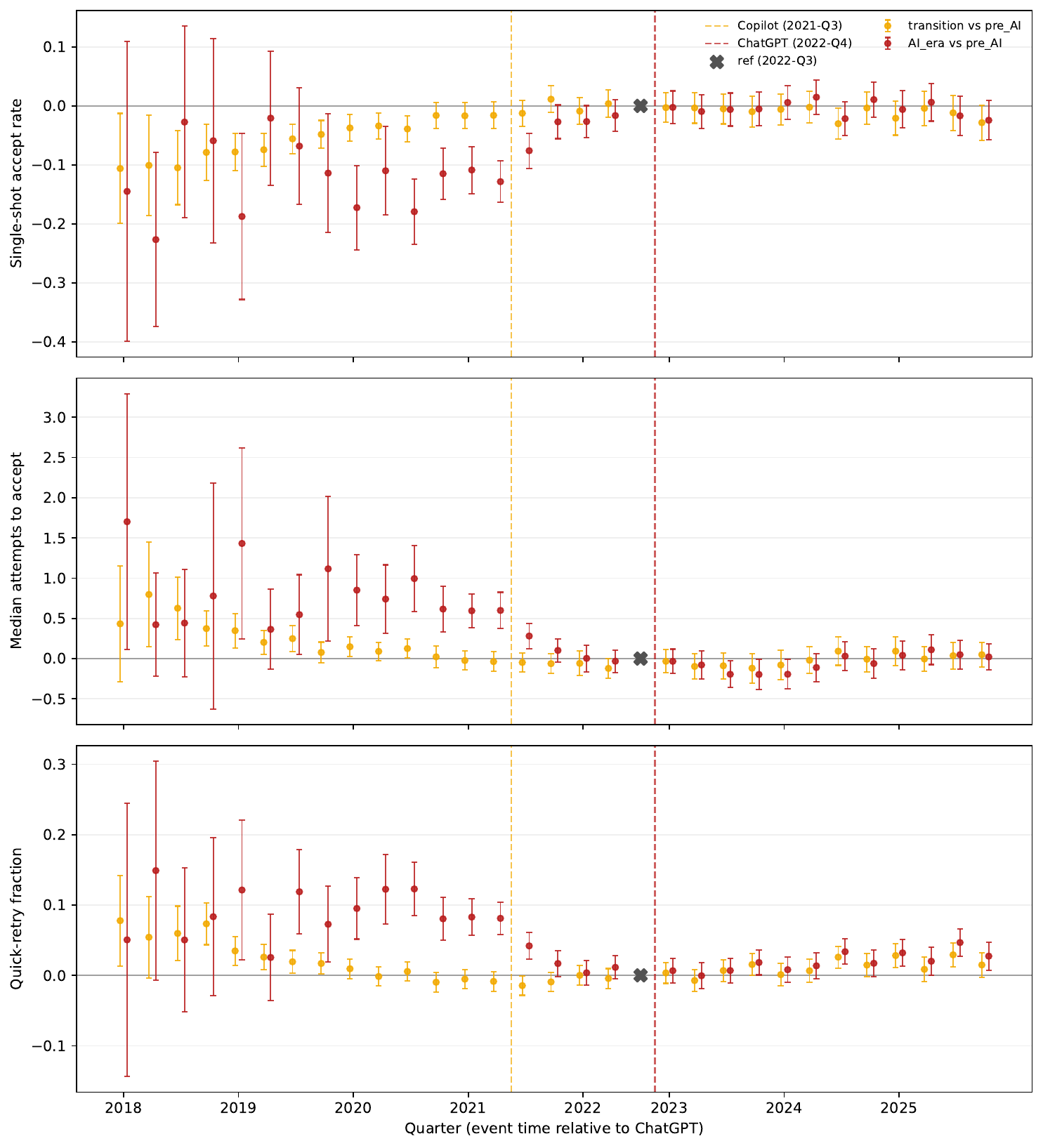}
\caption{\footnotesize Cohort event study of the three AI-prompt signature proxies (Equation~\ref{eq:a1_did} with quarterly event-time interactions). Each panel plots event-time coefficients for the transition (orange) and AI\_era (red) cohorts relative to the pre\_AI cohort; the reference period is 2022-Q3 (gray X, normalized to zero), the vertical dashed lines mark the first post-Copilot quarter (2021-Q3) and the ChatGPT release quarter (2022-Q4), and bands are 95\% confidence intervals (user-clustered standard errors). Neither cohort jumps at either release. Because the event study runs on the practice panel, whose submissions can predate a user's first rated contest, a few AI\_era users appear before 2021; these few-user leads are uninformative, so the testable pre-trend contrast is the transition cohort, which rejects parallel trends on all three proxies (Section~\ref{sec:results:patternA}).}
\label{fig:cohort_event_study}
\end{figure}

\subsection{COVID-19 disruption (2020--2021)}
\label{app:covid}
\label{sec:method:covid}

The empirical window straddles the COVID-19 pandemic. ICPC 2020 was postponed and held in 2021 with reduced attendance ($117$ teams in our rosters, against $134$--$138$ in 2018--2019); IOI 2020 was held virtually (participation in our rosters did not fall); multiple national qualification pipelines were disrupted across 2020 and 2021. The CF practice panel is the least exposed of the three samples to the pandemic disruption because it is continuous, while the contest panels run on an annual schedule; the cohort $\times$ post DiD on practice-pattern proxies (Section~\ref{sec:results:patternA}) absorbs year-level shocks through its quarter fixed effects, and within-user rating dynamics include 2020 and 2021 observations as part of the natural trajectory rather than as anomalies to be removed with indicator variables. The cross-subpopulation CF rating asymmetry (Section~\ref{sec:results:patternC}) is similarly insulated: the rating-change outcome $\Delta r_h$ is constructed as a pre-ChatGPT-versus-post-ChatGPT difference where both windows span multiple years, washing out year-specific shocks. The within-ICPC signature-shift test (Section~\ref{sec:results:patternB}) restricts the sample to contest years $\geq 2023$, which entirely post-date COVID, so the COVID years cannot directly contaminate the \texttt{shift\_pc1} coefficient. The quarterly cohort event study (Figure~\ref{fig:cohort_event_study}) does span 2020--2021. Its quarter fixed effects absorb the common COVID shock, and any cohort-differential COVID response is one more reason Section~\ref{sec:results:patternA} reads the cohort contrast as descriptive rather than causal.

\subsection{Data provenance and research ethics}
\label{app:provenance}

All data are collected from publicly accessible sources: submission metadata and rating histories through the documented CF public API and public contest pages, and ICPC results from \texttt{cphof.org} and IOI results from the contests' official public pages (\texttt{stats.ioinformatics.org}). Collection used rate-limited requests, and no login-gated content was accessed. CF usernames are public pseudonyms, and the roster linkage relies only on information the sources themselves publish (usernames and publicly listed contestant names). Working datasets are de-identified for analysis, results are reported only in aggregate, and no individual is named or identifiable in the paper. The study is secondary analysis of public records. We neither interact with nor intervene on the individuals observed, and we obtain no private identifiable information about them. The IRB of the author's institution determined that the study is not human subjects research (determination 202608190, August 25, 2026), and informed consent was not required.

\subsection{ICPC/IOI--CF linkage and sensitivity to the excluded name-only matches}
\label{app:linkage}

Candidate competitors from the ICPC and IOI rosters (2018--2025) are linked to CF accounts in two ways. ICPC contestants carry a CF username published on their Competitive Programming Hall of Fame profile, a direct link; the analysis ICPC roster is the 2{,}189 distinct accounts (2{,}784 contestant-year records, 88.0\% of ICPC contestant-years) that carry such a username. IOI contestants are linked by a name-and-country match to the IOI participant lists, yielding 1{,}164 accounts. The two sets overlap in 236 accounts, for 3{,}117 roster-linked CF users.

The name-and-country matcher processes 4{,}219 candidate persons and returns 1{,}592 matched to at least one CF username (1{,}036 ICPC, 467 IOI, 89 both). Each match carries a best-status confidence tier (verified $\geq 0.85$, likely 0.70--0.85, uncertain 0.55--0.70, rejected below 0.55), with counts 441 / 327 / 267 / 557 by person (441 / 338 / 330 / 1{,}106 across 2{,}215 match pairs). The matched set spans the competitive frontier: 15 World Champions and 86 gold, 114 silver, and 154 bronze medalists.

The account-level arithmetic connects these person-level counts to the panel. The roster-candidate scrapes collected $4{,}491$ CF accounts, targeting both link types: accounts named directly on cphof profiles and accounts proposed by the name-and-country matcher (a matched person can propose more than one candidate account, which is why the $2{,}215$ match pairs exceed the $1{,}592$ matched persons). Of the $4{,}491$, the $3{,}117$ confirmed roster links above are retained: $2{,}189$ ICPC direct-username accounts and $1{,}164$ IOI accounts whose name match is confirmed by an attached contest record, overlapping in $236$. The remaining $4{,}491 - 3{,}117 = 1{,}374$ accounts, scraped as candidates whose best match could not be confirmed against a contest record, are the excluded name-only matches of the next paragraph. Together with the $8{,}748$ accounts of the non-rostered expansion, the scrape universe is $4{,}491 + 8{,}748 = 13{,}239$ accounts, and the exclusions lead to the panel: $13{,}239 - 1{,}374$ (name-only) $- 1{,}408$ (outside the cohort window) $- 38$ (no practice-mode submissions) $= 10{,}419$.

Of the 3{,}117 roster-linked accounts, 2{,}494 are cohort-assignable (1{,}673 ICPC and 1{,}030 IOI contestants, 209 of whom competed in both): their first rated CF contest falls within the 2015--2024 cohort window of Section~\ref{sec:data:sample} in the main text; the remaining 623 fall outside the window and belong to no cohort. The linked contestants span 354 universities and 90 countries over 2018--2025. Three groups of scraped users are excluded from the analysis panel: the 1{,}374 lower-confidence name-only matches discussed below; 1{,}408 users whose first rated contest falls outside the cohort window (the 623 rostered contestants above and 785 non-rostered users); and 38 users with no practice-mode submissions in the analysis window of 2018-Q1 through 2025-Q4 excluding the 2022-Q4 boundary quarter (14 roster-linked and 24 non-rostered; the 14 reconcile the 2{,}494 cohort-assignable contestants with the panel's 2{,}480 roster-linked users). The panel is not otherwise filtered on any outcome variable.

The 1{,}374 users matched to a roster only by a lower-confidence name-and-country match, for which we cannot confirm a contest record or attach a contest score, are excluded from the analysis panel (Section~\ref{sec:data:sample}). The cross-subpopulation asymmetry of Section~\ref{sec:results:patternC} is insensitive to their handling: excluding them (the headline) gives a non-rostered slope of $-12.24$ and a roster-linked null of $-2.41$ (pooled interaction $-9.82$, $p = 0.048$); leaving them in the non-rostered subpopulation gives $-12.16$ and $-2.41$ (interaction $-9.75$, $p = 0.048$); and assigning them to the roster-linked subpopulation gives $-12.24$ and $-4.88$ (interaction $-7.36$, $p = 0.110$). The within-subpopulation slopes are stable across all three handlings; only the pooled-interaction $p$-value softens when the AI-era-heavy name-only matches are forced into the roster-linked arm, as expected if unverifiable matches contaminate the screened arm. Full results are provided in the replication package.

For reference, Table~\ref{tab:cohort_composition} reports the cohort composition of the CF analysis panel underlying the Section~\ref{sec:method:a1} DiD.

\begin{table}[h]
\centering
\caption{Cohort composition of the CF headline cohort $\times$ post DiD sample (Section~\ref{sec:method:a1}). All counts are over the three cohorts (pre\_AI, transition, AI\_era). Spanning $N$ counts users with $\geq 30$ practice submissions in both the pre-Copilot and post-ChatGPT windows. Per-user medians are computed on the practice-pattern panel after the 2018--2025 and non-2022-Q4 filters.}
\label{tab:cohort_composition}
\small
\begin{tabular}{l r r r r r r}
\toprule
Cohort & $N$ & Spanning $N$ & \multicolumn{2}{c}{Median rating} & \multicolumn{2}{c}{Median practice sub counts} \\
\cmidrule(lr){4-5} \cmidrule(lr){6-7}
& & & pre-ChatGPT & post-ChatGPT & pre-ChatGPT & post-ChatGPT \\
\midrule
pre\_AI & 3{,}347 & 2{,}061 & 1{,}874 & 1{,}874 & 422 & 49 \\
transition & 5{,}293 & 4{,}006 & 1{,}582 & 1{,}665 & 462 & 142 \\
AI\_era & 1{,}779 & 332 & 1{,}348\textsuperscript{$\dagger$} & 1{,}674 & 163 & 408 \\
\bottomrule
\end{tabular}

\vspace{0.5em}
\footnotesize\raggedright\textit{$\dagger$ Median over the 1{,}336 AI\_era users with at least one pre-ChatGPT month of rated activity in the monthly rating panel; the remaining 443 AI\_era users entered their first rated contest after ChatGPT and have no pre-ChatGPT rating to summarize. Cohort sample sizes are unbalanced: transition is the largest cohort (5{,}293) and AI\_era is the smallest (1{,}779), and the spanning sub-sample collapses AI\_era to 332 because most AI\_era users cannot yet supply 30 pre-Copilot submissions. The DiD uses the full analysis sample (10{,}419 users; see Section~\ref{sec:data:sample}), not the spanning sub-sample. Median rating is the per-user most recent rating in each window; median practice submissions are per-user totals on the practice-mode panel. Pre/post practice activity moves in opposite directions across cohorts (pre\_AI's median drops from 422 to 49 as veterans exit; AI\_era's median rises from 163 to 408 as new entrants accumulate post-ChatGPT activity); the $\log(1 + n_{h,t})$ control in the DiD absorbs this directly at the user-quarter level.}
\end{table}

\subsection{Sample summary and practice-shift tables}
\label{app:descriptives}

Table~\ref{tab:summary_stats} reports the headline summary statistics for the two contest panels, the CF panel, and the linkage; it is referenced from the main text.

\begin{table}[p]
\centering
\footnotesize
\renewcommand{\arraystretch}{0.92}
\caption{Summary statistics for the two contest panels, the CF panel, and the ICPC/IOI--CF linkage. Panels A and B are descriptive panels of the ICPC university-year and IOI country-year records; Panel C is the combined CF open pool panel (contest-record roster-linked users and the non-rostered expansion); Panel D summarizes the linkage connecting them. Definitions are given in the table note.}
\label{tab:summary_stats}
\begin{tabular}{l r}
\toprule
\multicolumn{2}{l}{\textit{Panel A. ICPC university-year panel}} \\
\midrule
University-year cells & 1{,}055 \\
Distinct universities & 394 \\
\quad spanning pre-AI and AI-era contest periods & 112 \\
Contest period: pre-AI / transition / AI-era & 272 / 249 / 534 \\
Team CF rating: mean / median / SD & 2{,}150 / 2{,}146 / 334 \\
Teams with all members CF-linked & 768 \,(72.8\%) \\
Universities by contest year, 2018--2025 & 138, 134, 117, 132, 124, 130, 141, 139 \\
\addlinespace
\multicolumn{2}{l}{\textit{Panel B. IOI country-year panel}} \\
\midrule
Country-year cells & 708 \\
Distinct countries & 103 \\
\quad spanning pre-AI and AI-era contest periods & 86 \\
Contest period: pre-AI / transition / AI-era & 353 / 89 / 266 \\
Delegation CF rating: mean / median / SD & 1{,}953 / 1{,}999 / 401 \\
Delegations with all contestants CF-linked & 251 \,(35.5\%) \\
Country-years by contest year, 2018--2025 & 88, 88, 88, 89, 89, 88, 93, 85 \\
\addlinespace
\multicolumn{2}{l}{\textit{Panel C. CF open pool panel}} \\
\midrule
Unique users & 10{,}419 \\
\quad of which spanning subset ($\geq$30 subs both windows) & 6{,}399 \,(61.4\%) \\
User-month observations & 793{,}920 \\
Cohort: pre-AI / transition / AI-era & 3{,}347 / 5{,}293 / 1{,}779 \\
\quad spanning subset, same order & 2{,}061 / 4{,}006 / 332 \\
Cohort CF rating (mean): pre-AI / transition / AI-era & 1{,}818 / 1{,}610 / 1{,}531 \\
Practice submissions per user (median) & 1{,}489 \\
Rated contests per user (median) & 54 \\
Source: roster-linked / non-rostered & 2{,}480 / 7{,}939 \\
\addlinespace
\multicolumn{2}{l}{\textit{Panel D. ICPC/IOI $\rightarrow$ CF linkage}} \\
\midrule
ICPC published-username linkage: accounts & 2{,}189 \\
IOI name-and-country linkage: accounts & 1{,}164 \\
Roster-linked CF users (ICPC $\cup$ IOI), all cohorts & 3{,}117 \,(236 in both; 2{,}494 cohort-assignable) \\
\bottomrule
\end{tabular}

\vspace{0.5em}
\footnotesize\raggedright\textit{Source: authors' calculations over the cleaned panels. Cohorts are assigned by first rated contest date: pre-AI (2015--2018), transition (2019 to the day before the Copilot release), and AI-era (Copilot release to 2024-12-31); users first rated outside 2015--2024 fall out of window and are excluded (Section~\ref{sec:data:sample}). Contest periods in Panels A and B group contest years: 2018--19, 2020--21, and 2022--25 for ICPC, and 2018--21, 2022, and 2023--25 for IOI. ``Spanning'' CF users have $\geq$30 practice submissions in both the pre-Copilot and post-ChatGPT windows; spanning universities and countries appear in both the pre-AI and the AI-era contest period. Cohort CF rating means are computed over all user-months within each cohort; practice-submission and rated-contest counts are medians over the 10{,}419-user analysis sample. The published-username linkage uses the CF username that the Hall of Fame lists on each ICPC contestant profile; Panel~A and the within-ICPC test (Section~\ref{sec:method:b1}) rely on it, while the name-and-country audit is the linkage route for IOI. Appendix~\ref{app:linkage} documents the candidate-matching confidence tiers and the sensitivity of the cross-subpopulation asymmetry to the 1{,}374 excluded name-only matches. Per-analysis estimation sub-samples are documented in Section~\ref{sec:data:sample}.}
\end{table}

\clearpage

\input{construct_validation_appendix}

\clearpage

%% file: construct_validation_appendix.tex
\section{Construct Validation of the AI-Prompt Signature}
\label{sec:construct-validation}

The cohort practice-pattern results (Section~\ref{sec:results:patternA}), the
cross-subpopulation asymmetry test (Section~\ref{sec:results:patternC}), and the
within-ICPC signature-shift test (Section~\ref{sec:results:patternB}) rest on the
AI-prompt signature: the behavioral composite of higher single-shot AC rate,
lower median attempts to AC, and lower quick-retry rate that we attribute to an
AI prompt-and-fix workflow. A skeptical reading is that this signature
captures programmer skill or an analytic work style rather than AI use
directly. This appendix bounds that concern with three tests (a cross-sectional test and a within-user test of discriminant validity, and a within-user event study) on the Codeforces panel restricted to submissions where
\texttt{author.participantType == "PRACTICE"} (the canonical practice
filter on the combined re-scraped panel; each test below reports its own sample).

The tests address two distinct alternatives to the AI-practice
interpretation. Subsection~\ref{ssec:cv-discriminant} tests whether the
signature is a proxy for skill, by examining its relationship to CF rating
in the strictly pre-shock window (2018-Q1 to 2022-Q3). Subsection~\ref{ssec:cv-event-study}
examines whether established practitioners exhibit a within-user
discontinuity in the signature at AI release dates, which would distinguish
a discrete AI-availability effect from a smooth skill-acquisition trend.
Subsection~\ref{ssec:cv-synthesis} synthesizes these
tests against the experimental causal anchor from \citet{ShenTamkin2026}.

\subsection{Discriminant validity against pre-shock skill}
\label{ssec:cv-discriminant}

If the signature were a tight proxy for CF rating, pre-shock high-rated
users would already exhibit the AI-prompt signature: higher single-shot
AC rates, lower median attempts to AC, and lower quick-retry rates than
lower-rated users. We test this directly using the pre-shock window
2018-Q1 through 2022-Q3 (strictly before the ChatGPT release on
2022-11-30).

\textbf{Cross-sectional test.} For each user with at least four active
pre-shock quarters ($n = 8{,}535$), we compute the user-level mean of
each signature component and the user-level mean pre-shock CF rating.
Pearson and Spearman correlations between the signature components and
mean rating are reported in Table~\ref{tab:cv-discriminant-cross}.

\begin{table}[h]
\centering
\caption{Pre-shock cross-sectional correlation: signature components vs.\ CF rating}
\label{tab:cv-discriminant-cross}
\begin{tabular}{lrrrrr}
\toprule
Signature component & $n$ users & Pearson $r$ & SE & $p$ & Spearman $r$ \\
\midrule
Single-shot AC rate            & 8{,}535 & $-0.091$ & $0.011$ & $<0.001$ & $-0.095$ \\
Median attempts to AC          & 8{,}535 & $+0.157$ & $0.011$ & $<0.001$ & $+0.175$ \\
Quick-retry rate               & 8{,}535 & $+0.147$ & $0.011$ & $<0.001$ & $+0.161$ \\
\bottomrule
\end{tabular}
\par\vspace{0.4em}
\footnotesize\raggedright \textit{Notes.} Pre-shock window: 2018-Q1 to 2022-Q3. User-level
means; users included if active in $\geq 4$ pre-shock quarters with
$\geq 5$ practice submissions per quarter. Filter:
\texttt{author.participantType == "PRACTICE"} on the combined re-scraped panel.
\end{table}

All three correlations have signs \emph{opposite} to the AI-prompt
prediction. In the pre-shock window, higher-rated users had \emph{lower}
single-shot AC rates, \emph{higher} median attempts to AC, and higher quick-retry rates than lower-rated users. Skill in the pre-shock
era was associated with attempting harder problems that required more
iteration, not with the prompt-and-fix pattern. Magnitudes are weak
($|r| < 0.20$) but the directional contrast with the AI-prompt prediction
is unambiguous.

\begin{figure}[h]
\centering
\includegraphics[width=0.95\textwidth]{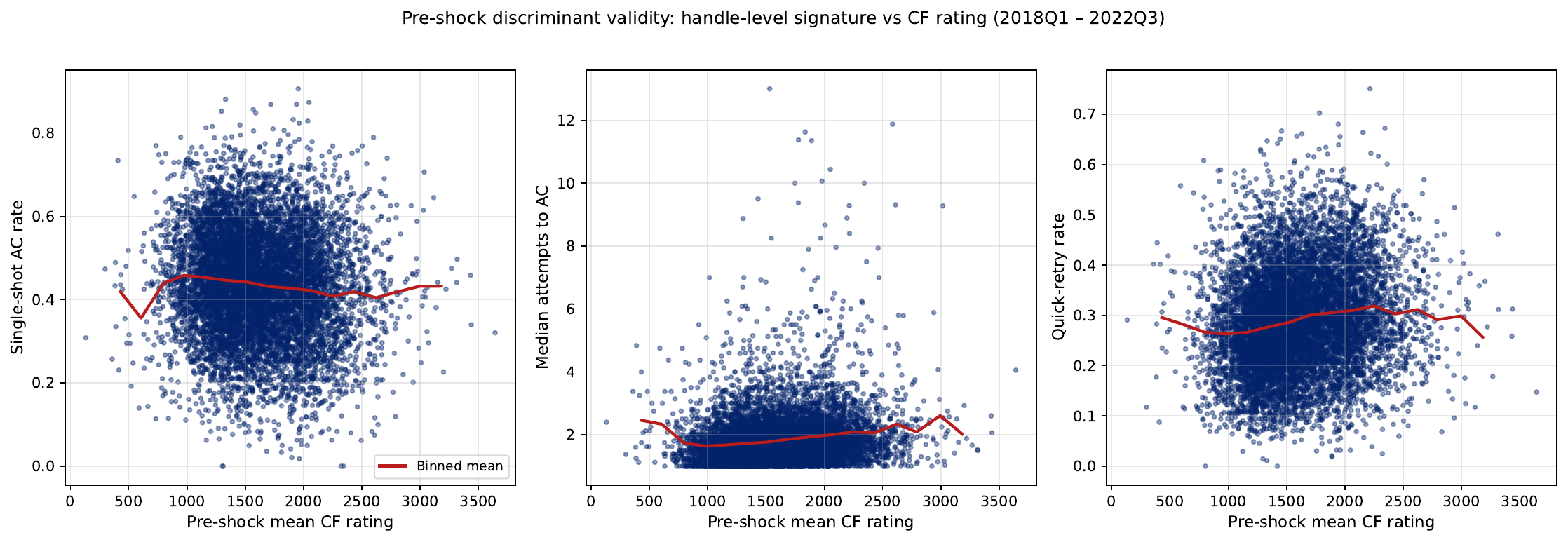}
\caption{Pre-shock discriminant validity: user-level mean signature
components vs.\ pre-shock mean CF rating ($n = 8{,}535$ users, 2018-Q1 to
2022-Q3). Red lines are binned means. The cross-sectional relationships
have signs opposite to the AI-prompt prediction: higher pre-shock rating
is associated with lower single-shot AC, higher median attempts, and
higher quick-retry rates --- inconsistent with a skill-proxy
interpretation of the signature.}
\label{fig:cv-discriminant}
\end{figure}

Figure~\ref{fig:cv-discriminant} plots the user-level scatter and binned
mean for each component against pre-shock rating. The binned-mean lines
make the directional contrast visible.

\textbf{Within-user dynamics.} The cross-sectional contrast could in
principle mask a within-user skill-on-signature relationship. We test this
with the specification
\begin{equation}
\label{eq:cv-within-user}
\text{signature}_{ht} = \alpha_h + \gamma_t + \delta \cdot \frac{\text{rating}_{ht}}{100} + \varepsilon_{ht},
\end{equation}
where $\alpha_h$ is a user fixed effect, $\gamma_t$ is a quarter fixed
effect, and $\delta$ is the within-user elasticity per 100 rating points.
The sample is the pre-shock quarterly panel; standard errors are clustered
at the user level. Table~\ref{tab:cv-discriminant-within} reports the
results.

\begin{table}[h]
\centering
\caption{Pre-shock within-user elasticity: signature vs.\ rating}
\label{tab:cv-discriminant-within}
\begin{tabular}{lrrrr}
\toprule
Signature component & $n$ obs & $\delta$ per 100 rating points & SE & $p$ \\
\midrule
Single-shot AC rate            & 80{,}184 & $+0.00434$ & $0.00045$ & $<0.001$ \\
Median attempts to AC          & 79{,}514 & $-0.03335$ & $0.00367$ & $<0.001$ \\
Quick-retry rate               & 80{,}203 & $+0.00110$ & $0.00036$ & $0.002$ \\
\bottomrule
\end{tabular}
\par\vspace{0.4em}
\footnotesize\raggedright \textit{Notes.} Specification~\eqref{eq:cv-within-user}: user and
quarter fixed effects; standard errors clustered at the user level. Pre-shock
window: 2018-Q1 to 2022-Q3.
\end{table}

Within users, the signature does shift with rating, but the magnitudes
are small. The single-shot AC elasticity is
$\delta = 0.0043$ per 100 rating points, meaning that a user would need to gain
roughly 1{,}350 rating points to produce a single-shot AC shift of magnitude
comparable to the AI-era cohort $\times$ post coefficient of $+0.0586$
documented in Section~\ref{sec:results:patternA}. The median-attempts component has an
elasticity of $-0.033$ per 100 rating points against the cohort coefficient of $-0.327$,
roughly a 10:1 ratio of cohort contrast to skill elasticity. Skill movements at the
magnitudes plausibly achievable within the 2022-Q4 to 2025-Q4 post-shock
window (typically tens to low hundreds of rating points for established users, not 1{,}000+) cannot reproduce the cohort contrast.

Taken together, the cross-sectional and within-user tests establish that
the signature is not a tight proxy for skill: the cross-sectional sign
runs opposite to the AI-prompt prediction, and the within-user
skill-on-signature elasticity is an order of magnitude too small to
explain the cohort contrast.

\subsection{Within-user stability around AI release}
\label{ssec:cv-event-study}

The remaining alternative explanation for the cohort contrast is a smooth-trend story:
practice patterns have been drifting toward the AI-prompt signature over time for
reasons unrelated to AI release, and we mistakenly attribute the cohort
contrast to AI exposure. A direct test is whether established users
exhibit a discrete change in their signature at known AI release dates.

\begin{figure}[h]
\centering
\includegraphics[width=0.95\textwidth]{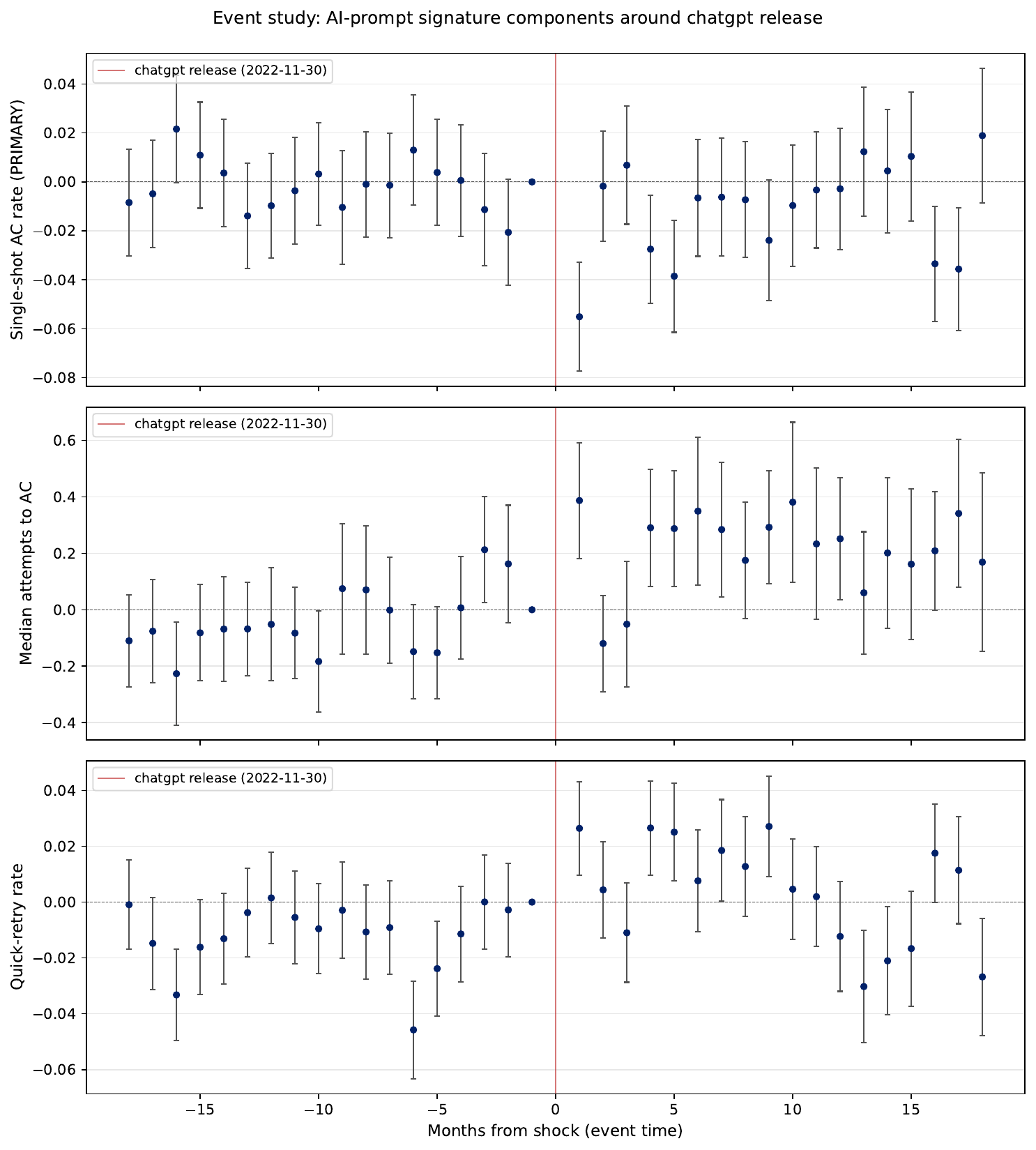}
\caption{Within-user event study at the ChatGPT release. Spanning panel
of $1{,}712$ users active both pre- and post-shock. Reference period:
$k = -1$ (one month before release, October 2022). Vertical red line marks
release. The release month is dropped, so no $k = 0$ coefficient
is estimated. The post-release movement that exists runs \emph{away} from the
AI-prompt signature (fewer single-shot accepts, more attempts); for the
single-shot rate, coefficients on the pre-shock lead months are individually
indistinguishable from zero.}
\label{fig:cv-event-study}
\end{figure}

We estimate a within-user event study around the ChatGPT release
(2022-11-30) on the spanning panel of users active both pre- and
post-shock ($n = 1{,}712$ users, $25{,}395$ user-month observations):
\begin{equation}
\label{eq:cv-event-study}
\text{signature}_{ht} = \alpha_h + \sum_{k = -18,\, k \notin \{-1,\, 0\}}^{+18} \beta_k \cdot \mathbf{1}[\text{event}_{ht} = k] + \varepsilon_{ht},
\end{equation}
where event time is months from 2022-11. User FE absorbs user-level
heterogeneity; the reference period is one month before release,
and the release month itself ($k = 0$) is dropped from estimation. Standard
errors are clustered at the user level. Figure~\ref{fig:cv-event-study}
plots the coefficient time series.

Because the release month is dropped, the profile is read from
the adjacent coefficients, and what movement exists after the release is
directed \emph{away} from the AI-prompt signature: single-shot acceptance falls
and median attempts rise at $k = +1$, with quick retries rising as well.
Established users show no shift \emph{toward} the signature at the release.
For the single-shot rate, coefficients on the pre-shock lead months are
individually indistinguishable from zero; the other two components have
scattered individually significant lead months (full results are provided in the replication package).

The Copilot release (2021-06-29) points the same way but with pre-shock
parallel-trends concerns (several pre-shock lead-month coefficients for the
single-shot rate are significantly positive). We therefore treat the Copilot
event study as inconclusive and relegate it to the replication package.

Cohort-stratified event studies on the ChatGPT shock show the same pattern: no within-user movement toward the signature among existing practitioners. Within the pre-AI
cohort ($n = 733$ users, $8{,}791$ user-month observations in the event window),
post-release movement again runs away from the signature or is
flat. Within the transition
cohort ($n = 775$ users, $14{,}460$ user-month observations), the post-shock pattern
matches the pre-AI cohort's direction, with no movement toward the signature at
the release. The AI-era stratum (the remaining 204 of the $1{,}712$
spanning-panel users, $2{,}144$ user-month observations) is not shown. AI-era
users enter rated competition after the Copilot release by construction, so
their lead windows before the ChatGPT release are short and thin, and the
event study was never run for that stratum.

\textbf{Interpretation.} Established practitioners did not shift their
practice patterns at the ChatGPT release. Two implications follow. First, the
smooth-trend alternative is bounded: if a smooth time trend were driving
the cohort contrast, established users would also drift toward the
AI-prompt signature over time, and they do not. Second, the cohort contrast must
originate primarily in cohort \emph{composition}: users who entered the
platform after AI's broad availability arrived with practice patterns that
differ from those of the pre-AI cohort. This is consistent with the paper's
skill-formation framework: AI exposure during the formation period shapes
practice habits; AI exposure among experienced practitioners does not change
theirs.

\subsection{Synthesis with experimental evidence}
\label{ssec:cv-synthesis}

The two subsections' tests bound two distinct alternatives to
the AI-practice interpretation. The discriminant test
(Subsection~\ref{ssec:cv-discriminant}) rules out the signature as a proxy
for \emph{static} skill: the cross-sectional sign runs opposite to the
AI-prompt prediction and the within-user skill-on-signature elasticity
is an order of magnitude too small to explain the cohort contrast.
The event-study test (Subsection~\ref{ssec:cv-event-study}) rules out
adoption by established practitioners: there is no within-user movement toward the signature at the ChatGPT release (the Copilot event study is inconclusive), which locates the cohort contrast in cohort composition rather than in existing-user adoption.

What these two tests establish is narrower than AI causation: the
cross-cohort signature difference is a genuine, cohort-compositional shift
in practice style, not an artifact of static skill or of established users
adopting AI. They do not establish that the cohort difference is
\emph{caused} by AI. As Section~\ref{sec:results:patternA} shows, the
cohort gradient does not coincide with the AI releases (the cohort event
study rejects parallel trends and shows no discontinuity), the AI-era and transition cohorts are indistinguishable at matched career stage, and the signature gap is largest on hard problems, where AI is weakest, rather than on the easy, templated problems AI handles best. The signature is therefore best read as a practice-style
proxy whose cross-cohort variation carries a secular and maturation
component; the AI interpretation is the leading reading, not an identified
effect.

The randomized experiment of \citet{ShenTamkin2026} on 52 software developers provides the complementary causal anchor. Their RCT randomly assigns AI
use during a learning task and finds that AI users score substantially
lower on a subsequent non-AI-aided assessment (a $17\%$ reduction,
Cohen's $d = 0.738$, $p = 0.010$), with no significant gain in completion time. In their exploratory clustering of usage patterns, participants in the low-engagement clusters (those delegating code generation or debugging to the AI) scored far below those in the cognitively engaged clusters, although the clusters are small and no formal test of where the effect concentrates is reported. That low-engagement, delegating way of working with AI is one of the usage patterns our signature tracks. Because AI use is randomly assigned in their design, the average effect on non-AI-aided performance is causally identified in that setting; the pattern-level contrast is descriptive. Our signature is the multi-year,
field-observational analog of the usage patterns they observe.

Taken together, the field tests and the experimental anchor establish that
the AI-prompt signature is a behavioral practice-style proxy consistent
with AI-assisted practice: the discriminant and event-study tests rule out
static skill and existing-user adoption, and the RCT of \citet{ShenTamkin2026} shows that AI use during learning lowers later unaided performance, with the lowest scores among participants who delegated to the AI. What they do not establish is
that the cross-cohort variation in the signature is AI-caused; the
career-stage-matched evidence in Section~\ref{sec:results:patternA} attributes
part of it to secular practice-norm change and novice maturation. We
therefore use the signature as a proxy for AI-style practice throughout,
and treat its AI interpretation as the leading but unverified reading
rather than a measured fact.

%% file: main.bbl
\begin{thebibliography}{}

\bibitem[\protect\citeauthoryear{Aaltonen and Stelmaszak}{Aaltonen and
  Stelmaszak}{2024}]{AaltonenStelmaszak2023}
Aaltonen, A. and M.~Stelmaszak (2024).
\newblock {The Performative Production of Trace Data in Knowledge Work}.
\newblock {\em Information Systems Research\/}~{\em 35\/}(3), 1448--1462.

\bibitem[\protect\citeauthoryear{Abdel-Karim, Pfeuffer, Carl, and
  Hinz}{Abdel-Karim et~al.}{2023}]{AbdelKarimEtAl2023}
Abdel-Karim, B., N.~Pfeuffer, K.~V. Carl, and O.~Hinz (2023).
\newblock {How AI-Based Systems Can Induce Reflections: The Case of
  AI-Augmented Diagnostic Work}.
\newblock {\em MIS Quarterly\/}~{\em 47\/}(4), 1395--1424.

\bibitem[\protect\citeauthoryear{Acemoglu, Kong, and Ozdaglar}{Acemoglu
  et~al.}{2026}]{AcemogluKongOzdaglar2026}
Acemoglu, D., D.~Kong, and A.~Ozdaglar (2026, February).
\newblock {AI, Human Cognition and Knowledge Collapse}.
\newblock NBER Working Paper 34910, National Bureau of Economic Research.

\bibitem[\protect\citeauthoryear{Alavi and Leidner}{Alavi and
  Leidner}{2001}]{AlaviLeidner2001}
Alavi, M. and D.~E. Leidner (2001).
\newblock {Research Commentary: Technology-Mediated Learning---A Call for
  Greater Depth and Breadth of Research}.
\newblock {\em Information Systems Research\/}~{\em 12\/}(1), 1--10.

\bibitem[\protect\citeauthoryear{Autor, Rodchenko, Martin, Iscenko, Strand,
  Pearl, and Ferere}{Autor et~al.}{2026}]{AutorEtAl2026}
Autor, D., T.~Rodchenko, J.~Martin, Z.~Iscenko, S.~Strand, D.~Pearl, and
  M.~Ferere (2026, September).
\newblock {Does AI Assistance Enhance or Erode Expertise? Evidence from a
  Three-Month Field Experiment in Patent Drafting}.
\newblock NBER Working Paper 35720, National Bureau of Economic Research.
\newblock Pre-registered, AEARCTR-0015823.

\bibitem[\protect\citeauthoryear{Baird and Maruping}{Baird and
  Maruping}{2021}]{BairdMaruping2021}
Baird, A. and L.~M. Maruping (2021).
\newblock {The Next Generation of Research on IS Use: A Theoretical Framework
  of Delegation to and from Agentic IS Artifacts}.
\newblock {\em MIS Quarterly\/}~{\em 45\/}(1), 315--341.

\bibitem[\protect\citeauthoryear{Bao, Huang, and Lin}{Bao
  et~al.}{2026}]{BaoHuangLin2024}
Bao, L., D.~Huang, and C.~Lin (2026).
\newblock {Can Artificial Intelligence Improve Gender Equality? Evidence from a
  Natural Experiment}.
\newblock {\em Management Science\/}~{\em 72\/}(1), 474--494.

\bibitem[\protect\citeauthoryear{Bapna, Langer, Mehra, Gopal, and Gupta}{Bapna
  et~al.}{2013}]{BapnaEtAl2012}
Bapna, R., N.~Langer, A.~Mehra, R.~Gopal, and A.~Gupta (2013).
\newblock {Human Capital Investments and Employee Performance: An Analysis of
  IT Services Industry}.
\newblock {\em Management Science\/}~{\em 59\/}(3), 641--658.

\bibitem[\protect\citeauthoryear{Bastani, Bastani, Sungu, Ge, Kabakc{\i}, and
  Mariman}{Bastani et~al.}{2025}]{BastaniEtAl2025}
Bastani, H., O.~Bastani, A.~Sungu, H.~Ge, {\"O}.~Kabakc{\i}, and R.~Mariman
  (2025).
\newblock {Generative AI Without Guardrails Can Harm Learning: Evidence from
  High School Mathematics}.
\newblock {\em Proceedings of the National Academy of Sciences\/}~{\em
  122\/}(26), e2422633122.

\bibitem[\protect\citeauthoryear{Beane}{Beane}{2019}]{Beane2018}
Beane, M. (2019).
\newblock {Shadow Learning: Building Robotic Surgical Skill When Approved Means
  Fail}.
\newblock {\em Administrative Science Quarterly\/}~{\em 64\/}(1), 87--123.

\bibitem[\protect\citeauthoryear{Belo, Ferreira, and Telang}{Belo
  et~al.}{2014}]{BeloFerreiraTelang2013}
Belo, R., P.~Ferreira, and R.~Telang (2014).
\newblock {Broadband in School: Impact on Student Performance}.
\newblock {\em Management Science\/}~{\em 60\/}(2), 265--282.

\bibitem[\protect\citeauthoryear{Ben-Porath}{Ben-Porath}{1967}]{BenPorath1967}
Ben-Porath, Y. (1967).
\newblock {The Production of Human Capital and the Life Cycle of Earnings}.
\newblock {\em Journal of Political Economy\/}~{\em 75\/}(4, Part 1), 352--365.

\bibitem[\protect\citeauthoryear{Boh, Slaughter, and Espinosa}{Boh
  et~al.}{2007}]{BohSlaughterEspinosa2007}
Boh, W.~F., S.~A. Slaughter, and J.~A. Espinosa (2007).
\newblock {Learning from Experience in Software Development: A Multilevel
  Analysis}.
\newblock {\em Management Science\/}~{\em 53\/}(8), 1315--1331.

\bibitem[\protect\citeauthoryear{Bondi and Johnson}{Bondi and
  Johnson}{2026}]{BondiJohnson2026}
Bondi, T. and G.~Johnson (2026).
\newblock {Skill Atrophy and AI Productivity Measurement}.
\newblock SSRN Working Paper 6169671, Social Science Research Network.

\bibitem[\protect\citeauthoryear{Borwankar, Khern-am nuai, Pocheptsova~Ghosh,
  and Kannan}{Borwankar et~al.}{2026}]{BorwankarEtAl2026}
Borwankar, S., W.~Khern-am nuai, A.~Pocheptsova~Ghosh, and K.~Kannan (2026).
\newblock {Unraveling the Impact: An Empirical Investigation of ChatGPT's
  Exclusion from Stack Overflow}.
\newblock {\em Information Systems Research\/}.
\newblock Articles in Advance.

\bibitem[\protect\citeauthoryear{Brynjolfsson, Li, and Raymond}{Brynjolfsson
  et~al.}{2025}]{BrynjolfssonLiRaymond2025}
Brynjolfsson, E., D.~Li, and L.~Raymond (2025).
\newblock {Generative AI at work}.
\newblock {\em Quarterly Journal of Economics\/}~{\em 140\/}(2), 889--942.

\bibitem[\protect\citeauthoryear{Budzy{\'n} et~al.}{Budzy{\'n}
  et~al.}{2025}]{BudzynEtAl2025}
Budzy{\'n}, K. et~al. (2025).
\newblock {Endoscopist deskilling risk after exposure to artificial
  intelligence in colonoscopy: a multicentre, observational study}.
\newblock {\em The Lancet Gastroenterology \& Hepatology\/}.

\bibitem[\protect\citeauthoryear{Chen and Chan}{Chen and
  Chan}{2024}]{ChenChan2024}
Chen, Z. and J.~Chan (2024).
\newblock {Large Language Model in Creative Work: The Role of Collaboration
  Modality and User Expertise}.
\newblock {\em Management Science\/}~{\em 70\/}(12), 9101--9117.

\bibitem[\protect\citeauthoryear{Cinelli and Hazlett}{Cinelli and
  Hazlett}{2020}]{CinelliHazlett2020}
Cinelli, C. and C.~Hazlett (2020).
\newblock {Making Sense of Sensitivity: Extending Omitted Variable Bias}.
\newblock {\em Journal of the Royal Statistical Society Series B\/}~{\em
  82\/}(1), 39--67.

\bibitem[\protect\citeauthoryear{Cowgill, Hern{\'a}ndez-Lagos, and
  Wright}{Cowgill et~al.}{2026}]{CowgillHernandezLagosWright2026}
Cowgill, B., P.~Hern{\'a}ndez-Lagos, and N.~L. Wright (2026).
\newblock {Does AI Cheapen Talk? Theory and Evidence from Global
  Entrepreneurship and Hiring}.
\newblock {\em Management Science\/}.
\newblock Articles in Advance.

\bibitem[\protect\citeauthoryear{Cui, Demirer, Jaffe, Musolff, Peng, and
  Salz}{Cui et~al.}{2026}]{CuiEtAl2026}
Cui, K.~Z., M.~Demirer, S.~Jaffe, L.~Musolff, S.~Peng, and T.~Salz (2026).
\newblock {The Effects of Generative AI on High-Skilled Work: Evidence from
  Three Field Experiments with Software Developers}.
\newblock {\em Management Science\/}.
\newblock Articles in Advance.

\bibitem[\protect\citeauthoryear{Cunha, Heckman, and Schennach}{Cunha
  et~al.}{2010}]{CunhaHeckmanSchennach2010}
Cunha, F., J.~J. Heckman, and S.~M. Schennach (2010).
\newblock {Estimating the Technology of Cognitive and Noncognitive Skill
  Formation}.
\newblock {\em Econometrica\/}~{\em 78\/}(3), 883--931.

\bibitem[\protect\citeauthoryear{Dell'Acqua}{Dell'Acqua}{2022}]{DellAcqua2022}
Dell'Acqua, F. (2022).
\newblock {Falling Asleep at the Wheel: Human/AI Collaboration in a Field
  Experiment on HR Recruiters}.
\newblock Working paper, laboratory for innovation science at harvard, Harvard
  Business School.

\bibitem[\protect\citeauthoryear{Dell'Acqua, McFowland, Mollick,
  Lifshitz-Assaf, Kellogg, Rajendran, Krayer, Candelon, and Lakhani}{Dell'Acqua
  et~al.}{2026}]{DellAcquaEtAl2026}
Dell'Acqua, F., E.~McFowland, E.~Mollick, H.~Lifshitz-Assaf, K.~Kellogg,
  S.~Rajendran, L.~Krayer, F.~Candelon, and K.~Lakhani (2026).
\newblock {Navigating the Jagged Technological Frontier: Field Experimental
  Evidence of the Effects of AI on Knowledge Worker Productivity and Quality}.
\newblock {\em Organization Science\/}~{\em 37\/}(2).
\newblock Earlier circulated as Harvard Business School Working Paper 24-013
  (2023).

\bibitem[\protect\citeauthoryear{Deming and Noray}{Deming and
  Noray}{2020}]{DemingNoray2020}
Deming, D.~J. and K.~Noray (2020).
\newblock {Earnings Dynamics, Changing Job Skills, and STEM Careers}.
\newblock {\em Quarterly Journal of Economics\/}~{\em 135\/}(4), 1965--2005.

\bibitem[\protect\citeauthoryear{Deng, Cheng, Ferreira, and Pavlou}{Deng
  et~al.}{2026}]{DengEtAl2025}
Deng, Z., Z.~A. Cheng, P.~Ferreira, and P.~A. Pavlou (2026).
\newblock {From Smartphones to Smart Students: Learning vs. Distraction Using
  Smartphones in the Classroom}.
\newblock {\em Information Systems Research\/}~{\em 37\/}(1), 276--293.

\bibitem[\protect\citeauthoryear{F{\"u}gener, Grahl, Gupta, and
  Ketter}{F{\"u}gener et~al.}{2021}]{FugenerEtAl2021Borgs}
F{\"u}gener, A., J.~Grahl, A.~Gupta, and W.~Ketter (2021).
\newblock {Will Humans-in-the-Loop Become Borgs? Merits and Pitfalls of Working
  with AI}.
\newblock {\em MIS Quarterly\/}~{\em 45\/}(3), 1527--1556.

\bibitem[\protect\citeauthoryear{F{\"u}gener, Grahl, Gupta, and
  Ketter}{F{\"u}gener et~al.}{2022}]{FugenerEtAl2021Delegation}
F{\"u}gener, A., J.~Grahl, A.~Gupta, and W.~Ketter (2022).
\newblock {Cognitive Challenges in Human--Artificial Intelligence
  Collaboration: Investigating the Path Toward Productive Delegation}.
\newblock {\em Information Systems Research\/}~{\em 33\/}(2), 678--696.

\bibitem[\protect\citeauthoryear{F{\"u}gener, Walzner, and Gupta}{F{\"u}gener
  et~al.}{2026}]{FugenerWalznerGupta2025}
F{\"u}gener, A., D.~D. Walzner, and A.~Gupta (2026).
\newblock {Roles of Artificial Intelligence in Collaboration with Humans:
  Automation, Augmentation, and the Future of Work}.
\newblock {\em Management Science\/}~{\em 72\/}(1), 538--557.

\bibitem[\protect\citeauthoryear{Gaessler and Piezunka}{Gaessler and
  Piezunka}{2023}]{Gaessler2023}
Gaessler, F. and H.~Piezunka (2023).
\newblock {Training with AI: Evidence from chess computers}.
\newblock {\em Strategic Management Journal\/}~{\em 44\/}(11), 2724--2750.

\bibitem[\protect\citeauthoryear{Galdin and Silbert}{Galdin and
  Silbert}{2025}]{GaldinSilbert2025}
Galdin, A. and J.~Silbert (2025).
\newblock {Making Talk Cheap: Generative AI and Labor Market Signaling}.
\newblock Working Paper 2511.08785, arXiv.

\bibitem[\protect\citeauthoryear{Hann, Roberts, and Slaughter}{Hann
  et~al.}{2013}]{HannRobertsSlaughter2013}
Hann, I.-H., J.~A. Roberts, and S.~A. Slaughter (2013).
\newblock {All Are Not Equal: An Examination of the Economic Returns to
  Different Forms of Participation in Open Source Software Communities}.
\newblock {\em Information Systems Research\/}~{\em 24\/}(3), 520--538.

\bibitem[\protect\citeauthoryear{Hicks}{Hicks}{1970}]{Hicks1970}
Hicks, J. (1970).
\newblock {Elasticity of Substitution Again: Substitutes and Complements}.
\newblock {\em Oxford Economic Papers\/}~{\em 22\/}(3), 289--296.

\bibitem[\protect\citeauthoryear{Horton and Tambe}{Horton and
  Tambe}{2025}]{HortonTambe2025}
Horton, J.~J. and P.~Tambe (2025).
\newblock {The Death of a Technical Skill}.
\newblock {\em Information Systems Research\/}~{\em 36\/}(3), 1799--1820.

\bibitem[\protect\citeauthoryear{Hou, Wang, Wang, Wang, and Yang}{Hou
  et~al.}{2025}]{HouEtAl2025}
Hou, J.~J., L.~Wang, G.~Wang, H.~J. Wang, and S.~Yang (2025).
\newblock {The Double-Edged Roles of Generative AI in the Creative Process:
  Experiments on Design Work}.
\newblock {\em Information Systems Research\/}.
\newblock Articles in Advance.

\bibitem[\protect\citeauthoryear{Jabarian and Reshidi}{Jabarian and
  Reshidi}{2025}]{JabarianReshidi2025}
Jabarian, B. and P.~Reshidi (2025).
\newblock {Choice as Signal: Designing AI Adoption in Labor Market Screening}.
\newblock SSRN Working Paper 5856382, Social Science Research Network.

\bibitem[\protect\citeauthoryear{Jussupow, Bauer, Heigl, Vogt, and
  Hinz}{Jussupow et~al.}{2026}]{JussupowEtAl2026}
Jussupow, E., K.~Bauer, R.~M. Heigl, B.~Vogt, and O.~Hinz (2026).
\newblock {The Indirect Disclosure Effect: How Disclosing Generative AI Use
  Impacts Human Creative Collaboration with AI}.
\newblock {\em Information Systems Research\/}.
\newblock Articles in Advance.

\bibitem[\protect\citeauthoryear{Jussupow, Spohrer, Heinzl, and
  Gawlitza}{Jussupow et~al.}{2021}]{JussupowEtAl2021}
Jussupow, E., K.~Spohrer, A.~Heinzl, and J.~Gawlitza (2021).
\newblock {Augmenting Medical Diagnosis Decisions? An Investigation into
  Physicians' Decision-Making Process with Artificial Intelligence}.
\newblock {\em Information Systems Research\/}~{\em 32\/}(3), 713--735.

\bibitem[\protect\citeauthoryear{Kanazawa, Kawaguchi, Shigeoka, and
  Watanabe}{Kanazawa et~al.}{2026}]{KanazawaEtAl2025}
Kanazawa, K., D.~Kawaguchi, H.~Shigeoka, and Y.~Watanabe (2026).
\newblock {AI, Skill, and Productivity: The Case of Taxi Drivers}.
\newblock {\em Management Science\/}~{\em 72\/}(2), 1376--1388.

\bibitem[\protect\citeauthoryear{Kim, Glaeser, Hillis, Kominers, and Luca}{Kim
  et~al.}{2024}]{KimEtAl2024}
Kim, H., E.~L. Glaeser, A.~Hillis, S.~D. Kominers, and M.~Luca (2024).
\newblock {Decision Authority and the Returns to Algorithms}.
\newblock {\em Strategic Management Journal\/}~{\em 45\/}(4), 619--648.

\bibitem[\protect\citeauthoryear{Kim, Mester, and Sun}{Kim
  et~al.}{2026}]{KimMesterSun2026}
Kim, K., N.~Mester, and G.~Sun (2026).
\newblock {AI and the Labor Market: A Worker's-Eye View}.
\newblock SSRN Working Paper 6240619, Social Science Research Network.

\bibitem[\protect\citeauthoryear{Knight, Mitrofanov, and Netessine}{Knight
  et~al.}{2026}]{KnightMitrofanovNetessine2026}
Knight, B., D.~Mitrofanov, and S.~Netessine (2026).
\newblock {Human-Algorithm Collaboration in Gig Work: The Role of Experience,
  Skill Level, and Task Complexity}.
\newblock {\em Information Systems Research\/}.
\newblock Articles in Advance.

\bibitem[\protect\citeauthoryear{Krakowski, Haftor, Luger, Pashkevich, and
  Raisch}{Krakowski et~al.}{2026}]{KrakowskiEtAl2025}
Krakowski, S., D.~Haftor, J.~Luger, N.~Pashkevich, and S.~Raisch (2026).
\newblock {Human-Centered Artificial Intelligence: A Field Experiment}.
\newblock {\em Management Science\/}~{\em 72\/}(1), 57--72.

\bibitem[\protect\citeauthoryear{Lee}{Lee}{2009}]{Lee2009}
Lee, D.~S. (2009).
\newblock {Training, Wages, and Sample Selection: Estimating Sharp Bounds on
  Treatment Effects}.
\newblock {\em Review of Economic Studies\/}~{\em 76\/}(3), 1071--1102.

\bibitem[\protect\citeauthoryear{Melumad and Yun}{Melumad and
  Yun}{2025}]{MelumadYun2025}
Melumad, S. and J.~H. Yun (2025).
\newblock {Experimental evidence of the effects of large language models versus
  web search on depth of learning}.
\newblock {\em PNAS Nexus\/}~{\em 4\/}(10), pgaf316.

\bibitem[\protect\citeauthoryear{Mirzayanov}{Mirzayanov}{2015}]{Mirzayanov2015}
Mirzayanov, M. (2015, October).
\newblock {Open Codeforces Rating System}.
\newblock Codeforces blog entry 20762.

\bibitem[\protect\citeauthoryear{{MIT Ad Hoc Committee}}{{MIT Ad Hoc
  Committee}}{2026}]{MITAICommittee2026}
{MIT Ad Hoc Committee} (2026, August).
\newblock {Report of MIT's Ad Hoc Committee on AI Use in Teaching, Learning,
  and Research Training}.
\newblock Technical report, Massachusetts Institute of Technology.
\newblock MIT Ad Hoc Committee on AI Use in Teaching, Learning, and Research
  Training; co-chairs Eric Klopfer and Sam Madden; dated August 13, 2026.

\bibitem[\protect\citeauthoryear{Narayanan, Balasubramanian, and
  Swaminathan}{Narayanan
  et~al.}{2009}]{NarayananBalasubramanianSwaminathan2009}
Narayanan, S., S.~Balasubramanian, and J.~M. Swaminathan (2009).
\newblock {A Matter of Balance: Specialization, Task Variety, and Individual
  Learning in a Software Maintenance Environment}.
\newblock {\em Management Science\/}~{\em 55\/}(11), 1861--1876.

\bibitem[\protect\citeauthoryear{Noy and Zhang}{Noy and
  Zhang}{2023}]{NoyZhang2023}
Noy, S. and W.~Zhang (2023).
\newblock {Experimental evidence on the productivity effects of generative
  artificial intelligence}.
\newblock {\em Science\/}~{\em 381\/}(6654), 187--192.

\bibitem[\protect\citeauthoryear{Oster}{Oster}{2019}]{Oster2019}
Oster, E. (2019).
\newblock {Unobservable Selection and Coefficient Stability: Theory and
  Evidence}.
\newblock {\em Journal of Business \& Economic Statistics\/}~{\em 37\/}(2),
  187--204.

\bibitem[\protect\citeauthoryear{Peng, Kalliamvakou, Cihon, and Demirer}{Peng
  et~al.}{2023}]{PengEtAl2023}
Peng, S., E.~Kalliamvakou, P.~Cihon, and M.~Demirer (2023).
\newblock {The Impact of AI on Developer Productivity: Evidence from GitHub
  Copilot}.
\newblock {\em arXiv preprint\/}.

\bibitem[\protect\citeauthoryear{Piccoli, Ahmad, and Ives}{Piccoli
  et~al.}{2001}]{PiccoliAhmadIves2001}
Piccoli, G., R.~Ahmad, and B.~Ives (2001).
\newblock {Web-Based Virtual Learning Environments: A Research Framework and a
  Preliminary Assessment of Effectiveness in Basic IT Skills Training}.
\newblock {\em MIS Quarterly\/}~{\em 25\/}(4), 401--426.

\bibitem[\protect\citeauthoryear{Rinta-Kahila, Penttinen, Salovaara, Soliman,
  and Ruissalo}{Rinta-Kahila et~al.}{2023}]{RintaKahilaEtAl2023}
Rinta-Kahila, T., E.~Penttinen, A.~Salovaara, W.~Soliman, and J.~Ruissalo
  (2023).
\newblock {The Vicious Circles of Skill Erosion: A Case Study of Cognitive
  Automation}.
\newblock {\em Journal of the Association for Information Systems\/}~{\em
  24\/}(5), 1378--1412.

\bibitem[\protect\citeauthoryear{Santhanam, Sasidharan, and Webster}{Santhanam
  et~al.}{2008}]{SanthanamSasidharanWebster2008}
Santhanam, R., S.~Sasidharan, and J.~Webster (2008).
\newblock {Using Self-Regulatory Learning to Enhance E-Learning-Based
  Information Technology Training}.
\newblock {\em Information Systems Research\/}~{\em 19\/}(1), 26--47.

\bibitem[\protect\citeauthoryear{Shan and Qiu}{Shan and
  Qiu}{2026}]{ShanQiu2025}
Shan, G. and L.~Qiu (2026).
\newblock {Examining the Impact of Generative AI on Users' Voluntary Knowledge
  Contribution: Evidence from a Natural Experiment on Stack Overflow}.
\newblock {\em Information Systems Research\/}~{\em 37\/}(2), 1021--1041.

\bibitem[\protect\citeauthoryear{Shaw and Nave}{Shaw and
  Nave}{2026}]{ShawNave2026}
Shaw, S.~D. and G.~Nave (2026).
\newblock {Thinking---Fast, Slow, and Artificial: How AI is Reshaping Human
  Reasoning and the Rise of Cognitive Surrender}.
\newblock Working paper, The Wharton School, University of Pennsylvania.

\bibitem[\protect\citeauthoryear{Shen and Tamkin}{Shen and
  Tamkin}{2026}]{ShenTamkin2026}
Shen, J.~H. and A.~Tamkin (2026).
\newblock {How AI Impacts Skill Formation}.
\newblock {\em arXiv preprint\/}.

\bibitem[\protect\citeauthoryear{Shihab, Hundhausen, Tariq, Haque, Qiao, and
  Mulanda}{Shihab et~al.}{2025}]{ShihabEtAl2025}
Shihab, M. I.~H., C.~Hundhausen, A.~Tariq, S.~Haque, Y.~Qiao, and B.~Mulanda
  (2025).
\newblock {The Effects of GitHub Copilot on Computing Students' Programming
  Effectiveness, Efficiency, and Processes in Brownfield Programming Tasks}.
\newblock In {\em Proceedings of the 2025 ACM Conference on International
  Computing Education Research (ICER 2025)}.

\bibitem[\protect\citeauthoryear{Singh, Tan, and Youn}{Singh
  et~al.}{2011}]{SinghTanYoun2010}
Singh, P.~V., Y.~Tan, and N.~Youn (2011).
\newblock {A Hidden Markov Model of Developer Learning Dynamics in Open Source
  Software Projects}.
\newblock {\em Information Systems Research\/}~{\em 22\/}(4), 790--807.

\bibitem[\protect\citeauthoryear{Song, Agarwal, and Wen}{Song
  et~al.}{2026}]{SongAgarwalWen2026}
Song, F., A.~Agarwal, and W.~Wen (2026).
\newblock {The Impact of Generative AI on Collaborative Open-Source Software
  Development: Evidence from GitHub Copilot}.
\newblock {\em Information Systems Research\/}.
\newblock Articles in Advance.

\bibitem[\protect\citeauthoryear{Spence}{Spence}{1973}]{Spence1973}
Spence, M. (1973).
\newblock {Job Market Signaling}.
\newblock {\em Quarterly Journal of Economics\/}~{\em 87\/}(3), 355--374.

\bibitem[\protect\citeauthoryear{Stadler, Bannert, and Sailer}{Stadler
  et~al.}{2024}]{StadlerBannertSailer2024}
Stadler, M., M.~Bannert, and M.~Sailer (2024).
\newblock {Cognitive ease at a cost: LLMs reduce mental effort but compromise
  depth in student scientific inquiry}.
\newblock {\em Computers in Human Behavior\/}~{\em 160}, 108386.

\bibitem[\protect\citeauthoryear{Stiglitz}{Stiglitz}{1975}]{Stiglitz1975}
Stiglitz, J.~E. (1975).
\newblock {The Theory of ``Screening,'' Education, and the Distribution of
  Income}.
\newblock {\em American Economic Review\/}~{\em 65\/}(3), 283--300.

\bibitem[\protect\citeauthoryear{Sundaresan and G{\"u}ler}{Sundaresan and
  G{\"u}ler}{2025}]{SundaresanGuler2025}
Sundaresan, S. and I.~G{\"u}ler (2025).
\newblock {Algorithmic Recommendation Tools and Experiential Learning in
  Clinical Care}.
\newblock {\em Organization Science\/}~{\em 36\/}(5), 1786--1802.

\bibitem[\protect\citeauthoryear{Tambe, Ye, and Cappelli}{Tambe
  et~al.}{2020}]{TambeYeCappelli2020}
Tambe, P., X.~Ye, and P.~Cappelli (2020).
\newblock {Paying to Program? Engineering Brand and High-Tech Wages}.
\newblock {\em Management Science\/}~{\em 66\/}(7), 3010--3028.

\bibitem[\protect\citeauthoryear{Wambsgan{\ss}, Janson, S{\"o}llner, Koedinger,
  and Leimeister}{Wambsgan{\ss} et~al.}{2025}]{WambsganssEtAl2024}
Wambsgan{\ss}, T., A.~Janson, M.~S{\"o}llner, K.~Koedinger, and J.~M.
  Leimeister (2025).
\newblock {Improving Students' Argumentation Skills Using Dynamic
  Machine-Learning--Based Modeling}.
\newblock {\em Information Systems Research\/}~{\em 36\/}(1), 474--507.

\bibitem[\protect\citeauthoryear{Wang and Lee}{Wang and
  Lee}{2026}]{WangLee2026}
Wang, L. and H.~C.~B. Lee (2026).
\newblock {How Costs Influence Preferences for Control in Generative Artificial
  Intelligence (GenAI): Human-Guided vs. GenAI-Based Delegated Search}.
\newblock {\em Information Systems Research\/}.
\newblock Articles in Advance.

\bibitem[\protect\citeauthoryear{Wiles, Munyikwa, and Horton}{Wiles
  et~al.}{2025}]{WilesMunyikwaHorton2025}
Wiles, E., Z.~Munyikwa, and J.~J. Horton (2025).
\newblock {Algorithmic Writing Assistance on Jobseekers' Resumes Increases
  Hires}.
\newblock {\em Management Science\/}~{\em 71\/}(12), 10144--10164.

\bibitem[\protect\citeauthoryear{Yao}{Yao}{2026}]{Yao2026Stranded}
Yao, S. (2026).
\newblock {Stranded Credentials: Keeping Online Reputation Systems Informative
  in the AI Era}.
\newblock {\em arXiv preprint\/}.
\newblock v2, September 2026.

\bibitem[\protect\citeauthoryear{Yin, Ngiam, Tan, and Teo}{Yin
  et~al.}{2025}]{YinEtAl2025}
Yin, J., K.~Y. Ngiam, S.~S.-L. Tan, and H.~H. Teo (2025).
\newblock {Designing AI-Based Work Processes: How the Timing of AI Advice
  Affects Diagnostic Decision Making}.
\newblock {\em Management Science\/}~{\em 71\/}(11), 9361--9383.

\bibitem[\protect\citeauthoryear{Zhang, Singh, and Ghose}{Zhang
  et~al.}{2019}]{ZhangSinghGhose2019}
Zhang, S., P.~V. Singh, and A.~Ghose (2019).
\newblock {A Structural Analysis of the Role of Superstars in Crowdsourcing
  Contests}.
\newblock {\em Information Systems Research\/}~{\em 30\/}(1), 15--33.

\end{thebibliography}
